\documentclass{aa}

\usepackage[varg]{txfonts} 
\usepackage{graphicx}
\usepackage{txfonts}
\usepackage{natbib}
\bibpunct{(}{)}{;}{a}{}{,} 
\usepackage[colorlinks,urlcolor=cyan,citecolor=blue,linkcolor=blue]{hyperref}
\usepackage{amssymb,amsmath}
\usepackage{array}
\usepackage{booktabs}

\usepackage{makecell}
\usepackage{csvsimple}
\usepackage{url}
\usepackage{subfigure}
\usepackage{siunitx}
\usepackage[x11names]{xcolor}
\usepackage{soul}

\usepackage{placeins}

\begin{document} 

\title{Impact of climate change on site characteristics of  eight major astronomical observatories using high-resolution global climate projections until 2050\\}
\subtitle{Projected increase in temperature and humidity leads to \\poorer astronomical observing conditions}

\author{ 
C.~Haslebacher\inst{1}
\and M.-E.~Demory\inst{2,6}
\and B.-O.~Demory\inst{3}
\and M.~Sarazin\inst{4}
\and P. L.~Vidale\inst{5}
}

\institute{Center for Space and Habitability and Department of Space Research \& Planetary Sciences, University of Bern, 3012, Bern, Switzerland \\\email{caroline.haslebacher@unibe.ch}
\and Institute for Atmospheric and Climate Science, ETH Zurich, 8092 Zurich, Switzerland\\\email{marie-estelle.demory@env.ethz.ch}
\and Center for Space and Habitability, University of Bern, 3012, Bern, Switzerland\\\email{brice.demory@unibe.ch}
\and European Southern Observatory, Garching, Germany
\and NCAS-Climate, Department of Meteorology, University of Reading, Reading, UK
\and Wyss Academy for Nature, University of Bern, 3012 Bern, Switzerland 
}

\titlerunning{Impact of climate change on site characteristics of eight major astronomical observatories}\authorrunning{C. Haslebacher et al.}


\abstract {
Sites for next-generation telescopes are chosen decades before the first light of a telescope. Site selection is usually based on recent measurements over a period that is too short to account for long-term changes in observing conditions such as those arising from anthropogenic climate change. For astronomical facilities with a typical lifetime of 30 years, it is therefore essential to be aware of climate evolution to optimise observing time.} 
{
In this study, we analyse trends in astronomical observing conditions for eight sites. Most sites either already host telescopes that provide in situ measurements of weather parameters or are candidates for hosting next-generation telescopes. For a fine representation of orography, we use the highest resolution global climate model (GCM) ensemble available provided by the high-resolution model intercomparison project and developed as part of the European Union Horizon 2020 PRIMAVERA project.} 
{
We evaluate atmosphere-only and coupled PRIMAVERA GCM historical simulations against in situ measurements and the fifth generation atmospheric reanalysis (ERA5) of the European centre for medium-range weather forecasts for the period 1979-2014. The projections of changes in current site conditions are then analysed for the period 2015-2050 using PRIMAVERA future climate simulations.} 
{
Over most sites, we find that PRIMAVERA GCMs show a good agreement in temperature, specific humidity, and precipitable water vapour compared to in situ observations and ERA5. The ability of PRIMAVERA to simulate those variables increases confidence in their projections. For those variables, the model ensemble projects an increasing trend for all sites, which will result in progressively poorer astronomical observing conditions compared to current conditions. 
On the other hand, no significant trends are projected for relative humidity, cloud cover, or astronomical seeing and PRIMAVERA does not simulate these variables well compared to observations and reanalyses. Therefore, there is little confidence in these projections.} 
{
Our results show that climate change will negatively impact the quality of astronomical observations and is likely to increase time lost due to bad site conditions. We stress that it is essential for astronomers to include long-term climate projections in their process for site selection and monitoring.
We show that high-resolution GCMs can be used to analyse the effect of climate change on site characteristics of next-generation telescopes.}

\keywords{Site testing - Telescopes - Earth - Methods: data analysis - Astronomical databases: miscellaneous}

\maketitle
\section{Introduction}\label{sec:intro}
Sites for ground-based telescopes are selected considering a wide variety of aspects, such as climatic conditions, proximity to other astronomical facilities, prevalent infrastructure, cultural environment, accessibility to resources, and economic implications \citep[e.g.][]{Schoeck2009}. Since prevailing climate conditions significantly influence the quality of astronomical observations made by ground-based telescopes \citep[e.g.][]{Falvey2016}, site selection studies are designed to find places with the most advantageous climate conditions.
The most important atmospheric parameters measured in a site selection process include astronomical seeing, cloud cover, precipitable water vapour (PWV), air temperature, and air humidity \citep{Schoeck2009, Vernin2011}. The process of building a next-generation telescope, starting with a site selection study and ending with the first light of a telescope, can take more than a decade. To give two examples, the whole process takes 20 years in the case of the European extremely large telescope (ELT; \citealp{ESOd}) and approximately 27 years for the thirty meter telescope (TMT; \citealp{Schoeck2009})\footnote{\url{https://www.tmt.org/page/timeline}, accessed on 2019 Apr 17}. If we approximate the typical lifetime of these telescopes for 30 years\footnote{\url{https://elt.eso.org/about/faq/}, accessed on 2021 Apr 23}, we implicitly expect the winner of the site selection study to stay the winner for over five decades, a timescale for which climate change might already play a role. 
 Therefore, every site survey brings the risk that site conditions measured during the survey do not reliably represent the long-term climate conditions \citep{Schoeck2009}. Astronomers are already worried about changes due to anthropogenic climate change and see a need to act \citep{Burtscher2021}.\footnote{Also see the open letter of Astronomers for Planet Earth: \url{https://astronomersforplanet.earth/open-letter/}, accessed on 2022 Jun 14}

The global mean surface temperature has already increased by 1.09 $\substack{+0.11\\-0.14}$~$^{\circ} C$ from 2011-2020 compared to the period from 1850-1900 \citep{IPCC2021}. 
Already now, anthropogenic climate change has an effect on extreme weather and climate events everywhere on Earth \citep{IPCC2021}.
It is a fact that anthropogenic green house gas emissions drive the rapid increase in global mean surface temperature \citep{IPCC2021}.

Studies considering the impact of climate change on site characteristics of ground-based telescopes started to emerge only recently. \citet{Hellemeier2019} analysed the site characteristics' cloud cover, 200-hPa-wind-speed, PWV, vertical wind velocity, and aerosol index of 15 well-established astronomical sites using the FriOWL database \citep{Graham2008}. The FriOWL database combines results from two reanalysis data sets: the ERA-40 reanalysis of the European centre for medium-range weather forecasts (ECMWF) covering the period from 1958-2002 \citep{Kallberg2004}, and the national centers for environmental prediction/national center for atmospheric research reanalysis covering the period since 1948 \citep{Kalnay1996}. A reanalysis assimilates archived observational data into a continuous weather simulation and is therefore a valuable tool for studying past climatic changes. 
To explain trends in long-term site conditions, especially in cloud cover, PWV, and 200-hPa wind speed, \citet{Hellemeier2019} explored the possibility of anthropogenic climate change as a cause.
However, the low horizontal resolution of ERA-40 of approximately 278 km limits the reliability of this study.

The first study, to the best of our knowledge, to look explicitly at the influence of climate change on ground-based telescopes was done by \citet{Cantalloube2020}. \citet{Cantalloube2020} conducted a study for Cerro Paranal on the impact of climate change on astronomical observations using reanalysis data from the fifth generation atmospheric reanalysis (ERA5; \citealp{Hersbach2020}) and the first atmospheric reanalysis of the 20th century (ERA-20C; \citealp{cisl_rda_ds626.0}) of ECMWF, and one global climate model (GCM) projection from the coupled model intercomparison project phase 6 (CMIP6), namely the Beijing climate centre BCC-CSM2-MR CMIP6 GCM \citep{Wu2019}.
A global climate model (also denoted as a general circulation model) simulates physical processes in subsystems of Earth such as the atmosphere or the ocean on a three-dimensional grid. 
They found an increase in near-surface air temperature on Cerro Paranal of 1.5~$^{\circ}$C between 1980 and 2020 assimilated by ERA5 and an increase of 4~$^{\circ}$C by 2100 projected by the CMIP6 GCM under the shared socioeconomic pathway SSP5-8.5 scenario \citep{Kriegler2017, Riahi2017}. The SSP5-8.5 scenario combines the worst-case scenario of five shared socioeconomic pathways (SSPs), which describe different possible developments of societal, demographic, and economic changes, with the worst-case representative concentration pathway (RCP) 8.5 of four RCPs defining the forcing in 2100 from 2.6 to 8.5~W/m$^2$.  \citet{Cantalloube2020} state that such an increase in temperature brings the active temperature control system of the four unit telescopes of the very large telescope (VLT) on Cerro Paranal to its limits more often, which would enhance dome seeing (\citealp{Tallis2020}; atmospheric turbulence inside the dome due to temperature differences inside and outside the dome).
\citet{Cantalloube2020} also analysed seeing measurements from 1986 to 2020 and found a positive trend in seeing due to an increased surface layer turbulence, either caused by climate change or constructional modifications and alterations of measurement devices and measurement locations. Higher temperature gradients, potentially arising from surface temperature increases, do indeed enhance convection and therefore turbulence close to the ground. Furthermore, \citet{Cantalloube2020} found indications that the area around Cerro Paranal may become drier in the future, although this result is uncertain due to the use of one GCM only and the coarse horizontal resolution of 110 km used by the BCC-CSM2-MR GCM \citep{Wu2019}.

Another study using ERA5 was done by \citet{Osborn2018a}. 
 \citet{Osborn2018a} find that ERA5 model data on 137 vertical pressure levels show strong agreement with the high-altitude resolution optical turbulence profiler, and their model is able to predict strong turbulence events, which is crucial because strong turbulence dominates instrument performance limitations of adaptive optics systems. Adaptive optics systems are used to correct the light wavefront distortions induced by the atmospheric turbulence at optical and infrared wavelengths in near-real time, which is an integral part of future 30-m-class telescopes.

Considering the typical timeline of next-generation telescopes from site selection to the end of lifetime, site selection processes as well as current telescopes  might need to address the challenges of changing climate conditions. 
As suggested by \citet{Graham2004} and already shown by \citet{Cantalloube2020}, climate models that project scenarios of anthropogenic radiative forcing into the future may prove useful.  
To address particular topographic textures as they often appear at astronomical sites, GCMs with the highest possible horizontal resolution are advantageous \citep{Moreno-Chamarro2022}. Even if the use of regional climate models (RCMs) would allow for even higher resolution, RCMs do not provide global coverage or global consistency. With RCM ensembles, such as the coordinated regional climate downscaling experiment (CORDEX; \citealp{Giorgi2009}), only a few sites would be covered by the computational domains.
Furthermore, it is indispensable to validate the agreement between model output and measurements for astronomical sites, especially since well-established astronomical sites are often located on highly elevated mountains that are not particularly well represented in climate models. 

In this study, we aim to assess the reliability of using global climate models to project future changes in site characteristics for astronomical observations.
To do so, we make use of six GCMs from the PRIMAVERA project \citep{Roberts2018}, which is the European contribution to the high-resolution model intercomparison project (HighResMIP; \citealp{Haarsma2016}).
We analyse the PRIMAVERA GCM simulations against in situ measurements and the ERA5 reanalysis product for eight well-established sites of major astronomical observatories around the world. Reanalysis products bridge the gap between measurements and global climate models because they provide homogeneous information across the globe. Here, ERA5 is chosen because it is the latest reanalysis product of ECMWF; it provides information over a long period and it has a spatial resolution similar to the PRIMAVERA GCMs. The evaluation of ERA5 against in situ measurements is therefore a necessary step for a careful evaluation of PRIMAVERA GCMs. 

The investigated sites are Mauna Kea on the island of Hawaii (USA), San Pedro M\'{a}rtir in Baja California in Mexico, the three Chilean sites Cerro Paranal, Cerro Tololo, and La Silla, La Palma on the Canary Islands (Spain), Sutherland in South Africa, and Siding Spring in Australia. 
We investigate air humidity, air temperature, PWV, cloudiness, and astronomical seeing parameters. After evaluating how well ERA5 and PRIMAVERA represent the present climate conditions of the eight sites, we use PRIMAVERA GCMs to project future conditions at these sites up to the year 2050. We focus on trends of mean changes and do not consider changes in climate variability or changes in the extremes, which are also to be expected \citep{IPCC2021}.

In Sect. \ref{sec:atmospheric_variables} we introduce the atmospheric variables we analysed. In Sect. \ref{sec:sites} we introduce each site in detail. In Sect. \ref{sec:Data} we describe in situ measurements, the ERA5 reanalyses, and the PRIMAVERA GCMs. In Sect. \ref{sec:methods} we present the general workflow for evaluating ERA5 and PRIMAVERA and the method employed for the trend analysis. 
In Sect. \ref{sec:results_ERA5_and_PRIMAVERA_evaluation} we discuss the results of the comparison between in situ measurements, ERA5, and PRIMAVERA GCMs and in Sect. \ref{subsubsection:Ensemble_Projections} the results of future trends projected by PRIMAVERA. The last Sect. \ref{sec:concl} concludes the study.
 
 \section{Atmospheric variables}\label{sec:atmospheric_variables}

 \subsection{Air temperature}
Air temperature influences the seeing \citep{Cavazzani2012}, specific humidity and with it, PWV and clouds \citep{McIlveen1992}. Air temperature is the easiest atmospheric variable to measure \citep{McIlveen1992}. Archived temperature data spanning a long time range are often provided by weather stations and long time series are available at observatories, often measured by a weather tower 2~m to 10~m above ground and sometimes enclosed in an instrument shelter. It is also the best modelled variable so far, and the most trustworthy for temperature projections \citep{IPCC2021}.
GCMs experimental data submitted to the CMIP6 intercomparison enabled the Intergovernmental Panel on Climate Change (IPCC) to state that anthropogenic carbon emissions driving climate change will cause the global mean surface temperature to rise at least until 2050 under all emission scenarios \citep{IPCC2021}. Under the highest green house gas (GHG) emission scenario (SSP5-8.5), the global surface temperature at the end of the century (2081-2100) is projected to be 3.3~$^{\circ}$ to 5.7~$^{\circ}$ higher compared to the average between 1850-1900.

\subsection{Air humidity}
For astronomical observatories, the relative humidity is an important parameter because whenever the relative humidity is too high, a telescope has to close its dome to protect sensitive parts such as the optical system or electrical wires from moisture, which could lead to oxidation or leaching \citep{Bradley2006}. Therefore, the relative humidity is often measured by observatories. As for the temperature, this is most often done with a weather tower that measures 2~m above the ground \citep[e.g.][]{Schoeck2009}. 

The relative humidity $RH$ depicts the ratio of the specific humidity $q$ to the saturation specific humidity $q_s$ in air for a specified temperature and pressure \citep{Peixoto1992}
\begin{equation} \label{eq:rel_hum_with_qs}
RH = 100\%\dfrac{q}{q_s} = 100\%\dfrac{e}{e_s},
\end{equation}
which is the same as the ratio of the water vapour pressure $e$ to the saturated vapour pressure $e_s$. The higher the values of relative humidity, the more saturated the air is, and saturation is defined as a relative humidity value of 100~\%.

The specific humidity $q$ is defined as the ratio of the water vapour mass $M_v$ of a specified volume of air to the total mass $M_{tot}$ of the same volume of air \citep{Peixoto1992} 
\begin{equation}
q = \dfrac{M_v}{M_{tot}}.
\end{equation}

Since at least 1970, there has been an increasing trend of near-surface specific humidity over land and ocean that can be attributed with medium confidence to human activities \citep{IPCC2021}. 
Since the temperature over land increases faster than the temperature above the ocean, near-surface relative humidity is expected to decrease over land, which has been observed since the year 2000 \citep{IPCC2021}. The result is regional drying due to increased evapotranspiration.

\subsection{Precipitable water vapour}
The PWV is the amount of water vapour in the atmosphere integrated over a unit column from the surface to the top of the atmosphere \citep{Peixoto1992}.  For the PWV, we integrate until the tropopause, because most weather events happen in the troposphere and most of the water is contained inside the troposphere.
Put another way, the PWV is the mass of liquid water if all the water vapour in a unit column of the atmosphere were condensed \citep{Peixoto1992}. It is given by 
\begin{equation}\label{Equation: PWV theory}
W(\theta, \phi, t) = \int_{0}^{p_0} q\dfrac{dp}{g},
\end{equation}
where $q$ is the specific humidity, $g$ is the gravitational acceleration, $p$ is the pressure and $p_0$ is the surface pressure. The PWV depends on the longitude $\theta$, the latitude $\phi$ and the time $t$ at which it is observed.

The PWV pertains to the important site characteristics of telescopes observing in the visible and infrared because water absorbs light over a wide range of wavelengths with an absorption maximum in the infrared. Not only does the light from an observed astronomical object get absorbed, but also thermal emissions of Earth itself can get backscattered into the telescope. Both contributions fluctuate in time due to a turbulent atmosphere. One example of an astronomical observation that depends on low PWV values is direct imaging, where the mid-infrared band is used preferentially because the contrast in flux between planet and star is optimal in the mid-infrared \citep{Pathak2021, Bessell2005}.
The temporal variation of PWV is in general relatively large due to evaporation and precipitation \citep[e.g.][]{Kalinnikov2017, Otarola2019}. For example at Paranal observatory on the timescale of an hour, the PWV fluctuates between 1~\% and 5.5~\% around the mean when measured in 120~s time intervals \citep{Otarola2019}.

The PWV content can be measured, for example with radiosondes, or it can be estimated, for example from measurements with an infrared radiometer that observes at 20~$\mu$m wavelength \citep[e.g.][]{Chapman2004}, with opacity measurements at 225 GHz \citep[e.g.][]{Dempsey2013}, or with 183 GHz radiometers \citep[e.g.][]{Kerber, Wiedner2001}. These measurements can be used by observatories with telescopes operating in the (sub)millimetre range to subtract spectra due to water vapour \citep{Chapman2004}.

PWV is expected to rise globally with increasing global mean surface temperature in the future \citep{Held2006, Trenberth2005}. This is backed up by an already observed increasing trend in total column water \citep{IPCC2021}.

\subsection{Cloud cover}\label{subsec:cloudcover}
Astronomical observations in the visible and infrared can only be carried out when the sky is clear, with as little cloudiness as possible. According to the European Southern Observatory (ESO; \citealp[]{Kerber2014}), a photometric night is defined by no visible clouds and transparency variations of the atmosphere below 2~\%, a clear sky is defined as a sky with less than 10~\% cloud cover and transparency variations below 10~\% and thin cirrus is the term for transparency variations above 10~\%. Methods were developed to assess the cloud coverage with infrared all-sky cameras that are sensitive to very thin clouds but not deceived by variations of PWV \citep{Kerber2014}.

Even though evaporation and precipitation must be in equilibrium globally, precipitation exceeds evaporation for the equatorial belt, where many clouds form, due to strong convection, and evaporation exceeds precipitation in the subtropical latitudes due to weak upward motion, which leads to a minimum in cloudiness
 \citep{Peixoto1992}. A maximum in cloud cover is found at latitudes near 50 - 60~$^{\circ}$ \citep{Peixoto1992}.
 
In GCMs, the dynamics of convection and cloud formation usually cannot be simulated explicitly because their horizontal resolution is too low. A horizontal grid spacing of a few kilometres would be necessary, whereas GCMs contributing to CMIP6 have a median horizontal atmospheric resolution of ca. 140~km \citep{IPCC2021}. 
GCMs therefore have to rely on the use of semi-empirical parameterisation schemes that generate large uncertainties in the simulation of such processes associated with cloud cover \citep[][and references therein]{Bony2015} 

Cloud interactions with the global climate will likely further amplify anthropogenic warming \citep{IPCC2021}. However, clouds are the most uncertain parameter in the assessment of global climate feedback, which gets more complicated due to a lack in the understanding of cloud forming processes and also because clouds need to be parameterised in GCMs due to the too low resolution of GCMs \citep{IPCC2021}.

 \subsection{Seeing}\label{subsec:Seeing}
The distortion of light wavefronts due to atmospheric turbulence is a well-studied issue in optical astronomy \citep{Roddier1981} and is caused by a variation in refractive index of the atmosphere along the line of sight due to temperature and density gradients. The Gladstone-Dale relation, 
\begin{equation}\label{eq:gladstone-dale}
\dfrac{n-1}{\rho} = \text{const}
\end{equation}
relates the refractive index $n$ to the density $\rho$ of a medium \citep{Gladstone1863}. In the framework of the Kolmogorov model \citep{Tatarski1961}, the local power spectrum of a well developed turbulence can be characterised by a single index of refraction structure coefficient $C_n^2$. The effect of the variations of $C_n^2$ along the line of sight is discussed in detail in Sect. \ref{sec:formulas_seeing}.
The alterations of a plane wavefront travelling through this turbulence, integrated along the line of sight, can be described by their coherence radius $r_0$ \citep{Fried1966}, for a wavelength $\lambda$ at zenith:
\begin{equation}\label{eq:Fried-parameter}
    r_0 = \left( 0.423 \cdot \int C_n^2(z) dz \cdot \left(\dfrac{2 \pi}{\lambda} \right)^2 \right)^{-3/5}.
\end{equation}
This 'Fried parameter' $r_0$ sets the sharpness of the image of a point source through a telescope limited by the atmosphere, usually described by its full width at half maximum (FWHM) in radian:
\begin{equation}\label{eq:seeing_FWHM}
    \epsilon = 0.976 \cdot \dfrac{\lambda}{r_0}.
\end{equation}
Astronomers characterise the atmospheric observing conditions by the so-called seeing, defined as the FWHM at a wavelength of $\lambda = 500$~nm expressed in arcseconds. 

The equivalence between seeing and image quality is only valid within the Kolmogorov turbulence model, which does not set upper limits to the scale of the turbulence in the inertial range. 
With the increase of the size of modern ground based telescopes, departures from this model were observed in the infrared on images sharper than predicted by the seeing. Obviously due to the deficit of large tilts within the wavefront, the observations could be explained by the introduction of an outer scale $L_0$ of the turbulence. The general Eq. \ref{eq:seeing_FWHM} is however suitable for studying climate impact on optical turbulence precisely because it is independent of the outer scale, which would depend on the telescope diameter. 
Deformable mirrors that can be adapted in real time to wave distortions due to atmospheric turbulence (adaptive optics), with which it is possible to get close to the diffraction limit
of the telescope, are nowadays part of a standard instrumentation for ground-based telescopes \citep{Davies2012}. The performance of adaptive optics correction depends on the instrument attached to the telescope (e.g. spectrograph, coronagraph) and on spatiotemporal properties of the wavefront.
 
 Since the seeing quality can be, depending on the orography, better on a mountain due to a higher chance of unperturbed laminar airflow on the elevated level of the mountain, many observatories are, despite other advantages, situated on mountains.
 Because impactors such as cars, trees or buildings have a significant effect on the seeing \citep[e.g.][]{Teare2000} but are not represented in GCMs, we do not expect GCMs to represent absolute seeing values realistically. However, we verify in this study whether GCMs are able to simulate the free atmosphere seeing.
 The free atmosphere seeing excludes the contribution from the ground and planetary boundary layer and starts at approximately 1-2 km above observatory level \citep{Osborn2018a}. The free atmosphere seeing therefore includes the tropopause level at 200~hPa, where, dependent on the latitude, a jet stream might prevail. At around 30~$^{\circ}$ latitude is the subtropical jet stream, so all sites in this study are influenced by the subtropical jet stream. For example in La Silla and Mauna Kea, the jet stream at 8 - 12 km altitude was found to be the dominant contribution to the free atmosphere seeing \citep{Vernin1986}. The lower the wind speed at 200 hPa, the better the seeing \citep{Vernin1986}, and seeing below 1 arcsecond requires wind speeds below 20~m~s$^{-1}$ at the tropopause. Trends in the strength of the subtropical jet stream could influence the free atmosphere seeing. In this context, the winter subtropical jet stream has accelerated and shifted polewards over the period 1979-2008 and extended regions in comparison to the period 1958-2008 \citep{Pena-Ortiz2013}. Ensemble means of future simulations project a strengthening of the southern mid-latitude jet stream under SSP5-8.5, while trends for the northern jet stream wind speeds are uncertain due to large natural internal variability, opposing changes in temperature gradients and identified simulation weaknesses \citep{IPCC2021}.

 Astronomical seeing is mainly measured with a Differential Image Motion Monitor (DIMM; \citealp{Sarazin1990}) or with a Multi-Aperture Scintillation Sensor (MASS; \citealp{Kornilov2003a}). The DIMM probes the entire atmosphere, starting at the level of the monitor. The MASS instrument returns an atmospheric six-layer turbulence profile excluding the lowest 0.5 km. 

\section{Description of major astronomical sites}\label{sec:sites}
\begin{figure}[thbp]
        \includegraphics[width=\columnwidth]{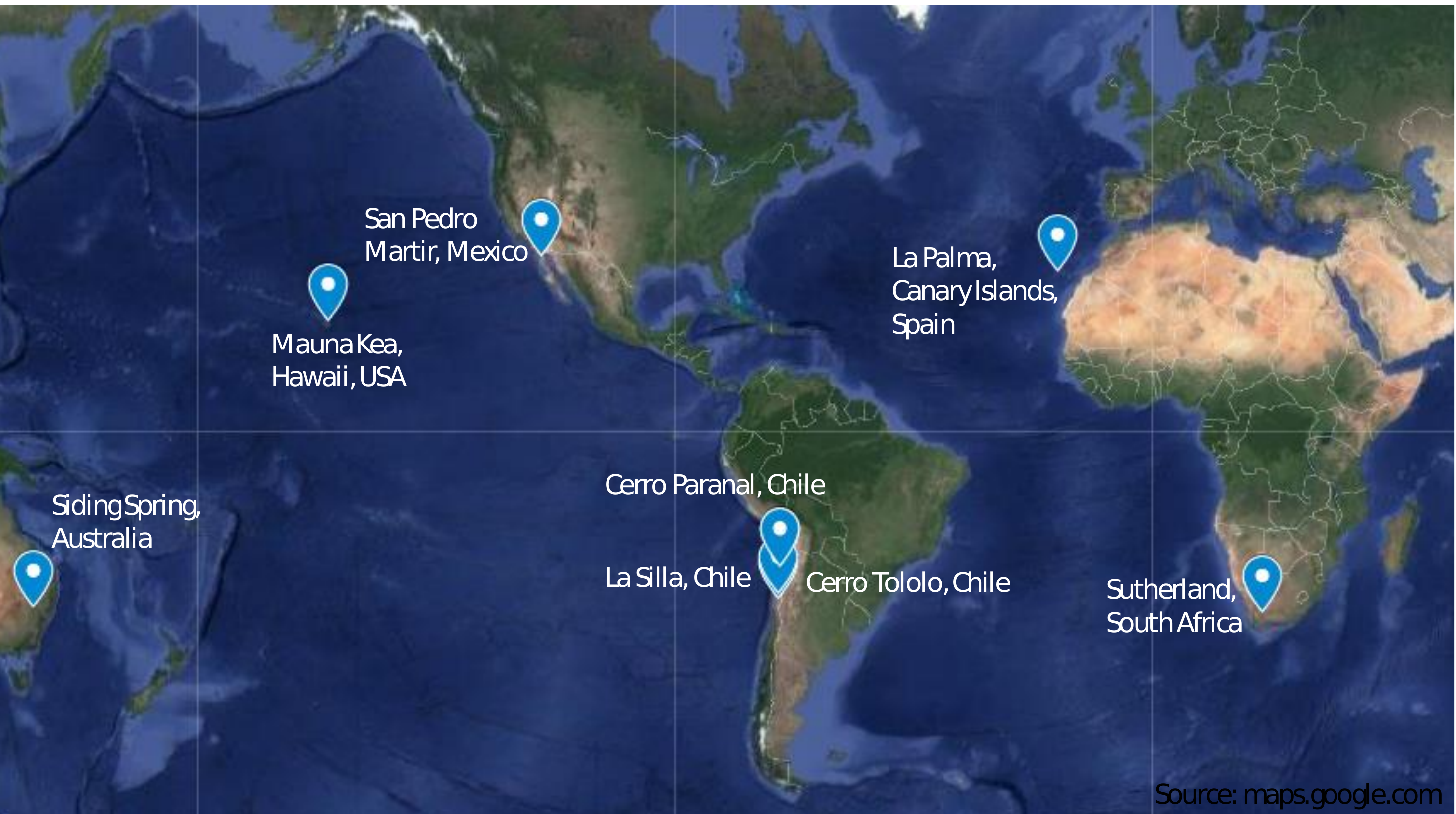}
        \caption{Map of sites evaluated in this study. (Generated with Google maps; \url{maps.google.com}, accessed on 2020 Sep 06)}
        \label{fig:Sites_map}
\end{figure}

We selected eight sites where one or more telescopes are already in operation (Table \ref{tab:Sites_data_overview} and Fig. \ref{fig:Sites_map}) and in situ measurements archived by the observatories are available. Furthermore, the selected sites have the potential to host next-generation telescopes.

An overview of the climatology of each selected site is presented hereafter. In general, all sites have an above average percentage of nights with clear skies. Also, observatories are keen to minimise light pollution to be able to observe dark skies, for example by choosing isolated sites.

\begin{table}[thbp]
        \centering
                \caption{Overview of sites evaluated in this study including longitude (Lon), latitude (Lat), elevation and pressure.}
            \label{tab:Sites_data_overview}
            \tiny
                \begin{tabular}{lccc} 
                
                        \hline\hline  
                        Observatory site & Lon/Lat [$^{\circ}$] & \makecell[c]{Elevation\\\relax [m]} & Pressure [hPa]\\                         
                        \hline
                        Mauna Kea (USA) & -155.47/19.82 & 4200 & 616.28 $\pm$ 1.70 \\ 
                        Cerro Paranal (Chile) & -70.40/-24.63  & 2635 & 743.66 $\pm$ 0.50 \\ 
                        La Silla (Chile) &  -70.73/-29.25 & 2400 &  771.06 $\pm$ 2.00 \\ 
                        Cerro Tololo (Chile) & -70.80/-30.17 & 2200  & 781.28 $\pm$ 2.66  \\ 
                        La Palma (Spain) & -17.89/28.76 & 2370 & 771.21 $\pm$ 3.10  \\ 
                        Siding Spring (Australia) & 149.07/-31.28 & 1165  & 891.56 $\pm$ 4.55  \\ 
                        Sutherland (South Africa) & 20.81/-32.38 & 1798 & 826.49 $\pm$ 1.50  \\
                        San Pedro M\'{a}rtir (Mexico)  & -115.46/31.04 &2800  & 732.70 $\pm$ 3.20 \\
                        \bottomrule
        \end{tabular}
        \tablefoot{Pressure: monthly mean in situ value $\pm$ standard deviation}.
\end{table}


\subsection{Mauna Kea}
The dormant volcano Mauna Kea in the north-east of the Hawaiian Big Island and north of the world's largest volcano Mauna Loa, is the highest elevated site, at 4200~m, that we investigate in this study. Between 1982-2011, measurements at the University of Hawaii 2.24-metre (UH2.2) telescope with a weather station 2~m above top of the 23~m high dome gave an average annual temperature of 1.3~$^\circ$ C and an average annual wind speed of 5.8~m~s$^{-1}$, with very low variability (standard deviation of 1.3~m~s$^{-1}$; \citealp{DaSilva2012}). 
The climate on top of Mauna Kea is very dry and often free of clouds, because the summit lies above the trade wind inversion at about 2000 m, which is a temperature inversion that traps moist air \citep{Ward2014}. However, there are seasonal and inter-annual variations in the quality of the observations, for instance due to the El Ni\~{n}o southern oscillation (ENSO) \citep{Lyman2020}, a quasi-periodic cycle that causes warming or cooling of the sea surface temperature of the eastern Pacific Ocean that impacts the global general circulation. Furthermore, \citet{Giambelluca1986} emphasised that the microclimatic conditions on Hawaii range from dry leeward areas to humid windward flanks.

The site hosts 13 telescopes, nine of which are observing in the optical and infrared, namely the 0.9-metre University of Hawaii (UH) Hilo educational telescope, the UH 2.2-metre telescope, the 3.0-metre NASA infrared telescope facility (IRTF), the 3.6-metre Canada-France-Hawaii telescope (CFHT), the 3.8-metre United Kingdom infra-red telescope (UKIRT), the two 10-metre W. M. Keck observatories Keck I and Keck II, the 8.3-metre Subaru telescope and the 8.1-metre Gemini northern telescope, three in the submillimetre range, namely the 10.4-metre Caltech submillimeter observatory (CSO), the 15-metre James Clerk Maxwell telescope (JCMT) and the 8x6-metre submillimeter array, and lastly the 25-metre very long baseline array (VLBA) in radio.\footnote{\url{http://www.ifa.hawaii.edu/mko-old/telescope_table.shtml}, accessed on 2022 Jan 25}.

\subsection{Chilean sites}
A VLT site selection study conducted in 1990 \citep{Grenon} chose an interdisciplinary approach to investigate the climate in Northern Chile using biogeography. In Chile, the further east, the more precipitation and clouds. This is why coastal sites in Chile are preferred for astronomical observatories. \citet{Grenon} found some evidence that, since 1940, Northern Chile got drier due to climate change. Cerro Tololo, La Silla and Cerro Pachon lie in the transition zone between the Atacama desert and the Chilean Matorral. This zone is characterised by a Mediterranean-type climate with sparse events of southern winter rains and a dry summer and autumn. The climate at Cerro Tololo is more humid than at Cerro Paranal, especially in southern winter. 
It is also in southern winter that most nights are not photometric \citep{Grenon}.

The subtropical jet stream is the dominant driver of the movement of the atmosphere for these Chilean sites, since the jet stream oscillates within a latitude range that includes Cerro Paranal, La Silla and Cerro Tololo. Its temporal variability is quite high: within a few day's time, the wind velocity can become up to four times weaker or stronger \citep{Sarazin2001}. In southern winter (July-September), jet stream wind speeds are higher than in southern summer (January-August).

\subsubsection{Cerro Paranal}\label{subsec:Paranal_site_description}
Cerro Paranal lies in one of the driest regions of the world, in the Atacama desert 12 km east of the South Pacific Ocean coast, with a typical relative humidity between 5 and 20 \% according to ESO\footnote{\label{ESOa}\url{http://www.eso.org/sci/facilities/paranal/astroclimate/site.html}, accessed on 2020 Aug 27}.
On average, the sky is photometric in 78 \% of the nights and PWV is less than 1 mm in 8.2~\% of the nights\footnote{see footnote \ref{ESOa}}. 

The optical very large telescope (VLT) is run by ESO on Cerro Paranal since 1998 with four 8.2-metre primary mirror unit telescopes and four auxiliary 1.8-metre primary mirror movable telescopes, each of which can be combined to form the VLT interferometer. Only 20 km east of Cerro Paranal lies Cerro Armazones, the mountain where the ELT is under construction, which will be the next biggest optical telescope with a primary mirror of 39-metre diameter whose first light is expected in 2026.

\subsubsection{La Silla}
La Silla observatory is located circa 500 km south of ESO Paranal and is operated by ESO since 1960. 
It shows a similar climatology than Paranal that is slightly wetter since La Silla is on the edge of the dry area of Chile \citep{Sarazin1988}. Also, the seasonal cycle of humidity in La Silla shows a larger amplitude than for Paranal \citep{Sarazin_VLTreport62}. Short-term bad weather associated with low pressure events coming from Antarctica occur in southern winter and hit La Silla harder than Paranal, for example in the form of snow storms.
At La Silla observatory, two 3.6-metre telescopes, namely the new technology telescope (NTT) and the ESO 3.6-metre telescope, are operated by ESO \citep{ESO2015}. Smaller telescopes hosted by La Silla observatory include the 2.2-metre Max-Planck telescope, the 1.2-metre Swiss telescope, the 1-metre Schmidt telescope, the 1.5-metre Euler telescope, the 0.6-metre REM telescope, the 0.6-metre TRAPPIST telescope and the 1.5-metre Danish telescope.

\subsubsection{Cerro Tololo}
Cerro Tololo is another site providing very good astronomical observing conditions in the region of Coquimbo, south to the Atacama desert. The Cerro Tololo inter-American observatory (CTIO) operates  the V\'{i}ctor M. Blanco 4-metre telescope and four 1-metre-class telescopes.\footnote{\url{https://noirlab.edu/science/programs/ctio/telescopes}, accessed on 2022 Jan 25} Additionally, the small and moderate aperture research telescope system (SMARTS) consortium on Cerro Tololo operates the SMARTS 1.5-metre, 1.3-metre, 1.0-metre and 0.9-metre telescope.
Approximately 10 km (beeline) south-east to Cerro Tololo lies Cerro Pachon, where the 4.1-metre southern astrophysical research (SOAR) and the 8.1-metre Gemini south telescope are located and where the Vera C. Rubin observatory is under construction with the planned 8.4-metre Simonyi survey telescope.

\subsection{La Palma}\label{subsubsec:site_la_palma}
The Spanish island 'Isla de la Palma' belongs to the Canary Islands. 
The total area of the island is small, ca. 700~km$^2$. On the highest point of the island, at 2370~m, lies the Roque de los Muchachos Observatory (ORM).
It houses 16 telescopes, among them the 10.4-metre optical and near infrared gran telescopio Canarias, the 4.2-metre optical and near infrared William Herschel telescope (WHT) and the 2.6-metre nordic optical telescope (NOT).\footnote{\url{https://www.iac.es/en/observatorios-de-canarias/roque-de-los-muchachos-observatory/telescopes-and-experiments}, accessed on 2022 Jan 25}
Roque de los Muchachos lies well above the tree line and the trade wind inversion based on ca. 1000~m, which prevents clouds from forming at ORM \citep{McInnes1974}. The convex profile of the Caldera de Taburiente directs air flow around the peak \citep{McInnes1974}.

\subsection{Siding Spring}
Siding Spring observatory (SSO) is located at the brink of a National Park near Coonabarabran in south-eastern Australia. It is the site with the lowest elevation (1165 m) out of the selected sites.
The climate at Siding Spring is drier and nights are colder compared to the rest of Australia\footnote{\label{Bonzle.}\url{http://www.bonzle.com/c/a?a=p&p=315007&d=w&s=siding\%20springs\%20obserbvatory&cmd=sp&c=1&x=149\%2E065885&y=\%2D31\%2E274767&w=40000&mpsec=0} accessed on 2020 Aug 27}.
One consequence of anthropogenic climate change that also threatens observatories is an increase in the frequency, intensity and duration of wildfires, which is expected with high confidence \citep{IPCC2021}. One example is the Wambelong bush fire in 2013 that destroyed facilities at the Siding Spring Observatory (SSO), but due to measures that were taken beforehand, it caused little damage to the telescopes \citep{AAO2013}. The site hosts 11 telescopes, among which is the 3.9-metre optical Anglo-Australian telescope (AAT), the 2.3-metre Australian national university (ANU) telescope, and the 1.2-metre UK Schmidt telescope (UKST).

\subsection{Sutherland}
Near the town of Sutherland in the semi-desert of the Great Karoo, 12 optical and infrared telescopes are operated, among which is the 10-metre segmented southern African large telescope (SALT). Temperatures are coldest in the town of Sutherland compared to the rest of South Africa. 
There is little precipitation due to a mountain chain surrounding the Great Karoo that works as a humidity trap.\footnote{\label{Karoo}\url{http://www.thegreatkaroo.com/index.php?page=climate_and_vegetation}, accessed on 2020 Aug 27} 
Almost the whole area of the Great Karoo has been classified as an astronomy advantage area (AAA) by the South African minister of science and technology, an area where regulations can be introduced to control communication, transport and other technology to support operation of specialised telescopes \citep{Walker2019}.

\subsection{San Pedro M\'{a}rtir}
The National Astronomical Observatory (OAN) is located on top of the Sierra de San Pedro M\'{a}rtir (SPM) in the north of the Mexican Baja California peninsula.  It hosts a 2.1-metre optical and infrared telescope and seven other 1-metre-class telescopes.\footnote{\url{https://www.astrossp.unam.mx/en/users/telescopes/2-1m-telescope}, accessed on 2022 Jan 25}
Compared to the three selected Chilean sites, the amount of precipitation is higher for the observatory located in SPM. Precipitation events are strong but rare \citep{Schoeck2009}.
The site offers low relative humidity and good seeing conditions \citep{Lopez2003}. Between 1984 and 2002, 63~\% of the nights were photometric \citep{Cruz-Gonzales2004a}.

\section{In situ, reanalysis, and model data}\label{sec:Data}

\subsection{In situ measurements}\label{subsection: in situ measurements}
Table \ref{Table: in situ data overview} provides details on the in situ measurements archived by the observatories. We filtered the in situ data before averaging with the following upper and lower cut-offs. The valid relative humidity as well as the valid cloud cover data lie between 0~\% and 100~\%. The valid temperature data lie between -20$^{\circ}$~C and 50$^{\circ}$~C. Every pressure value with an error of more than $\pm$100 hPa with respect to the mean value was filtered. The specific humidity is calculated from values that were filtered as just described. The valid range of seeing ranges from 0 to 10 arcseconds. The filters are tolerant because the measurement errors we found appeared to be very extreme, for example a temperature value of 2000~$^{\circ}$C or a seeing value of -1.2 arcseconds.

\begin{table*}[thbp]
        \centering
        \caption{Overview of in situ data. It is indicated from which observatory the data come from, the time period and the frequency of the measurements. }
        \label{Table: in situ data overview}
            
                \begin{tabular}{lllll} 
                        \hline\hline  
                        Site          & Data     & Observatory or telescope      & Time Period             & Frequency                                 \\ \hline
Mauna Kea     & Meteo\tablefootmark{a}              & CFHT                       & 2000-2019               & 1-10 min (24 hours)                       \\
              & Seeing\tablefootmark{b}             & MKWC                       & 2010-2019               & 2 min (night-time hours)                  \\
              & PWV\tablefootmark{c}                & JCMT                & 2009-2019               & 2 seconds (24 hours) \\
              & Weather loss        & Gemini South                    & 2014-2017               & 1 day (night-time)                        \\
Paranal       & Meteo, Seeing      & VLT                  & 2000-2019               & 1 min (24 hours)                          \\
              & PWV                & VLT                 & 2016-2019               & 2 min (24 hours)                          \\
              & photometric nights\tablefootmark{d}  & VLT                  & 1984-2017               & 1 month                                   \\
La Silla      & Meteo              & La Silla observatory & 2000-09/2008, 2011-2019 & 10 min (24 hours)                         \\
              & Seeing             & La Silla observatory & 2000-2019               & 1 min (24 hours)                          \\
              & PWV\tablefootmark{e}                & La Silla observatory & 2000-2007               & 3 h (24 hours)                            \\
              & photometric nights\tablefootmark{d} & VLT               & 1984-2017               & 1 month                                   \\
Cerro Tololo  & Meteo              & CTIO             & 2002-2018               & 5 min (24 hours)                          \\
              & Seeing             & CTIO                & 2004-2019               & 1 min (night-time)                        \\
  Cerro Pachon            & Weather loss    & Gemini North       & 2013-2017               & 1 day (night-time)                        \\
La Palma      & Meteo\tablefootmark{f}              & NOT                        & 2004-2019               & 5min (24 hours)                           \\
              & Seeing\tablefootmark{g}             & IAC (ING)                  & 2004-2019               & 3 min (night-time)                        \\
              & Weather loss\tablefootmark{h}        & TNG                        & 2012-2016               & 1 day (night-time)                        \\
Siding Spring & Meteo              & AAT                        & 2004-2019               & 10 min (24 hours)                         \\
              & Seeing\tablefootmark{i}             & AAT                        & 1993-2019               & by semester (night-time)                  \\
              & Weather loss         & AAT                        & 1993-2018               & 1 day (night-time)                        \\
Sutherland    & Meteo              & SALT                       & 2013-2019               & 1 min (24 hours)                               \\
SPM           & Meteo\tablefootmark{j}              & OAN                        & 2006-2019               & 15 min (24 hours)                         \\
              & Seeing\tablefootmark{k}             & TMT site testing campaign  & 10/2004-08/2008         & 1 min (night-time)                        \\
              & Pressure\tablefootmark{k}            & TMT site testing campaign  & 10/2004-08/2008         & 2 min (24 hours)                          \\
                        
                        \bottomrule
        \end{tabular}
        \tablefoot{If no web page is indicated in the following, we retrieved the in situ data via private communication with the indicated observatory. All web pages were accessed in the time period September 2019 to September 2020.
        \tablefoottext{a}{\url{http://mkwc.ifa.hawaii.edu/archive/wx/cfht/}}
        \tablefoottext{b}{\url{http://mkwc.ifa.hawaii.edu/current/seeing/mass/}}
        \tablefoottext{c}{constructed URL with 'YYYY' as placeholder for the year, 'MM' for the month and 'DD' for the day: \url{https://www.cadc-ccda.hia-iha.nrc-cnrc.gc.ca/data/pub/JCMT/YYYYMMDD.wvm}; column 10}
        \tablefoottext{d}{\url{https://www.eso.org/gen-fac/pubs/astclim/paranal/clouds/clouds-LPO.txt}}
        \tablefoottext{e}{\url{https://www.eso.org/gen-fac/pubs/astclim/lasilla/h2o/laseranal.dat}}
        \tablefoottext{f}{\url{http://www.not.iac.es/weather/archive/met/}}
        \tablefoottext{g}{\url{http://catserver.ing.iac.es/robodimm/}, \url{http://catserver.ing.iac.es/robodimm/robodimm.php?all=1}} 
        \tablefoottext{h}{\citep{Molinari2012, Molinari2014}}
        \tablefoottext{i}{\citep{Brookfield2020}}
        \tablefoottext{j}{\url{http://tango.astrosen.unam.mx/weather15/}, \citep{Plauchu-Frayn2020}}
        \tablefoottext{k}{\url{https://sitedata.tmt.org/Available_data/data_fields.html}}
        }
\end{table*}

\subsubsection{In situ measurements of meteo data}
For all sites, temperature, relative humidity and pressure are included in 'Meteo' in Table \ref{Table: in situ data overview}, except for SPM where the pressure came from the TMT site testing campaign \citep{Schoeck2009}. No meteorological data of La Silla are available in 2009 and 2010 due to technical issues during a replacement of the weather tower.
Meteo data are most often measured by an automatic weather station at a high frequency of 15 minutes at minimum (Table \ref{Table: in situ data overview}). In Siding Spring, meteo data are measured 1.5~m above ground, on La Silla, Cerro Paranal, Cerro Tololo and La Palma, 2~m above ground, at Sutherland and Mauna Kea, 3~m above ground and at SPM 6~m above ground.

\subsubsection{In situ measurements of PWV}
Not all observatories measure and archive the PWV. 
Measurements of PWV are only available for Mauna Kea, La Silla and Cerro Paranal. 
At Paranal, PWV is measured with the Low Humidity and Temperature Profiling microwave radiometer (LHATPRO;  \citealp{Kerber}). LHATPRO is specifically designed for low humidity environments with low PWV values as it is the case at Paranal. The device uses the strong water vapour emission line at 183 GHz for humidity profiling. The temperature is measured separately.
At la Silla, PWV is not measured on a regular basis but was retrieved for the VLT site selection process from sky radiance monitors \citep{Sarazin_VLTreport62} and later investigated with different techniques including archival optical spectra done with various spectrographs and radiosonde launches \citep{Querel2010}. The data used in this study were provided by ESO\footnote{\url{https://www.eso.org/gen-fac/pubs/astclim/lasilla/h2o/}, accessed on 2020 Aug 30} and are based on satellite measurements (Table \ref{Table: in situ data overview}).
On Mauna Kea, PWV is converted from opacity values measured by radiometers at a frequency of 225GHz and 183GHz.\footnote{\url{https://www.eaobservatory.org/jcmt/observing/weather-bands/}\label{MK-PWVconversion}, accessed on 2020 Aug 28}

\subsubsection{In situ measurements of cloud cover}\label{sec:insitu_cloud}
No observatory could provide cloud cover measurements. Instead, observatories on Mauna Kea, La Palma and Siding Spring record and archive the time lost due to bad weather at night (Table \ref{Table: in situ data overview}). This so-called weather loss could be triggered by clouds that absorb visible and infrared light, but there are also other reasons such as high wind speed or high humidity.\footnote{\label{url_atmospheric_restrictions}\url{https://www.eso.org/sci/facilities/paranal/sciops/At_Telescope.html} accessed on 2021 Jan 14}
ESO provides monthly fractions of photometric nights for Paranal and La Silla (Table \ref{Table: in situ data overview}).
We substituted weather loss data of Cerro Tololo with weather loss data from Cerro Pachon, since the distance between the two sites is only 10 km.

\subsubsection{In situ measurements of seeing}
In situ data of seeing come from measurements with a DIMM in Paranal, La Silla, Cerro Tololo and La Palma. For Paranal, data from a combined MASS-DIMM instrument are available only between 2016 and 2019. In situ data from SPM and Mauna Kea are measured with a MASS instrument.
For Siding Spring, because seeing is measured by different instruments and occasionally with a DIMM, seeing might be influenced by dome air conditioning.\footnote{Private communication with Dr. Daniel Cotton (AAT).} However, seeing measurements are only provided per semester (March - August, September - February).

\subsection{ERA5 reanalysis}\label{section:ERA5_reanalysis}
A reanalysis is a continuous weather simulation that assimilates archived observational data. 
The simulation of observational data via the laws of physics built into the weather model act as a physics-based interpolator and provide 4-dimensional data given by three spatial dimensions and a temporal dimension \citep{Parker2016}.

In this study, we consider a reanalysis for two reasons. First, a reanalysis builds a bridge between in situ observations and GCMs, as it provides homogeneous and continuous data over a long time period that enables the validation of GCMs. Second, a reanalysis can help verify the feasibility of the study. Its spatial resolution is either similar or higher than the resolution of GCMs so any technical issue encountered with the reanalysis may be similar with GCMs. Furthermore, a reanalysis provides data for sites where no observation is available and makes sites more inter-comparable. 

We use the latest ECMWF ERA5 reanalysis, which was made publicly available in early 2019 and then updated every 3 months \citep{Hersbach2018a, Hersbach2018}. According to \citet{IPCC2021}, ERA5 is the most reliable reanalysis, at least to assess climate trends. The ERA5 reanalysis provides hourly data on single or multiple pressure levels from 1979 onwards, and since 2021 preliminarily also from 1950 \citep{Bell2020}.
 Data are available for the atmosphere, the land surface and ocean waves of the Earth. The horizontal resolution of ERA5 is $0.25^{\circ} \times 0.25^{\circ}$ (latitude $\times$ longitude), where $0.25^{\circ}$ correspond to 27.75 km at the equator. The atmosphere is represented on 137 pressure levels, and outputs are  available on 37 pressure levels (Fig. \ref{fig:pressure_levels_climate_model_output}).
 The pressure level variables we use are air temperature, specific humidity, relative humidity, geopotential height and the two horizontal components of wind speed.
 The single level variables we use are total column water and total cloud cover. The total column water includes all kinds of water in the atmosphere such as water vapour, liquid water, cloud ice, rain and snow. The total column water is given as instantaneous values, which refer to one point in time and not to a temporal average.
For total cloud cover, monthly means of night-time values are considered for the comparison to in situ data as well as the comparison with PRIMAVERA.

In ERA5, data on single levels are computed on the model surface. The model surface elevation, also called orography, does not coincide with Earth's surface elevation at any specific location because of scale mismatch. The ERA5 reanalysis uses an interpolation from different surface elevation data sets to calculate the orography.\footnote{\label{ECMWF}\url{https://confluence.ecmwf.int/display/CKB/ERA5\%3A+data+documentation}, accessed on 2020 Aug 30} The orography of ERA5 is shown in Figs. \ref{Figure:Orography_MaunaKea} to \ref{Figure:Orography_SPM} for the eight sites and discussed in Sect. \ref{subsection:model_orography}. Due to its spatial resolution, the orography of ERA5 is smoother than Earth's surface elevation. As a consequence, elevations of the observatories are lower in ERA5 than observed, although they are generally located on the locally highest elevated grid cells of ERA5. This difference in elevation is an issue when evaluating surface data. To solve this problem, we considered data on pressure levels, which consequently have a pressure offset to the surface pressure unique to each grid cell, which only varies slightly in time.

\subsection{PRIMAVERA GCMs}\label{section: PRIMAVERA}
We use GCMs provided by the European Horizon 2020 PRIMAVERA project (PRocess-based climate sIMulation: AdVances in high-resolution modelling and European climate Risk Assessments)\footnote{\url{https://www.primavera-h2020.eu/}, accessed on 2020 Jun 07}. The PRIMAVERA project provides the European contribution to HighResMIP (\citealp{Haarsma2016}), endorsed by CMIP6. The initialisation of CMIP dates back to 1995 \citep{Meehl1995}, a project whose main objective is to compare the latest GCM simulations of the past, present and future, ran under controlled conditions, in order to assess changes in climate \citep{Meehl2000}. The latest CMIP Phase 6 (CMIP6) focusses on 1) Earth's response to radiative forcing, 2) systemic model biases and 3) future climate changes under the light of internal climate variability, predictability and scenario uncertainties \citep{Eyring2016}.

\begin{table*}[thbp]
        \caption{Overview of PRIMAVERA models.}
        \resizebox{\textwidth}{!}{%
                 
                \begin{tabular}{lllllll}
                        \hline\hline  
                        Model name                                & \makecell[l]{HadGEM3-\\GC31-HM} & \makecell[l]{EC-Earth3P\\-HR}        & \makecell[l]{CNRM-CM6\\-1-HR} & \makecell[l]{MPI-ESM1\\-2-XR} & \makecell[l]{CMCC-CM2\\-VHR4} & \makecell[l]{ECMWF-IFS\\-HR}          \\ \hline
                        Institute                                 & Met Office      & \makecell[l]{KNMI, SMHI,\\ BSC, CNR} & CERFACS       & MPI-M         & CMCC          & ECMWF                 \\
                        
                        Reference & \makecell[l]{\citet{HadGEM_Roberts}\\ \citet{HadGEM_Schiemann}} & \citet{EC-Earth} & \citet{CNRM} & \citet{MPI} & \citet{CMCC} & \citet{Roberts2017} \\[0.5cm]
                        
                        short name & HadGEM & EC-Earth & CNRM & MPI & CMCC & ECMWF\\
                        
                        \makecell[l]{Atmosphere \\ horizontal \\resolution \\(at 50N$^{\circ}$)} & 25km            & 36km                 & 50km          & 34km          & 18km          & \makecell[l]{25km (Output\\ at 50km)} \\[1cm]
                        \makecell[l]{Atmos \\dynamical \\ scheme \\(grid)}             & \makecell[l]{Grid point \\ (SISL, lat-long)} & \makecell[l]{Spectral (linear,\\ reduced \\Gaussian)} & \makecell[l]{Spectral (linear,\\ reduced \\Gaussian)} & \makecell[l]{Spectral \\(triangular, \\Gaussian)} & \makecell[l]{Grid point \\(finite volume, \\lat-long)} & \makecell[l]{Spectral (cubic \\octohedral, reduced\\ Gaussian)} \\[1cm]
                        \makecell[l]{Ensemble \\members}                   & r1i1p1f1        & r1i1p2f1             & r1i1p1f2      & r1i1p1f1      & r1i1p1f1      & r1i1p1f1     \\[0.5cm]
                        
                        \makecell[l]{historical \\simulations} &  \makecell[l]{coupled-past\\atmos-past}  & \makecell[l]{coupled-past\tablefootmark{a} }&   \makecell[l]{coupled-past\\atmos-past} &  \makecell[l]{coupled-past\\atmos-past} & \makecell[l]{coupled-past\\atmos-past} & \makecell[l]{coupled-past}\\[1cm]
                        
                        \makecell[l]{future \\simulations} & \makecell[l]{coupled-future\\atmos-future} & coupled-future\tablefootmark{a} & \makecell[l]{coupled-future\\atmos-future} & ... &\makecell[l]{coupled-future\\atmos-future} & coupled-future \\
                        \hline
        \end{tabular}}
        \tablefoot{ The terms 'coupled-past' and 'coupled-future' stand for the historical and future simulations of the coupled models and the terms 'atmos-past' and 'atmos-future' stand for the atmosphere-land models, respectively. The ensemble member identifiers 'ripf', e.g. r1i1p2f1, stand for realisation 'r', initialisation 'i', physics 'p' and forcing 'f' \citep{Taylor2018}. The realisation defines the initial conditions. The initialisation distinguishes between different initialisation procedures. The physics parameter indicates the physics schemes used.\\
                Table adapted and extended from \url{https://www.primavera-h2020.eu/modelling/our-models/}, accessed on 2020 Aug 06.\\
                \tablefoottext{a}{not used for cloud cover.}}

        \label{Table: PRIMAVERA overview}
\end{table*}

The primary goal of PRIMAVERA is to deliver a new generation of advanced high-resolution GCMs in order to provide detailed climate information at the regional and global scales, including extreme events. The horizontal resolution of PRIMAVERA GCMs is higher than those provided by the standard IPCC CMIP activities (CMIP3, CMIP5, CMIP6). Therefore, PRIMAVERA also enables to evaluate the effects of increasing spatial resolution on climate simulations.\footnote{see \url{https://www.primavera-h2020.eu/output/scientific-papers/} for a complete list of publications using PRIMAVERA data sets} Never before has a comparable study been conducted with an ensemble of GCMs that ensure such a high level of fidelity. A detailed overview of the PRIMAVERA models is given in Table \ref{Table: PRIMAVERA overview}.
Each GCM uses a different dynamical core with a grid spacing between 18 km and 50 km.
All monthly pressure level data are output on the same 19 vertical pressure levels (Fig. \ref{fig:pressure_levels_climate_model_output}).
 Almost all PRIMAVERA GCMs provide a historical and a future simulation (Table \ref{Table: PRIMAVERA overview}). The historical simulation covers the period from 1950 until 2014 and the future simulation goes from 2015 until 2050. There are two types of simulations: an atmosphere-land only (hereafter atmosphere-only) and a coupled atmosphere-ocean (hereafter coupled) climate simulation. While the coupled models allow for interactions between the ocean and the atmosphere, atmosphere-only simulations are constrained by observed sea surface boundary conditions (sea ice cover and sea surface temperature), which simplifies the representation of the climate system by removing interactions with the ocean and makes their results more comparable to observations and reanalyses. Moreover, atmospheric GCMs do not lack the ability to project anthropogenic forcings for the land and the atmosphere \citep{He2016}. The PRIMAVERA atmosphere-only simulations are forced by HadISST2.2.0.0 sea surface boundary conditions.\footnote{for details, refer to \url{https://hrcm.ceda.ac.uk/research/cmip6-highresmip/highresmip-protocol/} and \citealp{https://doi.org/10.22033/ESGF/input4MIPs.1221}}
 The PRIMAVERA atmosphere-only simulations provide an opportunity to decide whether performance issues arise from ocean-atmosphere coupling \citep{Moreno-Chamarro2022}. 

 PRIMAVERA coupled and atmosphere-only simulations use the same external radiative forcings, as described in the HighResMIP protocol \citep{Haarsma2016}. Those are based on observations for the historical period and on the shared socioeconomic pathway 5-8.5 (SSP5-8.5; \citealp{Kriegler2017}) for the future period, which is built upon the CMIP5 representative concentration pathway (RCP) 8.5. The RCP 8.5 \citep{Riahi2011} exhibits highest greenhouse gas emissions compared with the other existing RCPs. It demonstrates the worst case scenario and does not include any mitigation measures. If we follow RCP 8.5 until the end of the century, we would end up with a radiative forcing of 8.5 W/m$^2$.
 The two atmosphere-only simulations are called highresSST-present (hereafter: atmos-past) and highresSST-future (hereafter: atmos-future), since they make use of SST measurements. 
 The coupled climate model simulations are called hist-1950 (hereafter: coupled-past) and highres-future (hereafter: coupled-future).

 Climate model simulations approximate deterministic chaotic but non-linear and non-periodic processes by discrete systems, which makes them sensitive to initial conditions. Furthermore, coupled climate models have to overcome the difficulty of coupling two subsystems with very different timescales, defined as the time needed for a new equilibrium after a small perturbation, which are much shorter for the atmosphere than for the ocean \citep{Peixoto1992}. Therefore, PRIMAVERA coupled GCMs have a spin-up period of 50 years to ensure physical equilibrium \citep{Haarsma2016}.
 A major role for climate variability play quasi-periodic cycles such as ENSO or the quasi-biennial oscillation (QBO; e.g. \citealp{Baldwin2001}), which are simulated by GCMs. 
 In general and concluding from this and the previous section, we expect ERA5 to agree better with in situ observations than PRIMAVERA GCMs agrees with in situ observations and ERA5.

\section{Model evaluation and trend methods}\label{sec:methods}
\subsection{Data download and code}\label{sec:data_download_and_github}
We downloaded hourly ERA5 data \citep{Hersbach2018, Hersbach2018a} through the Copernicus climate change service climate data store \citep{CopernicusClimateChangeServiceC3S2020}, and PRIMAVERA data through the Earth system grid federation node.\footnote{\url{http://esgf-index1.ceda.ac.uk}, accessed on 2020 Sep 20}
The complete code we wrote for our analysis is available online on Github\footnote{\url{https://github.com/CarolineHaslebacher/Astroclimate-future-project}, accessed on 2021 Sep 22}.

\subsection{Evaluation of ERA5 and PRIMAVERA}\label{sec:methods_ERA5_and_PRIMAVERA_evaluation}
 For evaluating ERA5 and PRIMAVERA against in situ data, we considered the grid point that is closest to the location of the astronomical site for each individual climate model. There are several reasons for considering the nearest grid point instead of a possible weighted grid point: 1) for sites on islands or on the coast, we can select the nearest point on land, while the weighted nearest point method may give weights to points that are in the ocean; 2) for sites near mountain tops, the circulation may induce different climate whether the sites are on the windward or the leeward side of the mountain. This may not be so relevant at spatial resolution of CMIP6 models, but it starts to be more and more relevant as the resolution of the model increases, such as for PRIMAVERA. The nearest grid point therefore seems more appropriate. More information about the orography is provided in Sect. \ref{subsection:model_orography} and in the appendix \ref{appendix:orography}.
 For the analysis of the six PRIMAVERA GCMs (Table \ref{Table: PRIMAVERA overview}), we calculated a multi-model ensemble mean for each simulation (atmos-past, atmos-future, coupled-past, coupled-future) and we expect these ensemble means to outperform individual GCM simulations, based on  \citet{Gleckler2008}.
 
 For temperature, relative humidity and specific humidity, we considered ERA5 and PRIMAVERA pressure levels that show a compromise between best skill score, best representation of in situ orography (closest pressure level) and minimal absolute bias between models and observations (Table \ref{tab:methods_Plevs}).
While testing different pressure levels, we noticed that future trends were not sensitive to the choice of the pressure level, for example between 700 and 850 hPa for PRIMAVERA, the values of future trends are very similar (not shown). 

 We considered the 'total cloud cover' model variable to assess the cloudiness. This is a single level variable that includes clouds over the entire atmospheric column of the model. %

\subsubsection{General workflow}\label{sec:workflow}
To best represent night-time astronomical observations, it would be preferable to validate night-time ERA5 and PRIMAVERA against night-time in situ observations. However, preliminary analyses on the diurnal cycle using hourly data show that hourly ERA5 data are not able to represent the in situ diurnal cycle (Appendix \ref{appendix:diurnal_cycle}). This could be caused by the use of convective parameterisation and by the still too coarse horizontal resolution and discrepancies in orography of ERA5 compared to the real topography. 
Therefore and because we cannot expect GCMs to show a better match for the diurnal cycle than ERA5, we decided to evaluate monthly-mean data. 

Monthly ERA5 data are validated against in situ data by using a skill score classification (see Sect. \ref{sec:Model_skill_score} and Table \ref{Table:skill score classification}). That classification is also used to validate monthly historical PRIMAVERA GCM data against in situ and ERA5 data.
We investigated trends in future and historical PRIMAVERA simulations and in ERA5. 
Historical trend analysis allowed to set PRIMAVERA future projections into context. More details about the methods of trend analysis can be found in Sect. \ref{sec:methods_for_trend_analysis}.

\begin{table*}[thbp]
        \centering
        \caption{ERA5 and PRIMAVERA pressure levels used for temperature (T), relative humidity (RH) and specific humidity (SH) and vertical integration lower limits for PWV and seeing model (Eq. \ref{eq:Osborn_seeing}).}
        \label{tab:methods_Plevs}
            
                \begin{tabular}{lllllll} 
                        \hline\hline  
\multicolumn{1}{l}{Site} & \multicolumn{1}{l}{T, RH, SH}    & \multicolumn{1}{l}{T, RH, SH}    & \multicolumn{1}{l}{PWV}       & \multicolumn{1}{l}{PWV}         & \multicolumn{1}{l}{Seeing model} & \multicolumn{1}{l}{Seeing model} \\
\multicolumn{1}{l}{}                          & ERA5 & \multicolumn{1}{l}{PRIMAVERA} & ERA5 & \multicolumn{1}{l}{PRIMAVERA} & ERA5     & PRIMAVERA\tablefootmark{a}     \\ 
& [hPa] & [hPa]& [hPa]& [hPa]& [hPa]& [hPa] \\\hline
Mauna Kea                     & 600         & 600             & 600      & 600          & 800      & 850                   \\
Cerro Paranal                 & 750         & 700             & 750      & 700          & 900      & 850                   \\
La Silla                      & 775         & 850             & 775      & 700          & 825      & 850                   \\
Cerro Tololo                  & 800         & 850             & 775      & 850          & 825      & 887.5                 \\
La Palma                      & 800         & 850             & 775      & 700          & 1000     & 925               \\
Siding Spring                 & 900         & 925             & 900      & 925          & 950      & 925                   \\
Sutherland                    & 850         & 850             & 825      & 850          & 850      & 850                   \\
SPM                           & 750         & 700             & 750      & 700          & 850      & 925        
\\
                        \bottomrule
        \end{tabular}
        \tablefoot{\tablefoottext{a}{Median of individual model surface pressure values.}
        }
\end{table*}

\subsubsection{Calculation of specific humidity and PWV}\label{sec:formulas_SH_and_PWV}

Specific humidity is computed following the equation for the saturated vapour pressure $e_s$ \citep{Buck1981}
\begin{equation}
e_s = 6.1121 \cdot \exp\left(\left(18.678 - \dfrac{T}{234.5}\right)\left(\dfrac{T}{257.14 + T}\right)\right),
\end{equation}
where the temperature $T$ is in $^{\circ}$~C and  $e_s$ is in hPa.
With the formula for the saturation specific humidity \citep{Peixoto1992}
\begin{equation}
q_s = 0.622 \cdot \dfrac{e_s}{P}
\end{equation}
and Eq. \ref{eq:rel_hum_with_qs}, we arrive at a formula for specific humidity computed with relative humidity $RH$ (in decimal notation), temperature $T$ (in $^{\circ}$~C) and pressure $P$ (in hPa)
\begin{equation}\label{Formula: Specific humidity q(T, RH, P)}
q =  3.802 \cdot  RH \cdot \dfrac{1}{P} \exp\left(\left(18.678 - \dfrac{T}{234.5}\right)\left(\dfrac{T}{257.14 + T}\right)\right).
\end{equation}
We use this formula to calculate specific humidity using relative humidity, pressure and temperature in situ data measured by observatories.

The PWV is computed directly from Eq. \ref{Equation: PWV theory} by replacing the integral with the sum over all discrete pressure levels
\begin{equation}\label{Formula: PWV numerical integral of q}
PWV = \dfrac{1}{g} \sum_{i=1}^{N}q_i \cdot \Delta p_i,
\end{equation}
where $p_i$ is the pressure (in hPa) and $q_i$ is the specific humidity  (in kg~kg$^{-1}$) on pressure level $i$. We use the Euler forward method to compute $\Delta p_i$.

Specific humidity values are integrated from the surface pressure (Table \ref{tab:methods_Plevs}) of the observatory to the top of the atmosphere (250 hPa), following equation \ref{Formula: PWV numerical integral of q}. This integral is called the 'integrated PWV' hereafter. Moreover, we verified that ERA5 integrated PWV is equivalent to ERA5 total column water (as described in Sect. \ref{section:ERA5_reanalysis}), which confirms that total column water is mostly composed of water vapour in long term means \citep{Peixoto1992}.

\subsubsection{Calculation of astronomical seeing}\label{sec:formulas_seeing}
The contribution of the $C_n^2$ profile that varies the most is the ground layer contribution, which consists of the boundary layer (0~km to 1-2~km altitude) and the surface layer (0~m to 50~m altitude), whereas the free atmosphere contribution (1-2~km to 20~km altitude) appears to be similar independent of the geographical site location \citep{Abahamid2004}. 
\citet{Racine2005} shows that surface layer turbulence is the dominant cause of seeing at low elevations above ground, decreasing  with a scale height of about 3.5~m. He has proposed a model vertical distribution of the turbulence which can fit the median seeing values of major telescopes over the world only based upon altitude and elevation. The GCMs used in our study do not have the required resolution to model surface layer turbulence, so we must assume that the top of large telescopes enclosures is naturally high enough above ground (15~m or more) so that the surface turbulence layer does not influence the seeing at the telescope level. Moreover, even in open enclosures such as the VLT that uses passive flushing, the outside local turbulence cannot easily enter at the level of the observing floor (10~m above ground) partly because of the wind screen on the front and of the smaller size ventilation slots on the back.

We introduce two different approaches to estimate seeing with ERA5 and PRIMAVERA output: The first approach, identified as the '200-hPa-wind-speed seeing' hereafter, uses the correlation between wind speed at 200 hPa, where strong winds prevail, and the free atmosphere seeing \citep{Vernin1986, Racine2005, Sarazin2013}. This correlation, which is based on physical shear, is expected to exist at all sites, modulated locally by the intensity and frequency of appearance of the jet stream. The turbulence integral can be split into a lower and an upper contribution:
\begin{equation}
    \int_{0 km}^{20 km} C_n^2(z) dz = \int_{0 km}^{2 km} C_n^2(z) dz + \int_{2 km}^{20 km} C_n^2(z) dz.
\end{equation}
The upper contribution set between 2~km and 20~km is essentially the free atmosphere seeing, which is dominated by wind speeds at 200 hPa, whereas the lower contribution set between 0~km and 2~km represents the surface layer and planetary boundary layer seeing. For wind speeds greater than 1~m~s$^{-1}$, wind shear produces wind speeds that are approximately proportional to turbulent velocities \citep{Mahrt2013}. Empirically, \citet{Vernin1986} find:
\begin{equation}\label{eq:200-hPa-wind-speed-seeing}
   \int_{2 km}^{20 km} C_n^2(z) dz \approx A \cdot (u_{200 hPa}^2 + v_{200 hPa}^2),
\end{equation}
where $u$ and $v$ are the horizontal components of the wind velocity and $A$ is a calibration constant.

The second approach, hereafter referred to as the 'seeing model', is based on \citet{Osborn2018a}. Their formula for the $C_n^2$ is based on temperature, wind speed and pressure of an atmospheric vertical column and is a combination of formulas introduced by \citet{Tatarskii1971}, \citet{Masciadri1999, Masciadri2017}, and the Gladstone relation \citep{Gladstone1863}. An estimation of the optical turbulence refractive index structure constant $C_n^2$ is given by
\begin{equation}\label{eq:Osborn_seeing}
    C_n^2(z) = k \cdot \left( \dfrac{80 \cdot 10^{-6} \cdot P(z)}{T(z) \Theta(z)} \right)^2 \cdot L(z)^{4/3} \cdot \left(\dfrac{d\Theta}{dz} \right)^2,
\end{equation}
where $P(z)$ is the pressure, $\Theta(z)$ is the potential temperature, $T(z)$ is the temperature at the altitude $z$ and $k$ is a calibration factor, which depends on the atmospheric stability \citep{Masciadri1999}. The calibration factor was determined with the mean of in situ seeing data. For Sutherland, we used a mean of 1.32~arcsec \citep{Catala2013}.
The scale of the largest energy input $L(z)$, also called outer scale \citep{Masciadri1999}, is defined by
\begin{equation}
    L(z) = \sqrt{\dfrac{2S^2}{\dfrac{g}{\Theta(z_i)} \left| \dfrac{d\Theta}{dz} \right| }},
\end{equation}
where the vertical wind shear $S$
\begin{equation}
    S = \left( \left( \dfrac{du}{dz} \right)^2 + \left( \dfrac{dv}{dz} \right)^2 \right) ^{1/2}
\end{equation}
 is the square root of the turbulent kinetic energy.
The potential temperature $\Theta$ is given by
\begin{equation}
    \Theta = T \left(\dfrac{P_0}{P} \right)^{R/c_p}. 
\end{equation}
The variable $R$ is the gas constant and $c_p$ is the specific heat at constant pressure of air ($R/c_p = 0.286$). The variable $P_0$ is the reference pressure at sea level for which we choose $P_0 = 1000$~hPa.

The advantage of this numerical turbulence model is that GCM output can be used for the calculation. However, \citet{Osborn2018a} only validated the seeing model for Cerro Paranal. 
For integrating $C_n^2$ following Eq. \ref{tab:methods_Plevs}, we use the Euler forward numerical scheme and integrate to the pressure level defined in Table \ref{tab:methods_Plevs} to include the turbulence ground layer of the model.
To calculate the seeing, we used equations \ref{eq:Fried-parameter} - \ref{eq:seeing_FWHM} for both approaches. We expect to gain complementary insights from these two different approaches.

\subsubsection{Model orography}\label{subsection:model_orography}
The model surface elevation (also called orography) is different for all models because they all use different numerical grids (Table \ref{Table: PRIMAVERA overview}). 
 Figures \ref{Figure:Orography_MaunaKea} - \ref{Figure:Orography_SPM} show the orography of each PRIMAVERA model as well as the orography of the ERA5 reanalysis for all eight sites.
 Differences in mountain heights between ERA5 and PRIMAVERA GCMs are attributed to different numerical schemes, different spatial resolutions and to  different reference satellite data sets used before being interpolated into the model grids.
 
The MPI, CNRM, ECMWF and EC-Earth PRIMAVERA GCMs use spectral methods that make use of orthogonal spherical harmonics (Table \ref{Table: PRIMAVERA overview}). Despite valuable advantages, spectral models are not able to resolve sharp discontinuities because they use a finite amount of wave numbers. This leads to the creation of unnatural gravity waves (over- and undershooting), called Gibbs oscillations or Gibbs ripples \citep{Navarra1994}. To avoid Gibbs oscillations, the orography is commonly smoothed out \citep[e.g.][]{Vanniere2019}, but more aggressive methods have been found recently, moving ripples from the far ocean closer to the coast \citep{Wedi2014}.

\begin{figure}[thbp!]
        \resizebox{\hsize}{!}{\includegraphics{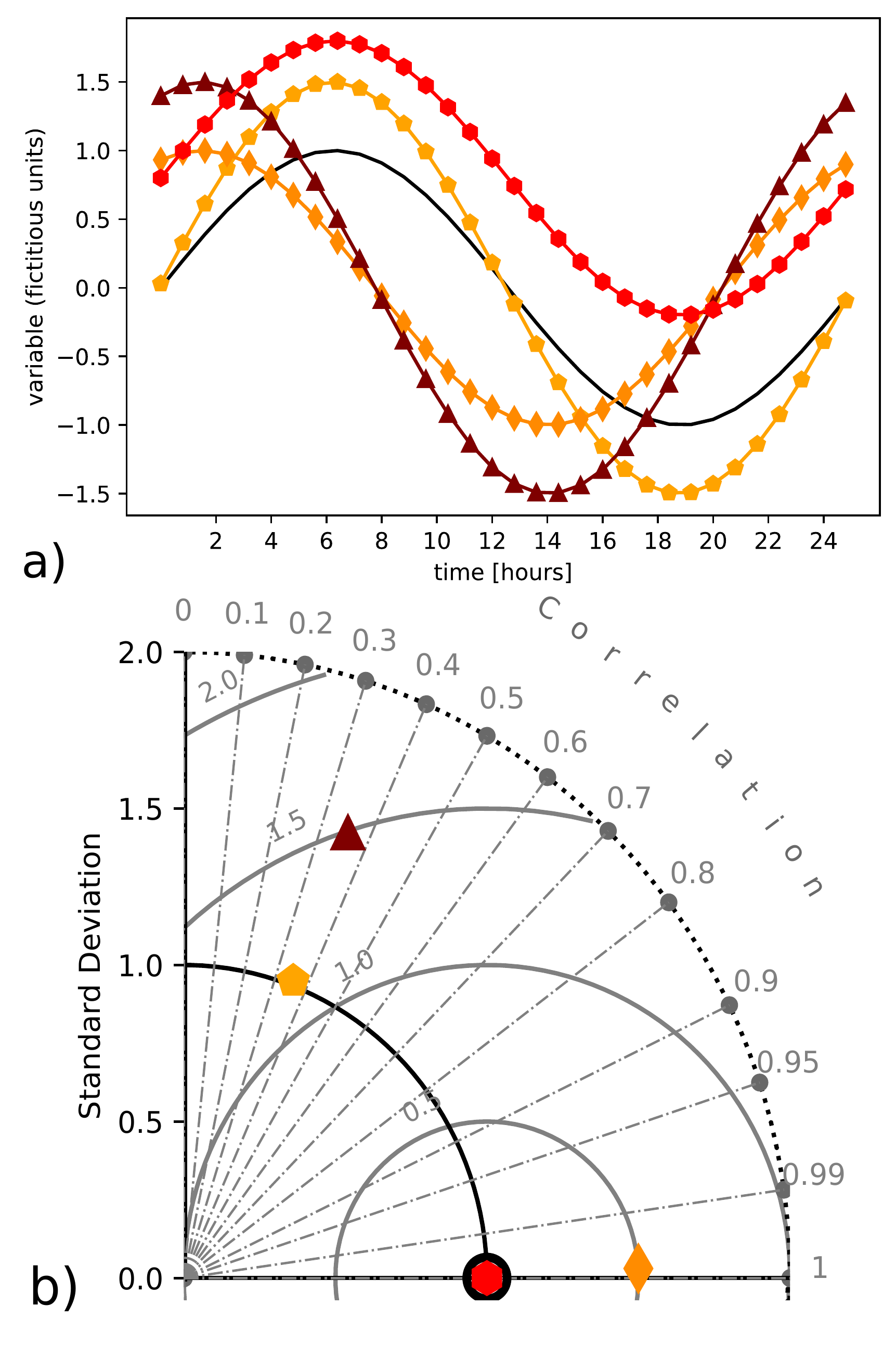}}
        \caption{Example of model performance assessment. a) Fictitious diurnal cycle with b) corresponding Taylor diagram. The solid black line in a) represents the ground truth observation, whereas the solid coloured lines in a) represent the test data. One test dataset shows an offset (or bias) from the ground truth (filled red hexagons), one overestimates the amplitude of the diurnal cycle (filled orange pentagons), one demonstrates a delay in phase (filled dark-orange diamonds) and one that includes both biases (filled maroon triangles. In the Taylor diagram in b), the black circle on the x-axis represents the ground truth observation. The displayed sine-functions are all made up of $\mathrm{y} = \mathrm{a} \cdot \sin(\mathrm{x} + \mathrm{b}) + \mathrm{c} $ with parameters a, b and c as follows:\\ 
        black solid: a $=0$, b $=0$, c $=0$\\
        hexagon (red): a $=0$, b $=0$, c $=0.8$\\ pentagon (orange): a $=1.5$, b $=0$, c $=0$\\ diamond (dark-orange): a $=0$, b $=1.2$, c $=0$\\ up-triangle (maroon): a $=1.5$, b $=1.2$, c $= 0$.}
        \label{fig:Taylor_example_plots_sine}
\end{figure}

All PRIMAVERA orographies as well as the ERA5 orography exhibit a lower elevation than the elevation of the observatory on the nearest grid point for all sites. This is reasonable since the observatories are located on highly elevated terrain.

Observatories located near coastlines or on an island also present a challenge for our analyses. The land-sea mask (contour lines in Figs. \ref{Figure:Orography_MaunaKea} - \ref{Figure:Orography_SPM}) defines the frontier between land and sea. For ERA5, HadGEM and CNRM, the land-sea mask has grid points with fractional values, while for EC-Earth, MPI, ECMWF and CMCC only integer values of 1 for land points and 0 for sea points are used. The ERA5 and PRIMAVERA land-sea masks coincide well with the coastlines (Figs. \ref{Figure:Orography_MaunaKea} - \ref{Figure:Orography_SPM}) in all sites but La Palma.

The island of La Palma is a special case due to its small size. It is treated very differently in all models (Fig. \ref{Figure:Orography_La_Palma}). MPI does not show La Palma as an island (no surface elevation and no land point). ECMWF and EC-Earth show elevated grid points classified as ocean where the island of La Palma lies, so effectively, at the location of La Palma, there is a mountain of water. These are examples of overshooting Gibbs ripples.
ERA5, CNRM and the two CMCC and HadGEM grid point GCMs show a surface elevation classified as land. However, the island is smoothed: its elevation is much lower (by 2174~m at minimum), while its size is much larger (by several grid points). 

In the case of Mauna Kea, which is located on the Hawaiian Big Island that is larger than La Palma, all models show a surface elevation classified as land. Moreover, all nearest grid points are within the coastline boundaries of Big Island, which is not the case for La Palma. However, as for La Palma, all models represent a much smoother orography than the real topography around Mauna Kea.

Despite the horizontal grid spacings of 18 to 50 km, it is challenging for high-resolution PRIMAVERA GCMs  to represent the sites accurately. Based on this assessment, we decided not to include any CMIP5 and CMIP6 GCMs, which are resolved on much coarser horizontal resolutions of 70 to 400~km  and for which the Hawaiian Big Island and La Palma are not represented.

\subsubsection{Model skill score}\label{sec:Model_skill_score}

To evaluate the skills of the GCMs, we use \citet{Taylor2001} diagrams, which combine correlation coefficient, centred root mean square error (RMSE) differences and standard deviation of GCMs and in situ observations. 
Figure \ref{fig:Taylor_example_plots_sine} shows a fictitious diurnal cycle with its corresponding Taylor diagram. The observational data (black solid line in Figure \ref{fig:Taylor_example_plots_sine}.a) constitute the reference dataset used to evaluate the test data (coloured lines). In the example plot, we have four test data sets, the first is simply offset from the ground truth by a constant, the second overestimates the amplitude of the diurnal cycle, the third demonstrates a delay in phase and the fourth includes both biases of the second and third case. First of all, we note that the Taylor diagram (Fig. \ref{fig:Taylor_example_plots_sine}.b) does not account for the bias of the test dataset compared to the reference dataset, so in the first case where the test data set is shifted on the y-axis in Fig. \ref{fig:Taylor_example_plots_sine}.a, the point in the Taylor diagram is the same. In the Taylor diagram (Fig. \ref{fig:Taylor_example_plots_sine}.b), the overestimation of the amplitude is seen as an increase in standard deviation, but reflects perfect correlation: it lies on the x-axis of the Taylor diagram. The third dataset showcases a decreased correlation due to the phase delay, but exhibits the same standard deviation as the observation. Finally, the fourth dataset shows the same correlation as the third dataset and the same standard deviation as the second dataset.

In general, the shorter the distance to the reference point in the Taylor diagram, the better the model skill. There are different approaches to quantify the skill of a model. \citet{Taylor2001} suggests two formulas for the calculation of a skill score, one emphasising the correlation and the other highlighting the pattern variation. In this study, we use the formula that puts emphasis on the correlation
\begin{equation}\label{eq:skillScore}
\centering
        S = \dfrac{4(1 + R)^4}{(\hat{\sigma_f} + 1/\hat{\sigma_f})^2(1 + R_0)^4},
\end{equation}
with $R$ being the correlation coefficient and $\sigma_f$ the standard deviation of the GCM.
The hat on top of the standard deviation denotes the normalisation. The parameter $R_0$ accounts for the fact that no model is able to match observations exactly. It therefore describes the highest possible correlation. Here, $R_0$ is set to 0.995 arbitrarily. The closer to 1 the skill score $S$, the better the skill of the model. 

For this study, we define a guideline for classifying the skill score between model data and observational data (Table \ref{Table:skill score classification}). The guideline is based on requirements for the correlation and the model standard deviation out of which the skill score is calculated using Eq. \ref{eq:skillScore}.

\begin{table}[thbp!]
        \centering
        \caption{Classification of the skill score based on requirements of the correlation (corr) between climate model data and observational in situ data and the standard deviation (std dev) of the model.}
        \label{Table:skill score classification}
        \begin{tabular}{llll}
            \hline \hline
                classification & corr  & std dev & skill score                 \\ \hline
                excellent      & $> 0.95$  & $< \pm 0.05$    & $> 0.91$ \\
                good           & $0.80 - 0.94$   & $\pm (0.04 - 0.20) $  & $0.63 - 0.90$ \\
                mediocre       & $0.60 - 0.79$   & $\pm (0.21 - 0.40) $  & $0.32 - 0.62$ \\
                poor           & $< 0.60$ & $> \pm 0.40$   & $< 0.32$    \\  
                \bottomrule
        \end{tabular}
        \tablefoot{The requirements for the model standard deviation are displayed as deviations from the normalised reference standard deviation, which is always 1.}

\end{table}

\subsection{Methods for trend analysis}\label{sec:methods_for_trend_analysis}
We analyse past and future trends of monthly time series of ERA5 and PRIMAVERA ensemble mean with Bayesian inference. The Bayesian approach allows to approximate the trend per decade while receiving posterior distributions and credible intervals. This way, we make use of all the data available. The credible intervals allow us to decide whether the trends are strong and certain (small credible intervals) or whether the trends are weak and uncertain and not larger than internal climate variabilities (large credible intervals).

We approximate the trends in a linear way which smooths out climate variability since we focus on the mean trends between 2015 and 2050. We chose to display trends in unit per decade to have a natural timescale on which climate change might already be visible, compared to trends expressed in unit per year.

\begin{figure*}[thbp]
        \centering 
        \includegraphics[width=17cm]{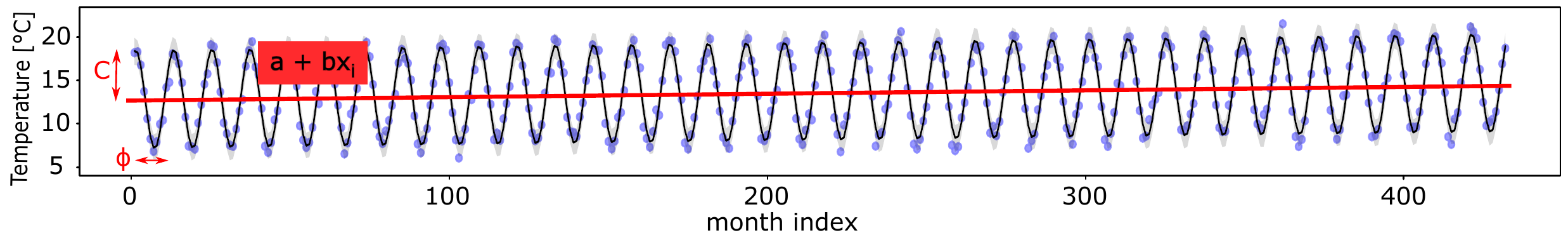}
        \caption{Conceptual visualisation of Bayesian trend analysis with the example of a temperature time series. The monthly data points (blue filled circles) are approximated by a sine with amplitude $C$ and phase $\phi$, combined with a linear function with intercept $a$ and slope $b$, so that a linear increase, decrease or stagnation of an inferred seasonal cycle with progressing month index $x_i$ can be modelled (black line). With this Bayesian approach, sampling from the posterior distribution gives back an estimate of uncertainty, for which we chose an 89~\% percentile credible interval (shaded region).}
        \label{fig:Bayesian_approach_explained}
\end{figure*}

We use a Gaussian probability density function as the Bayesian likelihood function  for our data points
\begin{equation}
    h_i \sim \mathcal{N}(\mu_i,\sigma),
\end{equation}
with the standard deviation $\sigma$ chosen as a weak prior with a uniform distribution with non-zero values between 0 and 10. For the mean $\mu_i$, we chose to infer a sine-shaped seasonal cycle, which could increase, decrease or stagnate linearly with the monthly time series. This progression with time, namely the historical and future yearly trends, is implemented in the model by an added linear function with slope $b$ and intersect $a$. With this, we arrive at 
\begin{equation}\label{Eq:bayes_model}
    \mu_i = a + b\cdot x_i + C \cdot \sin\left(\dfrac{2\pi}{12} \cdot (x_i + \phi)\right),
\end{equation}
where the slope $b$, the amplitude $C$ and the phase $\phi$ were chosen to have weak Gaussian prior distributions, which work with every dataset, centred around 0 with standard deviations of 3, 6 and 6, respectively. The offset $a$ has a weak Gaussian prior distribution centred around the mean of the monthly time series and a standard deviation of 5. The parameter $x_i$, the predictor variable, is the index of the month, starting at zero. 
As an uncertainty interval, we take an 89~\% percentile credible interval of predicted values sampled from the posterior distribution. 
The described Bayesian approach is visualised in Fig. \ref{fig:Bayesian_approach_explained}. In simple words, the Bayesian algorithm searches for a set of parameters $a$, $b$, $C$, $\phi$ and $\sigma$ that, given that Eq. \ref{Eq:bayes_model} needs to be used, fit the data the best.

For the computation, we apply a Hamiltonian Monte Carlo (HMC) algorithm with 1000 samples \citep{Betancourt2017}. We monitor the convergence of the chain by checking if the Rubin convergence diagnostic $\hat{R}$ \citep{Gelman1992} is within 2~\% deviation around 1. If this is not the case, we increase the number of samples to 4000 and run 2 chains.

We applied Bayesian inference using the rethinking package in R.\footnote{\url{https://github.com/rmcelreath/rethinking}, accessed on 2020 Aug 25} 
To allow for reproducibility of the results, we uploaded samples from the posterior distribution to the Github repository as well as a list of simulations for which the Hamiltonian Monte Carlo sampling output chains did not converge optimally.

\section{Evaluation of PRIMAVERA GCMs}\label{sec:results_ERA5_and_PRIMAVERA_evaluation}

\subsection{Current site conditions seen by ERA5 and PRIMAVERA GCMs}
In this section, we focus on the analysis of the PRIMAVERA multi-model ensemble mean for atmos-past and coupled-past simulations (Table \ref{Table: PRIMAVERA overview}), using the skill score classification (Table \ref{Table:skill score classification}). This evaluation aims to provide a basis for assessing the reliability of PRIMAVERA future projections for every site and variable.

The analysis of Taylor diagrams introduced in Sect. \ref{sec:Model_skill_score} shows that, in most cases, the PRIMAVERA multi-model ensemble mean tends to outperform the PRIMAVERA models taken individually. This is in line with previous studies and might be due to a smoothing of individual features \citep[e.g.][]{Gleckler2008, Reichler2008}.

\subsubsection{Temperature}\label{sec:results_T_evaluation}

\paragraph{Seasonal cycle and time series}
The seasonal cycle of temperature (Fig. \ref{Figure: Temperature T - Time series, seasonal cycle and taylor diagram}) is well represented by ERA5 and PRIMAVERA for all sites. For Mauna Kea (Fig. \ref{Figure: Temperature T - Time series, seasonal cycle and taylor diagram}.a), for which the seasonal cycle shows the biggest discrepancies among all sites, all individual PRIMAVERA models as well as ERA5 underestimate the seasonal temperature amplitude. Possible reasons are summarised  in Sect. \ref{sec:discussion_PRIMAVERA_current_site_conditions} after we got an overview of all variables.
Extra-annual oscillations such as ENSO are not represented in the seasonal cycle, but we see peaks in temperature in the time series on Mauna Kea (Fig. \ref{Figure: Temperature T - Time series, seasonal cycle and taylor diagram}.a), the Chilean sites (Fig. \ref{Figure: Temperature T - Time series, seasonal cycle and taylor diagram}.b, c and d), Siding Spring (Fig. \ref{Figure: Temperature T - Time series, seasonal cycle and taylor diagram}.f) and SPM (Fig. \ref{Figure: Temperature T - Time series, seasonal cycle and taylor diagram}.h) that are possibly linked to ENSO, since ERA5, in situ data and most of the time also PRIMAVERA atmosphere-only simulations coincidentally show a peak. Such a peak can be observed for example in the year 2003, 2009 or 2015, which are indeed El Ni\~{n}o years.

\paragraph{ERA5 versus in situ}
Averaged over all sites, ERA5 shows an excellent skill score of 0.95 (Fig. \ref{fig:skill_score_and_trend_analysis}.a and Table \ref{tab:appendix_T_skill_score_classification}).
The site with the highest skill score is Sutherland with an excellent skill score of 1.00 (Fig. \ref{fig:skill_score_and_trend_analysis}.a and Table \ref{tab:appendix_T_skill_score_classification}). This means that 1) in situ data are of high quality, 2) observations that were fed into ERA5 have a good temporal and horizontal resolution for the grid point nearest to Sutherland and 3) the nearest grid point is a valid choice, at least for the temperature. This is also true for the other sites that scored excellent, namely Paranal, La Silla, La Palma, Siding Spring and SPM .
The site with the worst skill score is Cerro Tololo with a good skill score of 0.84 (Fig. \ref{fig:skill_score_and_trend_analysis}.a and Table \ref{tab:appendix_T_skill_score_classification}), which could be attributed to the downward trend in in situ data that is not captured by ERA5 (Fig. \ref{Figure: Temperature T - Time series, seasonal cycle and taylor diagram}.d), on which we elaborate further in Sect. \ref{sec:results_RH_evaluation}. Despite the proximity of Cerro Tololo to La Silla, we do not see a similar decreasing temperature trend  in in situ data of La Silla (Fig. \ref{Figure: Temperature T - Time series, seasonal cycle and taylor diagram}.c).
Just a slightly bit higher scores Mauna Kea with a skill score of 0.86, which might indicate that either 1) in situ data are not of high quality, 2) observations that were fed into ERA5 do not have a sufficient temporal and horizontal resolution for the grid point nearest to Mauna Kea, 3) data assimilation on the ERA5 grid is difficult for Mauna Kea or 4) the nearest grid point does not represent the conditions at the observatory well, at least for the temperature. Regarding the fourth point, we tested other grid points close to the observatory and did not get an improvement. We conclude that ERA5 is not as successful in representing Mauna Kea than it is for most other sites.

\paragraph{PRIMAVERA versus ERA5}
Averaged over all sites, PRIMAVERA shows a good skill score of 0.79 (Fig. \ref{fig:skill_score_and_trend_analysis}.a and Table \ref{tab:appendix_T_skill_score_classification}). This is the highest average skill score over all variables, which is in good agreement with the fact that temperature is the best modelled variable \citep{IPCC2021}.
The site that scores highest is Siding Spring with an excellent skill score of 0.93 (Fig. \ref{fig:skill_score_and_trend_analysis}.a and Table \ref{tab:appendix_T_skill_score_classification}). This is the highest skill score reached by PRIMAVERA in this study.
As discussed previously in Sect. \ref{section: PRIMAVERA}, we expect that PRIMAVERA GCMs show lower skill scores than the reanalysis ERA5 and we see this confirmed for temperature (Fig. \ref{fig:skill_score_and_trend_analysis}.a and Table \ref{tab:appendix_T_skill_score_classification}).
The site with the lowest skill score is Mauna Kea with a mediocre skill score of 0.60 (Fig. \ref{fig:skill_score_and_trend_analysis}.a and Table \ref{tab:appendix_T_skill_score_classification}). Possible causes are summarised in Sect. \ref{sec:discussion_PRIMAVERA_current_site_conditions}. Cerro Paranal shows a similar skill score of 0.61 (Fig. \ref{fig:skill_score_and_trend_analysis}.a and Table \ref{tab:appendix_T_skill_score_classification}). In the Taylor diagram (Fig. \ref{Figure: Temperature T - Time series, seasonal cycle and taylor diagram}.b, Taylor diagram), we see that the standard deviation closely matches with ERA5 and in situ observations, but the correlation of 0.77 is lower compared to other sites. The correlation does not improve when the analyses are performed over eastward grid point neighbours that show higher elevation (not shown; see Fig. \ref{Figure:_Orography_Cerro_Paranal} grid points).

\subsubsection{Relative humidity}\label{sec:results_RH_evaluation}
\paragraph{Seasonal cycle and time series}
The seasonal cycle of relative humidity (Fig. \ref{Figure: Relative Humidity RH - Time series, seasonal cycle and taylor diagram}) reveals a more complex behaviour than the seasonal cycle of temperature (Fig. \ref{Figure: Temperature T - Time series, seasonal cycle and taylor diagram}).
For example in the seasonal cycle of SPM, the relative humidity rapidly rises in July, which is also captured by PRIMAVERA and ERA5 (Fig. \ref{Figure: Relative Humidity RH - Time series, seasonal cycle and taylor diagram}.h). 
In the seasonal cycle of Siding Spring, a decrease in relative humidity after 2014 is not fully captured by ERA5 (Fig. \ref{Figure: Relative Humidity RH - Time series, seasonal cycle and taylor diagram}.f). This bias is not accounted for in the Taylor diagram, which only considers monthly data until 2014 (Fig. \ref{Figure: Relative Humidity RH - Time series, seasonal cycle and taylor diagram}.f).

The in situ time series for Cerro Tololo (Fig. \ref{Figure: Relative Humidity RH - Time series, seasonal cycle and taylor diagram}.d) shows an increase in relative humidity, whereas ERA5 shows a slight decrease. The increasing trend in in situ data is confirmed by an equivalent trend observed by the weather station of the Gemini South Observatory on Cerro Pachon\footnote{Analysis is based on data spanning the years 2008-2020 requested via private communication with the Gemini South Observatory.} (not shown).
While a drying would be expected according to \citep{Grenon}, an increase in relative and specific humidity could be linked to the construction of the 'Embalse Puclaro' storage lake\footnote{lon: -70.85$^{\circ}$, lat: -30.00$^{\circ}$}, which was built in the proximity of Cerro Tololo and filled in 2002 \citep{Ribeiro2015}. This hypothesis is supported by the scarcity of ERA5 observational data in the region of Cerro Tololo and the Puclaro storage lake. The closest land weather station is situated at a longitude of -71.20~$^{\circ}$ and latitude of -29.92~$^{\circ}$, which is approximately 45~km to the west of the storage lake\footnote{Analysis (not shown) is based on data request through the Copernicus Helpdesk at ECMWF \url{https://confluence.ecmwf.int/site/support}, Accessed on 2021 Aug 02)}. Nested climate models by \citet{Bischoff-Gauss2006} confirm changes in temperature and relative humidity 4~km downwind of the storage lake. Even though Cerro Tololo is situated approximately 20 km southward, it might still be influenced. The decreasing temperature (Fig, \ref{Figure: Temperature T - Time series, seasonal cycle and taylor diagram}.d) on Tololo would be in agreement with the enhancement of latent and sensible heat fluxes in the east due to the storage lake \citep{Bischoff-Gauss2006}. However, since in situ data measurements on Tololo start in 2002, we cannot compare to a period without storage lake. Nevertheless, these differences between ERA5 and in situ data on Cerro Tololo stem from a regional effect.
On the other hand, PRIMAVERA does not show any obvious trend and simulates a double-peaking seasonal cycle  that is not seen in the seasonal cycles of in situ or ERA5 data (Fig. \ref{Figure: Relative Humidity RH - Time series, seasonal cycle and taylor diagram}.d) that unlikely stems from an elevation mismatch between ground truth elevation and PRIMAVERA orography, since all individual models show the same structure in the seasonal cycle (not shown).

In contrast to other sites, the in situ seasonal cycle of Mauna Kea does not show a clear seasonal variability (Fig. \ref{Figure: Relative Humidity RH - Time series, seasonal cycle and taylor diagram}.a), which was also found with data from another observatory on Mauna Kea \citep{DaSilva2012}, but the coupled PRIMAVERA ensemble mean shows a peak in September, which explains the negative correlation in the Taylor diagram (Fig. \ref{Figure: Relative Humidity RH - Time series, seasonal cycle and taylor diagram}.a). The atmosphere-only PRIMAVERA simulation, which is constrained by SST measurements, exhibits a smaller amplitude of the seasonal cycle and a higher skill score than the coupled simulation (Fig. \ref{Figure: Relative Humidity RH - Time series, seasonal cycle and taylor diagram}.a), which indicates that the coupling of ocean to atmosphere poses additional problems for Mauna Kea. More reasons why PRIMAVERA does not exhibit the seasonal and interdecadal features seen in observations are discussed in Sect. \ref{sec:discussion_PRIMAVERA_current_site_conditions}. 

\paragraph{ERA5 versus in situ}
Overall, the skill scores of ERA5 show an average value of 0.78, which classifies as good (Fig. \ref{fig:skill_score_and_trend_analysis}.c and Table \ref{tab:appendix_RH_skill_score_classification}). 
For Sutherland, which scores highest, the skill score is excellent with 0.96 (Fig. \ref{fig:skill_score_and_trend_analysis}.c and Table \ref{tab:appendix_RH_skill_score_classification}). Again, we can argue for the same three points as in the above Sect. \ref{sec:results_T_evaluation}.
The site with the lowest skill score of 0.57 is Cerro Tololo (Fig. \ref{fig:skill_score_and_trend_analysis}.c and Table \ref{tab:appendix_RH_skill_score_classification}). It is the only site with a mediocre skill score, since all the sites except Sutherland score good. Reasons discussed in the previous paragraph might explain this.
\paragraph{PRIMAVERA versus ERA5}
On average, over all sites, PRIMAVERA shows a mediocre skill score of 0.39 (Fig. \ref{fig:skill_score_and_trend_analysis}.c and Table \ref{tab:appendix_RH_skill_score_classification}).
The site with the highest skill score is SPM with a mediocre skill score of 0.51 (Fig. \ref{fig:skill_score_and_trend_analysis}.c and Table \ref{tab:appendix_RH_skill_score_classification}).
The site that scores lowest is Sutherland with a poor skill score of 0.21 (Fig. \ref{fig:skill_score_and_trend_analysis}.c and Table \ref{tab:appendix_RH_skill_score_classification}). However, differences between the sites are small (Fig. \ref{fig:skill_score_and_trend_analysis}.c and Table \ref{tab:appendix_RH_skill_score_classification}).
Overall, the underperformance of PRIMAVERA with respect to the performance of ERA5 for relative humidity lets us conclude that PRIMAVERA is not able to represent monthly relative humidity mean values satisfyingly. 

\subsubsection{Specific humidity}\label{sec:results_SH_evaluation}
\paragraph{Seasonal cycle, time series and vertical profile}
Since the in situ specific humidity was calculated with Eq. \ref{Formula: Specific humidity q(T, RH, P)}, the time series and seasonal cycles show influences of both in situ temperature and in situ relative humidity (Fig. \ref{Figure: Relative Humidity RH - Time series, seasonal cycle and taylor diagram} and Fig. \ref{Figure: Specific Humidity SH - Time series, seasonal cycle and taylor diagram}).

The comparison of the median vertical profile of specific humidity between ERA5 and PRIMAVERA shows a good agreement for all sites (Fig. \ref{Figure:SH_vertical_profile}). Therefore, we expect a similarly good agreement between ERA5 and PRIMAVERA for PWV. 
\paragraph{ERA5 versus in situ}
On average, ERA5 has a good skill score of 0.89 (Fig. \ref{fig:skill_score_and_trend_analysis}.e and Table \ref{tab:appendix_SH_skill_score_classification}).
The site with the highest skill score is Sutherland with an excellent skill score of 0.99 (Fig. \ref{fig:skill_score_and_trend_analysis}.e and Table \ref{tab:appendix_SH_skill_score_classification}). 
The lowest skill score is still good with a value of 0.82 and belongs to Mauna Kea. We notice that the spread of skill scores for ERA5 is as small as for the temperature.

\paragraph{PRIMAVERA versus ERA5}
On average, PRIMAVERA has a good skill score of 0.66 (Fig. \ref{fig:skill_score_and_trend_analysis}.e and Table \ref{tab:appendix_SH_skill_score_classification}).
The site with the highest skill score is Siding Spring with 0.82 (Fig. \ref{fig:skill_score_and_trend_analysis}.e and Table \ref{tab:appendix_SH_skill_score_classification}).
The spread of PRIMAVERA skill scores is larger than for temperature due to Mauna Kea, which shows the worst skill score of 0.23 (Fig. \ref{fig:skill_score_and_trend_analysis}.e and Table \ref{tab:appendix_SH_skill_score_classification}). Possible explanations are summarised in Sect. \ref{sec:discussion_PRIMAVERA_current_site_conditions}.

These projected local increases for the eight selected sites are in good agreement with globally increasing near-surface humidity.

\subsubsection{PWV}\label{sec:results_PWV_evaluation}
\paragraph{Seasonal cycle and time series}
As expected due to the close link given by Eq. \ref{Equation: PWV theory}, the PWV time series and seasonal cycle show similarities with specific humidity such as maxima during the same months (Figs. \ref{Figure: Specific Humidity SH - Time series, seasonal cycle and taylor diagram} and \ref{Figure: PWV - Time series, seasonal cycle and taylor diagram}). Less pronounced similarities exist between temperature and PWV and specific humidity due to enhanced evaporation and higher saturation specific humidity with higher temperature.

For Paranal, the in situ data are compared to future simulations of PRIMAVERA because PWV measurements are only measured since 2016 (Fig. \ref{Figure: PWV - Time series, seasonal cycle and taylor diagram}.b). Therefore, the adequate representation of the seasonal cycle (Fig. \ref{Figure: PWV - Time series, seasonal cycle and taylor diagram}.b) and the mediocre skill score of PRIMAVERA (Fig. \ref{fig:skill_score_and_trend_analysis}.g and Table \ref{tab:appendix_PWV_skill_score_classification})  support the credibility of PRIMAVERA future simulations.

\paragraph{ERA5 versus in situ}
On average, ERA5 shows a good skill score of 0.70 (Fig. \ref{fig:skill_score_and_trend_analysis}.g and Table \ref{tab:appendix_PWV_skill_score_classification}). The skill score classification of ERA5 integrated PWV compared with in situ data ranges from mediocre for Mauna Kea and La Silla to excellent for Paranal (Fig. \ref{fig:skill_score_and_trend_analysis}.g and Table \ref{tab:appendix_PWV_skill_score_classification}). For the other sites, no in situ data are available. The excellent result for ERA5 at Paranal observatory strengthens confidence in the measurements performed by LHATPRO.
\paragraph{PRIMAVERA versus ERA5}
On average, PRIMAVERA shows a mediocre skill score of 0.61 (Fig. \ref{fig:skill_score_and_trend_analysis}.g and Table \ref{tab:appendix_PWV_skill_score_classification}).
The site with the best skill score is SPM with a good skill score of 0.86 (Fig. \ref{fig:skill_score_and_trend_analysis}.g and Table \ref{tab:appendix_PWV_skill_score_classification}). Sutherland and Siding Spring score good as well. The three Chilean sites and the two sites situated on islands La Palma and Mauna Kea all score mediocre.
The worst site is Mauna Kea with a mediocre skill score of 0.34 (Fig. \ref{fig:skill_score_and_trend_analysis}.g and Table \ref{tab:appendix_PWV_skill_score_classification}). Possible reasons are summarised in Sect. \ref{sec:discussion_PRIMAVERA_current_site_conditions}.

Over sites where no PWV measurements are available (i.e. Cerro Tololo, La Palma, Siding Spring, Sutherland and SPM), we propose that PRIMAVERA is able to reliably simulate PWV because: 1) the differences in the specific humidity skill scores of PRIMAVERA versus in situ and PRIMAVERA versus ERA5 are small (Fig. \ref{fig:skill_score_and_trend_analysis}.e, blue versus red bars), 2) ERA5 matches good or excellent with in situ specific humidity (Fig. \ref{fig:skill_score_and_trend_analysis}.e) and can therefore be considered as a surrogate of observations and 3) the PWV skill scores of PRIMAVERA versus ERA5 are mediocre or good over sites without in situ PWV, which are comparable to specific humidity skill scores (Fig. \ref{fig:skill_score_and_trend_analysis}.e and g).

\subsubsection{Cloud cover}\label{sec:results_CloudCover_evaluation}
\paragraph{Seasonal cycle and time series}
We find that the seasonal cycle of in situ data corresponds with the seasonal cycle of ERA5 and PRIMAVERA for all sites except Mauna Kea, where ERA5 agrees to some degree with PRIMAVERA, and La Palma, where PRIMAVERA does not show a minimum in July in contrast to ERA5 and in situ data (Fig. \ref{Figure:total_cloud_cover - Time series, seasonal cycle and taylor diagram}).  
In the Taylor diagram of La Palma (Fig. \ref{Figure:total_cloud_cover - Time series, seasonal cycle and taylor diagram}.e), the coupled CNRM and atmosphere-only HadGEM simulations lie close to ERA5, and all individual GCMs are scattered widely, which likely reflects the very different representations of orography. The CNRM and HadGEM models allow for fractional values in the land-sea mask (Sect. \ref{section: PRIMAVERA}) and make up for two out of three models that classify the island of La Palma as land (Fig. \ref{Figure:Orography_La_Palma}.a, e). When we analyse the seasonal cycle of these two models, they indeed show a minimum cloud cover in July (not shown).

\paragraph{ERA5 versus in situ}
On average, ERA5 nighttime cloudiness shows a mediocre skill score of 0.50 when compared to in situ data (Fig. \ref{fig:skill_score_and_trend_analysis}.i and Table \ref{tab:appendix_cloudcover_skill_score_classification}), but no in situ data are available for Sutherland and SPM. This average is a satisfying result given that in situ data include reasons for down-time other than clouds (Sect. \ref{sec:insitu_cloud}).
The site with the highest skill score is Siding Spring with a good skill score of 0.72 (Fig. \ref{fig:skill_score_and_trend_analysis}.i and Table \ref{tab:appendix_cloudcover_skill_score_classification}).
Mauna Kea is the site that exhibits the lowest skill score with a poor value of 0.13 (Fig. \ref{fig:skill_score_and_trend_analysis}.i and Table \ref{tab:appendix_cloudcover_skill_score_classification}). We only considered the year 2014 for comparison for Mauna Kea (Fig. \ref{Figure:total_cloud_cover - Time series, seasonal cycle and taylor diagram}.a). However, for Cerro Tololo, where in situ data also comes from the Gemini observatory (from Gemini South on Cerro Pachon), we compare two years (2013, 2014), but we get a good match between ERA5 and in situ data (Fig. \ref{fig:skill_score_and_trend_analysis}.i and Table \ref{tab:appendix_cloudcover_skill_score_classification}). Therefore, the poor skill score is likely not due to unreliable in situ data, but must be caused by an inadequate representation of cloud cover in ERA5 for the grid point nearest to the observatory, the great distance to the continent and microclimatic conditions. Tests (not shown) by shifting the selected ERA5 grid point to the highest elevated grid point on Hawaii (Fig. \ref{Figure:Orography_MaunaKea}.g) only increased the skill score from 0.13 to 0.16, while it increases from 0.13 to 0.21 if only the 'high cloud cover' (above 6~km) variable is considered. These results still classify as poor (Table \ref{Table:skill score classification}).

\paragraph{PRIMAVERA versus ERA5}
On average, PRIMAVERA shows a mediocre skill score of 0.33 when compared to 24-hour ERA5 data (Fig. \ref{fig:skill_score_and_trend_analysis}.i and Table \ref{tab:appendix_cloudcover_skill_score_classification}). This makes cloud cover the variable with the lowest average skill score and highlights the difficulty in simulating cloud cover, which is essentially due to the coarse grid resolution that does not enable the explicit representation of cloud processes \citep{Bony2015}. 
The site with the highest skill score is Cerro Tololo with a good skill score of 0.69, which is as high as the skill score of ERA5 (Fig. \ref{fig:skill_score_and_trend_analysis}.i and Table \ref{tab:appendix_cloudcover_skill_score_classification}).
The skill scores of PRIMAVERA are lowest for La Palma (Fig. \ref{fig:skill_score_and_trend_analysis}.i and Table \ref{tab:appendix_cloudcover_skill_score_classification}). We already discussed possible reasons in the first paragraph of this section.

\subsubsection{Seeing}\label{sec:results_seeing_evaluation}

\begin{figure*}[thbp]
    \centering
        \includegraphics[height=0.9\textheight]{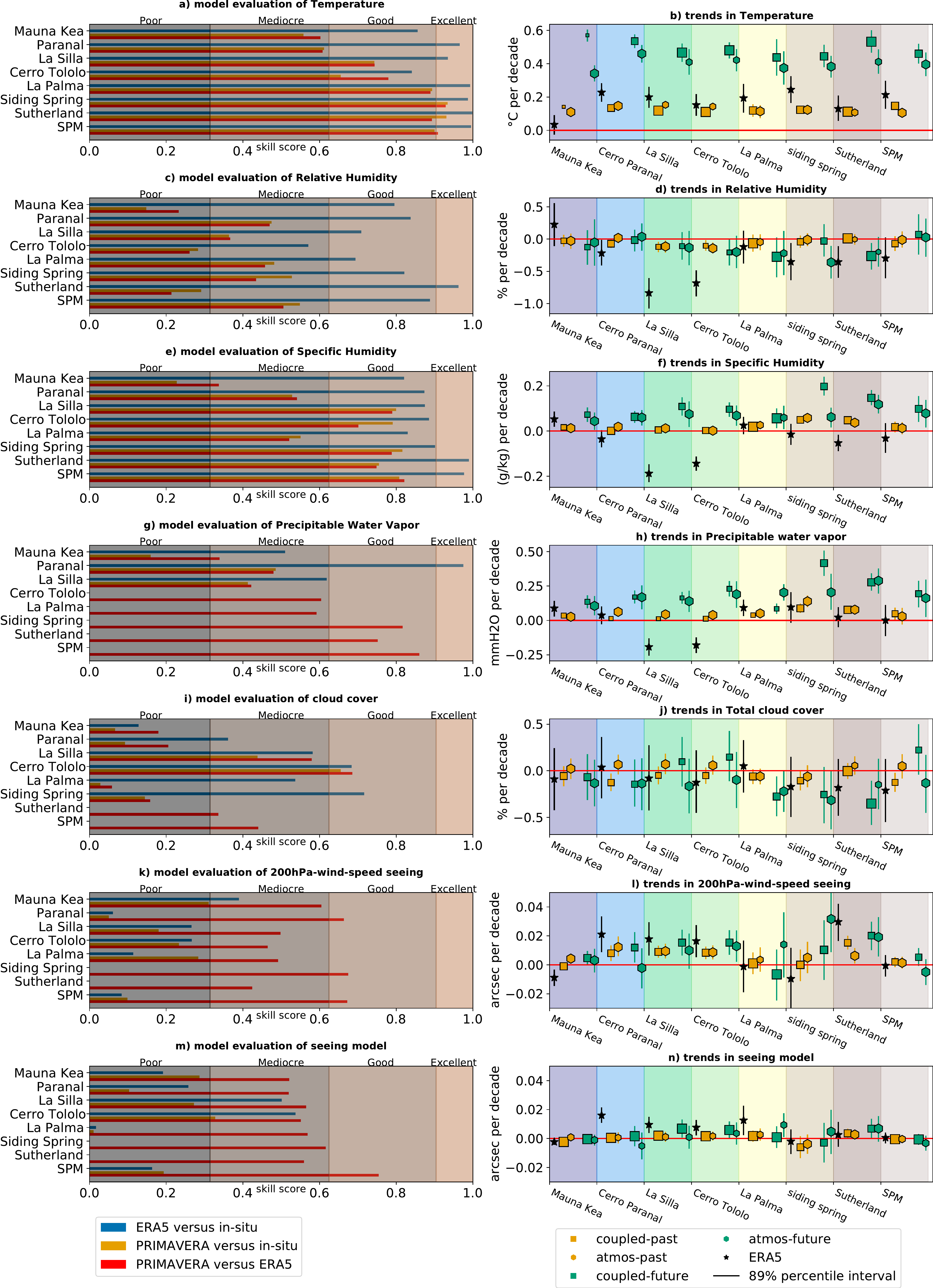}
        \caption{Skill scores (left) and trends (right) for all sites and all variables: temperature (a-b), relative humidity (c-d), specific humidity (e-f), PWV (g-h), total cloud cover (i-j), 200hPa wind-speed seeing (k-l) and seeing model (m-n). Skill scores (Sect. \ref{sec:Model_skill_score}) are shown with their classification (Table \ref{Table:skill score classification}) for the time-intersecting ERA5 versus in situ (blue), time-intersecting PRIMAVERA versus in situ (dark-yellow) and PRIMAVERA versus ERA5 (1979-2014, red). Trends (in units per decade) are shown for ERA5 (1979-2019, black stars), PRIMAVERA historical (1950-2014, dark-yellow markers) and PRIMAVERA future simulations (2015-2050, green markers). Coupled and atmosphere-only simulations are shown with squares and diamonds, respectively. The larger the markers, the higher the skill scores. The error bars show the 89th percentile credible interval of the Bayesian analysis (Sect. \ref{sec:methods_for_trend_analysis}). The horizontal red line marks the zero line (with no trend). The data used to produce this Figure can be found in Tables \ref{tab:appendix_T_skill_score_classification} - \ref{tab:appendix_tab_wind_speed_seeing_trends}.}
        \label{fig:skill_score_and_trend_analysis}
\end{figure*}

\paragraph{Seasonal cycle and time series}
The Taylor diagrams (Figs. \ref{Figure:Seeing_model - Time series, seasonal cycle and taylor diagram} and \ref{Figure:Seeing_200hPa - Time series, seasonal cycle and taylor diagram}) show a separation between the 200-hPa-wind-speed seeing and the seeing model, caused by a generally lower standard deviation of the seeing model.
The extremely poor match of ERA5 seeing model versus in situ data of La Palma (Fig. \ref{fig:skill_score_and_trend_analysis}.m) stems from a misrepresentation of the seasonal cycle (Fig. \ref{Figure:Seeing_model - Time series, seasonal cycle and taylor diagram}.e): 
ERA5 shows a maximum in July and August, whereas in situ data shows a maximum in February. This might be explained with the in situ variability of the temperature inversion layer that leads to best seeing values measured in July and August due to a stronger and lower than usual temperature inversion layer caused by strong trade winds \citep{Munoz-Tunon1997}. This enhances the suppression of local convection which leads to clouds below the temperature inversion. Now, since the seeing model integrates well below the inversion layer, it does not take this seasonal phenomenon into account. The opposite is the case: we see an enhanced contribution from the surface winds compared to other sites (Fig. \ref{Figure:Cn2_vertical_profile}).
Also, no individual PRIMAVERA model shows a better match than the ensemble means (Fig. \ref{Figure:Seeing_model - Time series, seasonal cycle and taylor diagram}.e, Taylor diagram). Based on the slightly better matching 200-hPa-wind-speed seeing data sets for La Palma (Figs. \ref{Figure:Seeing_model - Time series, seasonal cycle and taylor diagram}.e and \ref{Figure:Seeing_200hPa - Time series, seasonal cycle and taylor diagram}.e, Taylor diagram, and Fig. \ref{fig:skill_score_and_trend_analysis}.k and m), we conclude that discrepancies for the seeing model arise from an inadequate representation of the island in the models (Sect. \ref{subsection:model_orography} and Fig. \ref{Figure:Orography_La_Palma}) and a dominant local in situ turbulence ground layer that is poorly linked to global climate. In fact, seeing on La Palma is mostly defined by surface winds which are highly perturbed by the nearby caldera \citep{Vernin1994}.

To explain differences between the 200-hPa-wind-speed and the seeing model approaches, we investigate the vertical profile of the refractive index structure constant $C_n^2$ (Sect. \ref{subsec:Seeing}) computed for ERA5 and PRIMAVERA following Eq. \ref{eq:Osborn_seeing} (Fig. \ref{Figure:Cn2_vertical_profile}). 
We would expect the vertical profile of $C_n^2$ to show a dominant contribution from the ground layer and varying smaller contributions from the free atmosphere \citep{Osborn2018a}. However, \cite{Osborn2018a} highlight the importance of considering more than 100 vertical levels for the computation of $C_n^2$, while 19 vertical pressure levels are provided for PRIMAVERA and 37 for ERA5. Especially the ground layer turbulence cannot be represented adequately with such a small number of vertical levels \citep{Osborn2018a}, and even fine details in topography can have a significant influence on ground layer turbulence \citep{Osborn2018a, Teare2000}. Therefore, it is unsurprising that the $C_n^2$ profiles of PRIMAVERA and ERA5 do not show a dominant contribution of the median ground layer turbulence (Fig. \ref{Figure:Cn2_vertical_profile}). 
To gain deeper insights, we investigate the vertical profiles of northward and eastward wind components of PRIMAVERA and ERA5 (Figs. \ref{u_v_t_z_MaunaKea} - \ref{u_v_t_z_SPM}). In both ERA5 and PRIMAVERA, the eastward wind shows a global maximum at 200 hPa reaching 20~m~s$^{-1}$ for all sites, which is caused by the subtropical jet stream (Figs. \ref{u_v_t_z_MaunaKea} - \ref{u_v_t_z_SPM}). This similarity among sites is expected from \citet{Abahamid2004} (Sect. \ref{sec:formulas_seeing}). The northward wind shows a different profile for each site. Absolute northward winds do not exceed 10~ms$^{-1}$, which indicates that the eastward wind component (jet stream) is the dominant contribution at 200 hPa (Figs. \ref{u_v_t_z_MaunaKea} to \ref{u_v_t_z_SPM}). We conclude that the spatial variability in the vertical $C_n^2$ profile is dominated by local variabilities in the northward wind speed and the ground layer contribution and temporal variations of the eastward wind speed.

Furthermore, there might be a bias in in situ measurements due to challenges in measuring the seeing. For example at Paranal observatory, it might be that the positive trend in seeing seen in Figure \ref{Figure:Seeing_model - Time series, seasonal cycle and taylor diagram}.b is influenced by the construction of various telescopes between 1994 and 2003, which could have led to increased surface turbulence \citep{Cantalloube2020, Sarazin2008}. Moreover, if we compare the more sophisticated DIMM-MASS seeing values measured at Paranal observatory between 2016 and 2019 to ERA5, the skill score of the seeing model would increase from 0.26 (poor) to 0.49 (mediocre; not shown). However, the better skill is likely not only due to the different measurement method with the combination of DIMM and MASS, but also increased by the new location of the DIMM-MASS instrument, which was mounted at a location undisturbed by telescopes \citep{Cantalloube2020}.

\paragraph{ERA5 versus in situ}
On average, ERA5 has a poor skill score of 0.20 for the 200-hPa-wind-speed seeing and a poor skill score of 0.28 for the seeing model (Fig. \ref{fig:skill_score_and_trend_analysis}.k and m and Table \ref{tab:appendix_wind_speed_seeing_skill_score_classification} and \ref{tab:appendix_seeing_model_skill_score_classification}). Likely due to local effects, ERA5 is not representing in situ seeing measurements well.
The site with the highest skill score for the 200-hPa-wind-speed seeing is Mauna Kea with a mediocre skill score of 0.39 (Fig. \ref{fig:skill_score_and_trend_analysis}.k and Table \ref{tab:appendix_wind_speed_seeing_skill_score_classification}). Mauna Kea is also the site with the lowest seeing values and the highest altitude. The $C_n^2$ profile (Fig. \ref{Figure:Cn2_vertical_profile}) reveals that ERA5 shows higher $C_n^2$ values at the jet stream level for Paranal and La Silla compared to Mauna Kea. This goes well together with the fact that the jet stream is less prominent at Mauna Kea compared to Paranal and La Silla, which are closer to -30~$^{\circ}$ latitude. Therefore, we conclude that very low seeing values are better fit with ERA5.
The site with the lowest skill score for the 200-hPa-wind-speed seeing is Paranal with a poor skill score of 0.06 (Fig. \ref{fig:skill_score_and_trend_analysis}.k and Table \ref{tab:appendix_wind_speed_seeing_skill_score_classification}). If we compare ERA5 to MASS-DIMM data from Paranal, we get an equally poor skill score of 0.07 (not shown).
We conclude that the ERA5 200-hPa-wind-speed seeing is not suitable to represent in situ seeing data. As we elaborated in the previous paragraph, the vertical profile justifies the 200-hPa-wind-speed seeing approach. However, since the 200-hPa-wind-speed seeing does not account for the ground layer but is an approximation for the free atmosphere seeing, we would need to compare it to in situ measurements of the free atmosphere. The closest we have is the MASS data of Mauna Kea and SPM. Together with the high elevation of the Mauna Kea observatory, the exclusion of the surface layer by MASS might explain why Mauna Kea is the only site that exhibits a mediocre skill score. On the other hand, it does not explain the poor skill score of SPM.
The site with the highest skill score for the seeing model is Cerro Tololo with a mediocre skill score of 0.54 (Fig. \ref{fig:skill_score_and_trend_analysis}.m and Table \ref{tab:appendix_seeing_model_skill_score_classification}). 
The site with the lowest skill score for the seeing model is La Palma with a poor skill score of 0.02 (Fig. \ref{fig:skill_score_and_trend_analysis}.m and Table \ref{tab:appendix_seeing_model_skill_score_classification}). We already discussed possible explanations for this poor skill score in the previous paragraph.

\paragraph{PRIMAVERA versus ERA5}
On average, PRIMAVERA shows a mediocre skill score of 0.56 for the 200-hPa-wind-speed seeing and a mediocre skill score of 0.58 for the seeing model (Fig. \ref{fig:skill_score_and_trend_analysis}.k and m and Table \ref{tab:appendix_wind_speed_seeing_skill_score_classification} and \ref{tab:appendix_seeing_model_skill_score_classification}).
The site with the highest skill score for the 200-hPa-wind-speed seeing is Siding Spring with a good skill score of 0.68 (Fig. \ref{fig:skill_score_and_trend_analysis}.k and Table \ref{tab:appendix_wind_speed_seeing_skill_score_classification}).
The site with the lowest skill score for the 200-hPa-wind-speed seeing is Sutherland with a mediocre skill score of 0.42 (Fig. \ref{fig:skill_score_and_trend_analysis}.k and Table \ref{tab:appendix_wind_speed_seeing_skill_score_classification}).
The site with the highest skill score for the seeing model is SPM with a good skill score of 0.75 (Fig. \ref{fig:skill_score_and_trend_analysis}.m and Table \ref{tab:appendix_seeing_model_skill_score_classification})
The sites with the lowest skill score for the seeing model are Mauna Kea and Paranal both with a mediocre skill score of 0.52 (Fig. \ref{fig:skill_score_and_trend_analysis}.m and Table \ref{tab:appendix_seeing_model_skill_score_classification}).
For both the 200-hPa-wind-speed seeing and the seeing model approach and in contrast to the other studied variables, ERA5 shows lower skill scores on average than PRIMAVERA. This means that PRIMAVERA agrees better with ERA5 than ERA5 represents in situ data. On the one hand, this is due to the low vertical resolution of ERA5, on the other hand, it indicates the challenge of measuring seeing reliably.

\begin{figure*}[thbp]
        
        \centering
        \includegraphics[height=0.89\textheight]{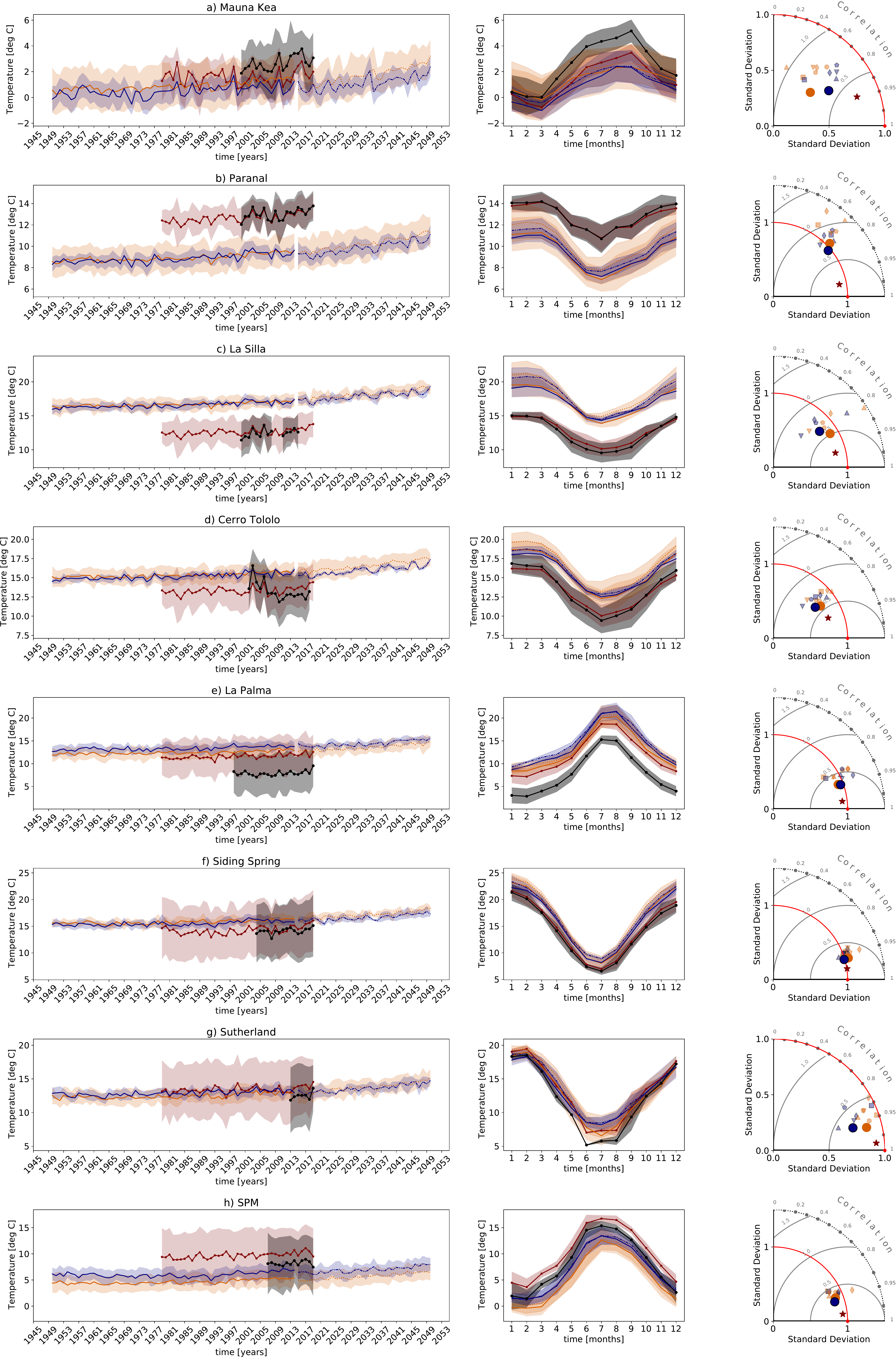}
        \caption{Time series (left column), seasonal cycle (middle column) and Taylor diagram (right column) for temperature (T). The shaded uncertainty regions are standard deviations. The time series, seasonal cycle and Taylor diagram contain in situ data (black filled circles in time series and seasonal cycle; red filled circle on x-axis of Taylor diagram), ERA5 data (maroon filled stars) and PRIMAVERA data for coupled-past (orange solid), coupled-future (orange dotted), atmos-past (blue solid) and atmos-future (blue dotted). In the Taylor diagram, apart from the ensemble mean (filled circles), we show the individual PRIMAVERA GCMs, namely HadGEM3-GC31-HM (pentagon), EC-Earth3P-HR (hexagon), CNRM-CM6-1-HR (square), MPI-ESM1-2-XR (diamond), CMCC-CM2-VHR4 (up-triangle) and ECMWF-IFS-HR (down-triangle). The selected pressure levels are given in Table \ref{tab:methods_Plevs}. The seasonal cycle and the Taylor diagram use monthly time-intersecting data (in situ measurements, ERA5 and historical PRIMAVERA). The future seasonal cycle is plotted for the time period 2015-2050.}
        \label{Figure: Temperature T - Time series, seasonal cycle and taylor diagram}
\end{figure*}

\begin{figure*}[thbp]
        \begin{center}
        \includegraphics[height=0.98\textheight]{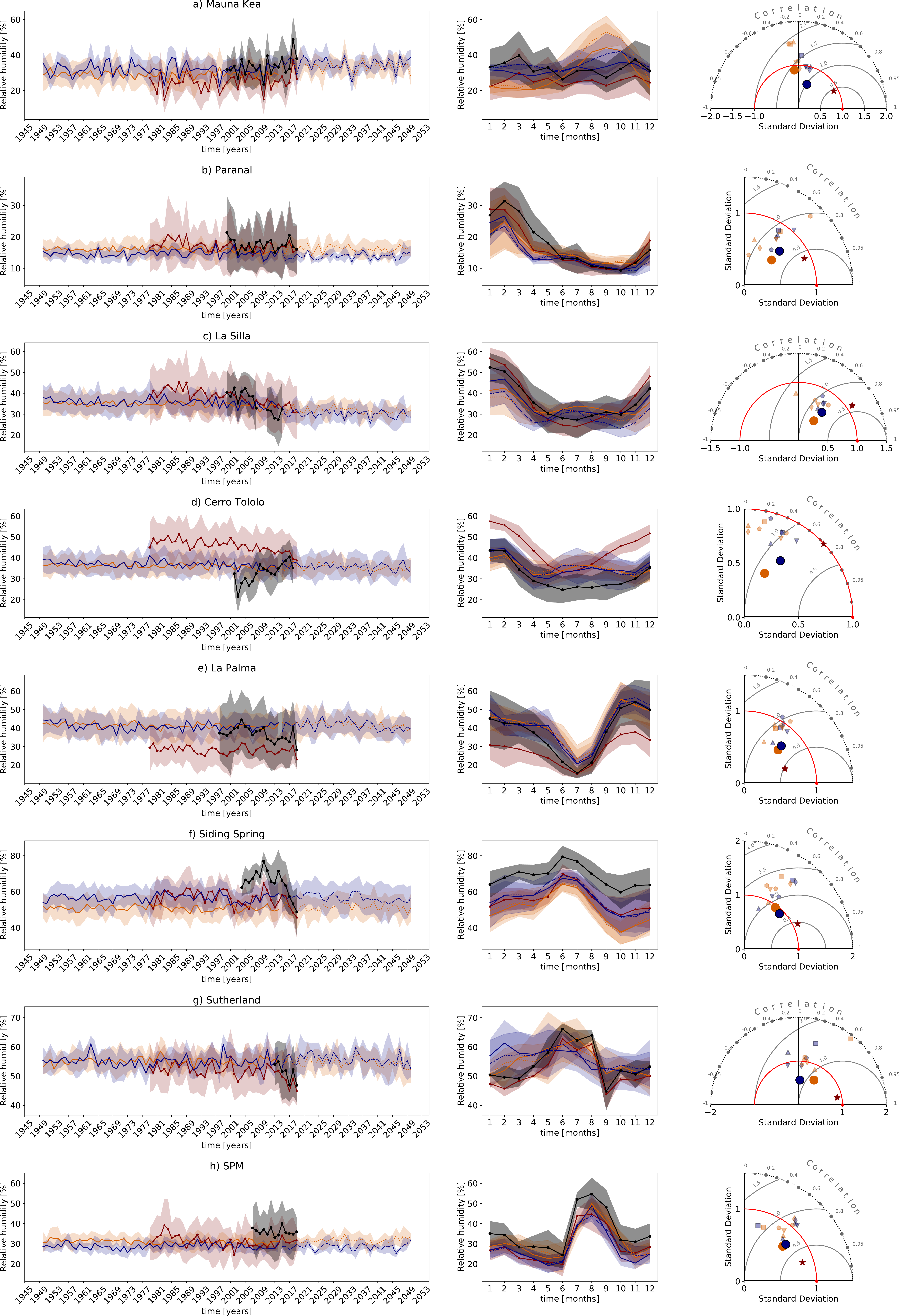}
        \caption{Same as Fig. \ref{Figure: Temperature T - Time series, seasonal cycle and taylor diagram}, but for relative humidity (RH).}
        \label{Figure: Relative Humidity RH - Time series, seasonal cycle and taylor diagram}
        \end{center}
\end{figure*}

\begin{figure*}[thbp]
        \begin{center}
        \includegraphics[height=0.98\textheight]{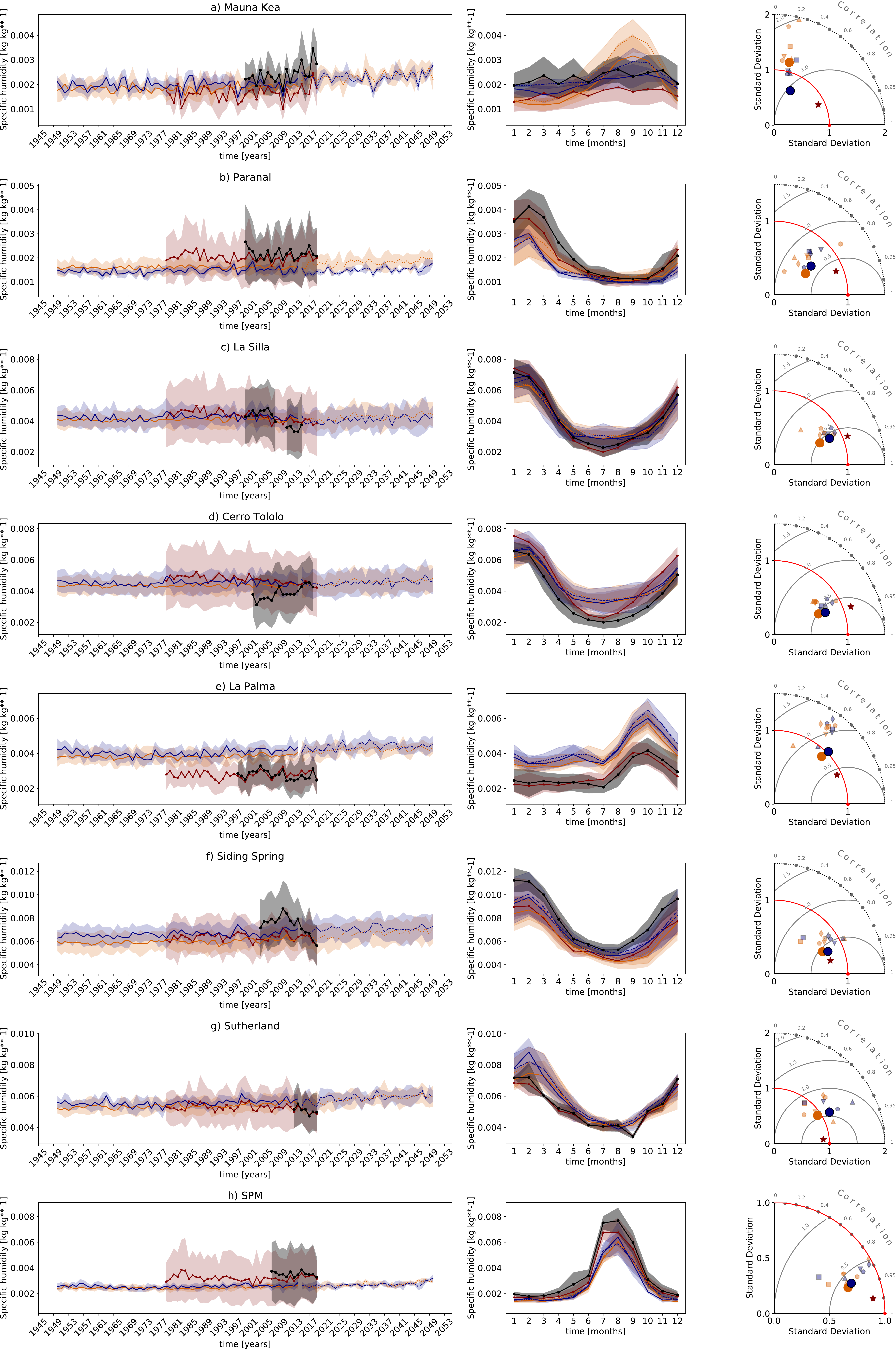}
        \caption{Same as Fig. \ref{Figure: Temperature T - Time series, seasonal cycle and taylor diagram}, but for specific humidity (SH).}
        \label{Figure: Specific Humidity SH - Time series, seasonal cycle and taylor diagram}
        \end{center}
\end{figure*}

\begin{figure*}[thbp]
        \begin{center}
        \includegraphics[height=0.98\textheight]{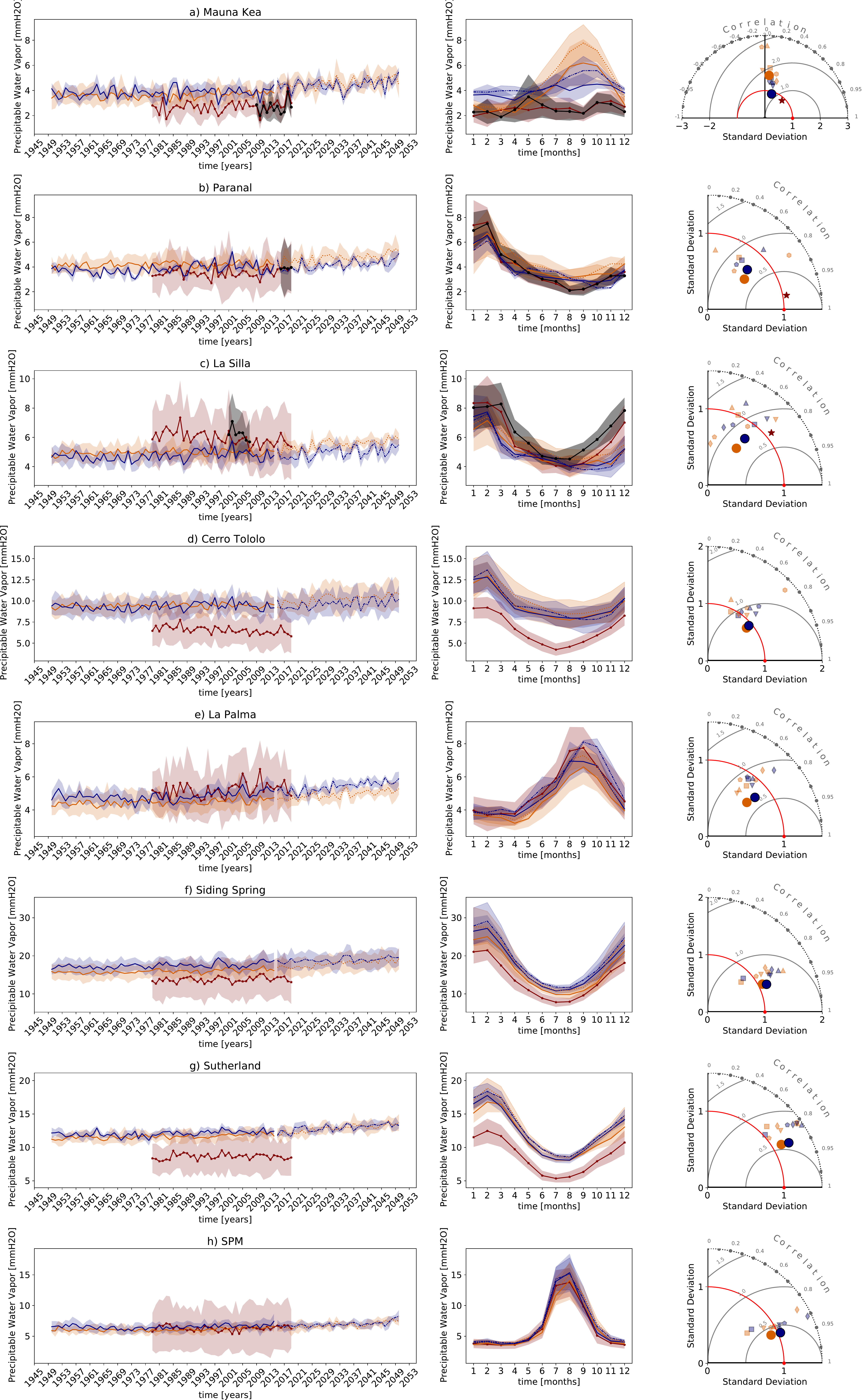}
        \caption{Same as Fig. \ref{Figure: Temperature T - Time series, seasonal cycle and taylor diagram}, but for PWV. For Paranal, the future simulations are taken for the seasonal cycle and the Taylor diagram since in situ data starts only in 2016.}
        \label{Figure: PWV - Time series, seasonal cycle and taylor diagram}
        \end{center}
\end{figure*}

\begin{figure*}[thbp]
        \begin{center}
        \includegraphics[height=0.98\textheight]{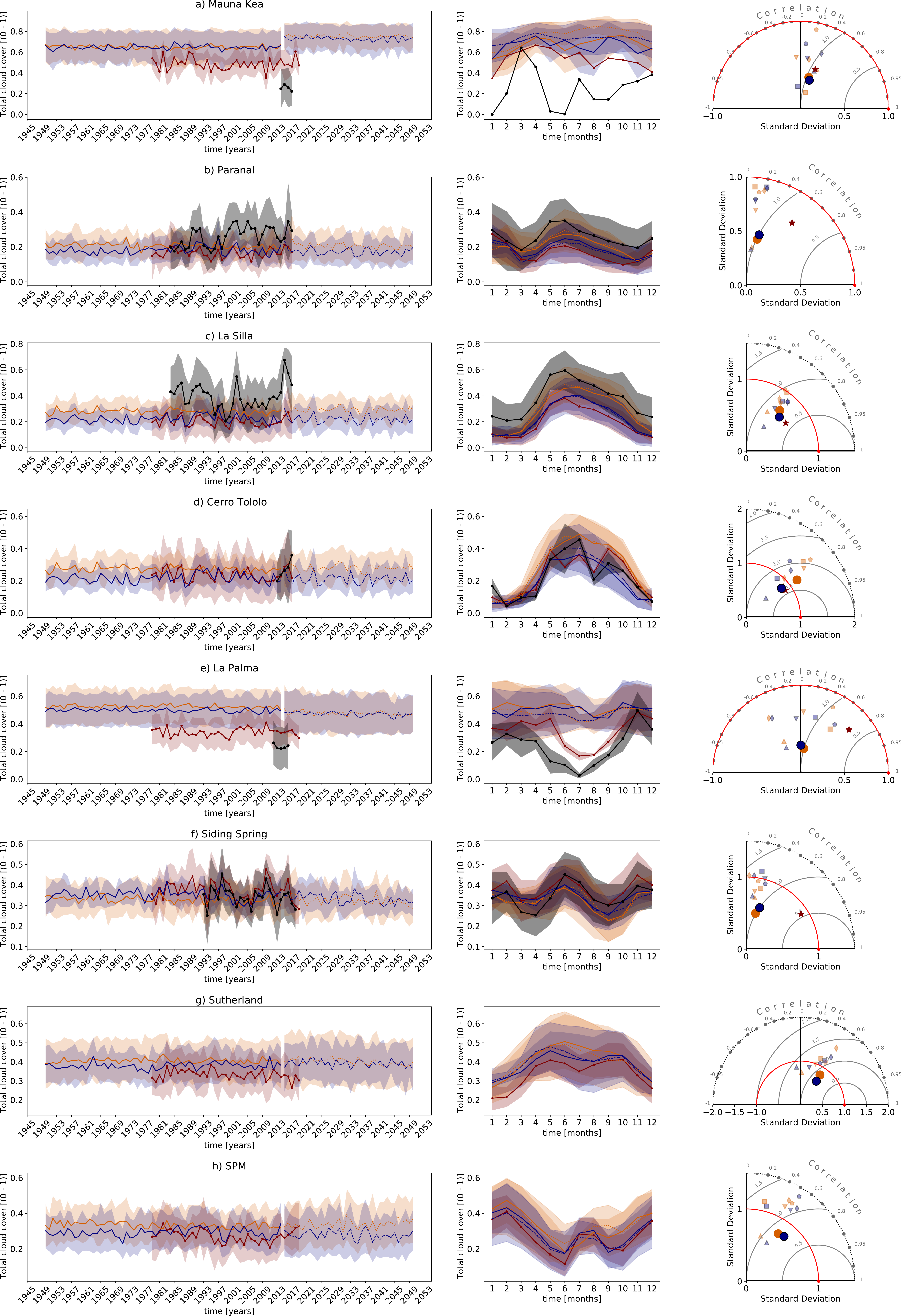}
        \caption{Same as Fig. \ref{Figure: Temperature T - Time series, seasonal cycle and taylor diagram}, but for total cloud cover.}
        \label{Figure:total_cloud_cover - Time series, seasonal cycle and taylor diagram}
        \end{center}
\end{figure*}

\begin{figure*}[thbp]
\begin{center}
        \includegraphics[height=0.98\textheight]{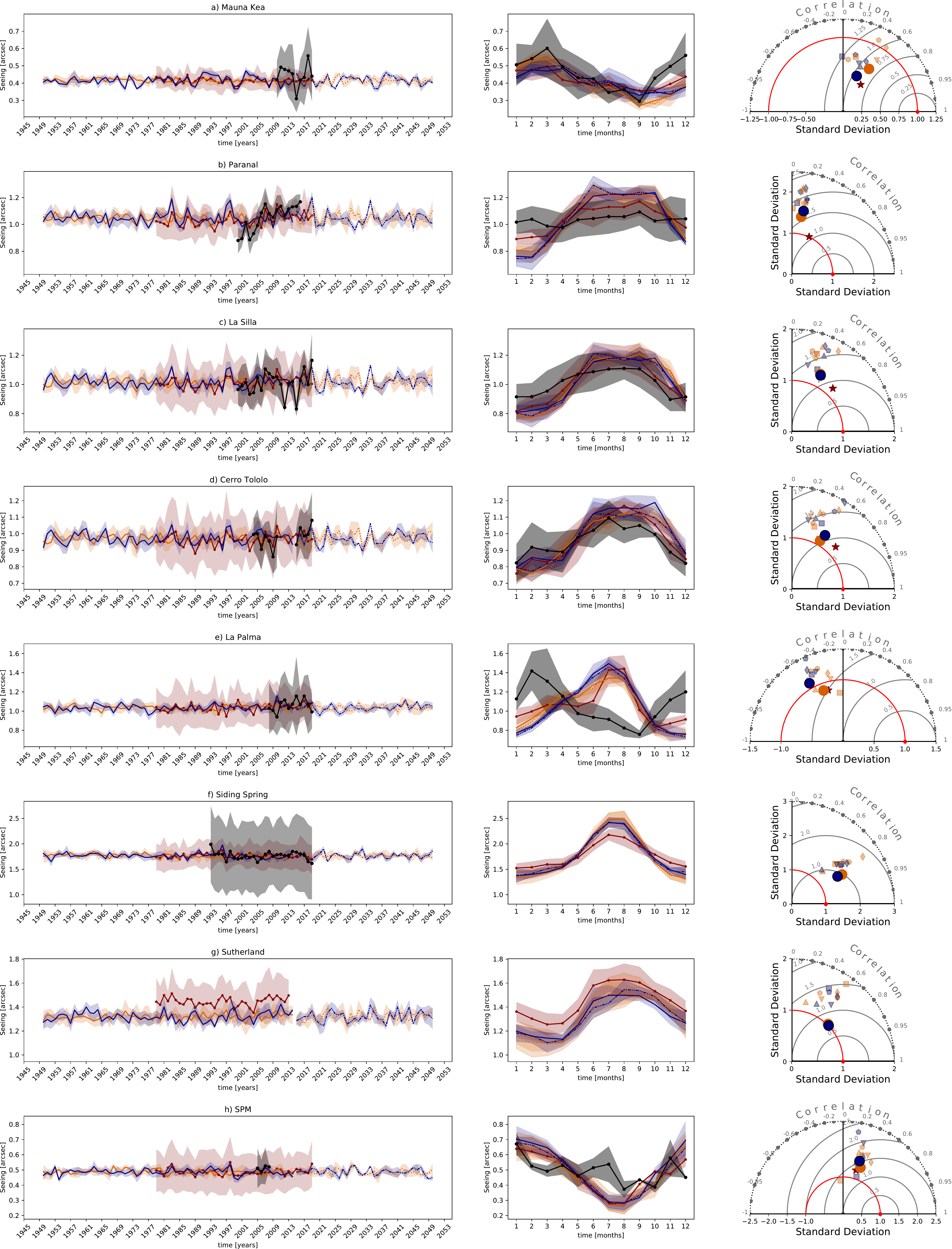}
        \caption{Same as Fig. \ref{Figure: Temperature T - Time series, seasonal cycle and taylor diagram}, but for the seeing model.
        For Siding Spring (f), only yearly in situ data are available, which is therefore neither included in the seasonal cycle nor in the Taylor diagram.}
        \label{Figure:Seeing_model - Time series, seasonal cycle and taylor diagram}
\end{center}
\end{figure*}

\begin{figure*}[thbp]
\begin{center}
        \includegraphics[height=0.98\textheight]{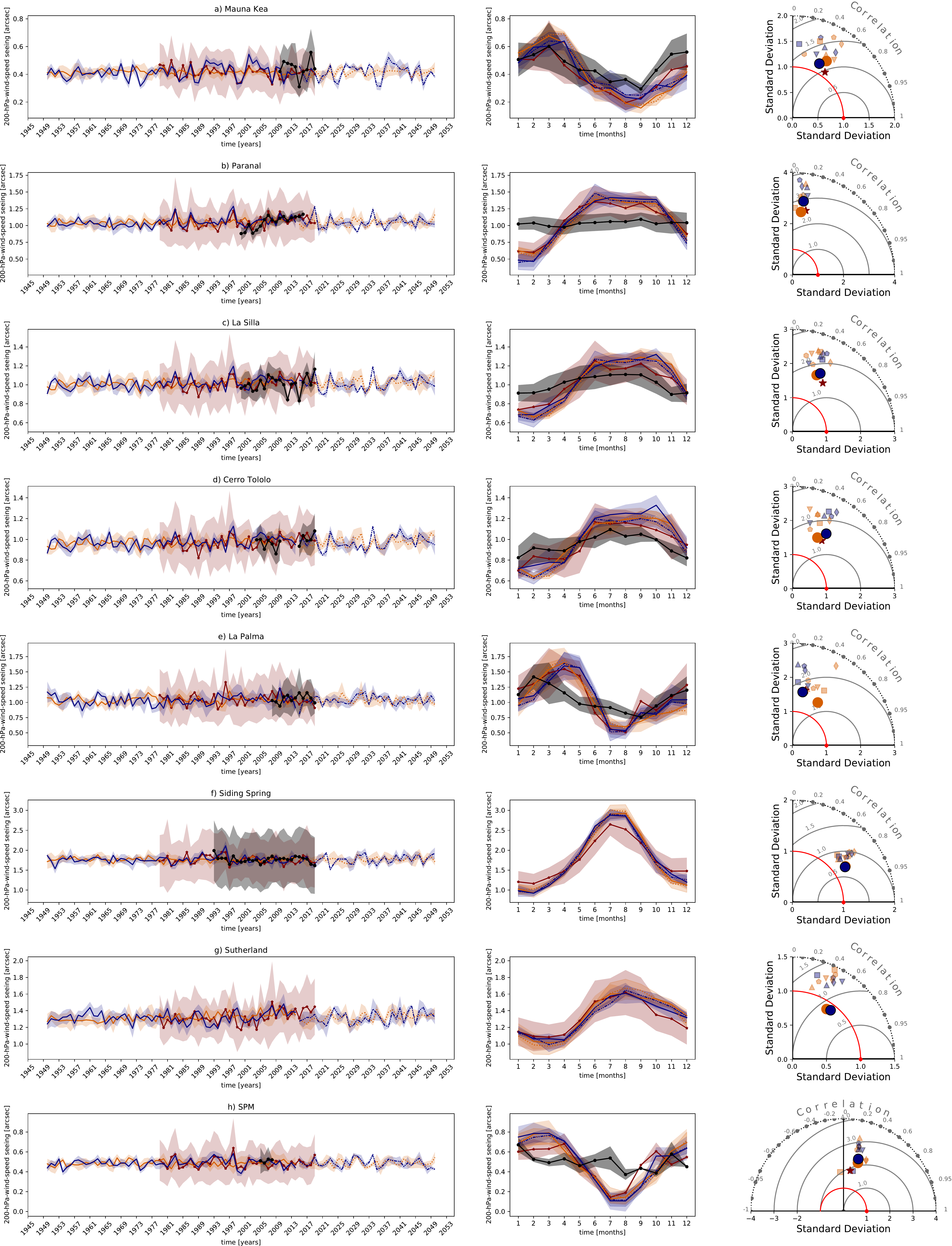}
        \caption{Same as Fig. \ref{Figure: Temperature T - Time series, seasonal cycle and taylor diagram}, but for the 200-hPa-wind-speed seeing.
        For Siding Spring (f), only yearly in situ data are available, which is therefore neither included in the seasonal cycle nor in the Taylor diagram.}
        \label{Figure:Seeing_200hPa - Time series, seasonal cycle and taylor diagram}
\end{center}
\end{figure*}

\subsection{Discussion of adequacy of methods}
\subsubsection{In situ data}
As shown in this study, in situ observations notably allow a thorough evaluation of reanalyses and climate models, which is an important step to assess the trustworthiness of future climate projections. We want to emphasise that it is advantageous for astronomical sites to archive continuous and high-quality in situ observations, with which it is possible to monitor ongoing changes and analyse past trends in site conditions. We found that too few observatories maintain such an archive that is easily accessible.

\subsubsection{ERA5}
The ERA5 reanalysis performs well over most sites. The skill scores of ERA5 versus in situ are always higher than the PRIMAVERA versus ERA5 skill scores, except for seeing. The good agreement of ERA5 with in situ data shows that ERA5 is a suitable data set for analysing past climate and validating GCMs. However, missing or incongruous in situ data, for example down-time data instead of cloud cover measurements or missing monthly seeing data for siding spring, complicate an exhaustive evaluation of ERA5.

\subsubsection{PRIMAVERA}\label{sec:discussion_PRIMAVERA_current_site_conditions}
It would be possible to improve the reliability of PRIMAVERA projections by considering a weighted multi-model mean based on the skill scores of each model \citep{Stocker}. This is not done in our approach, in which all GCMs are given the same weight. More tests and assessments whether a weighted approach performs better together with a more robust skill metric would be highly recommended \citep{Stocker}.

The PRIMAVERA multi-model ensemble mean performs well for most sites and variables. The PRIMAVERA skill scores averaged over all sites classify as good for temperature and specific humidity and as mediocre for relative humidity, PWV, cloud cover and the two seeing approaches. 
The atmosphere-only simulations show higher skill scores than the coupled simulations in 71 \% of all cases. This result is expected since atmosphere-only simulations are constrained by SST measurements. The differences are, however, only small.

The site that shows the best agreement between PRIMAVERA and ERA5 throughout all variables is SPM with an average skill score of 0.71.
The site that shows the worst agreement of PRIMAVERA versus ERA5 is Mauna Kea with an average of 0.40. Mauna Kea is located on the Big Island of Hawaii with steep orography that is not well represented in most PRIMAVERA GCMs because of their still too low resolution. The model orography is rather smooth and exhibits differences in elevation greater than 2000 m compared to observations. Moreover, Big Island lies in the middle of the North Pacific Ocean with a distance of approximately 4000~km to America and more than 7000~km to Australia. The difficulty of GCMs to simulate climate on the island of Hawaii could also be due to the challenging simulation of the boundary trade wind layer regime \citep{Wyant2010}. To address this question, we have compared PRIMAVERA data to the 3 km horizontal grid spacing Hawaii Regional Climate Model (HRCM), based on the Weather Research and Forecasting (WRF) model V3.3 \citep{Zhang2012}. We find that the HRCM is able to represent the diurnal cycle of temperature adequately compared to in situ data (not shown). If we average monthly, the skill score of HRCM 2-metre temperature against in situ data over Mauna Kea is not improved compared to PRIMAVERA (not shown). However, since PRIMAVERA has already a good skill score for temperature, monthly means of other variables could still improve a lot with a regional model such as the HRCM.
Furthermore, the microclimates of Hawaii \citep{Giambelluca1986} highlight the need for a high-resolution model that is able to represent such fine geographical variations. The PRIMAVERA GCMs, with their horizontal grid spacing of 25-50 km, are still too coarse to represent all aspects of the climate on Mauna Kea realistically.

It becomes clear from our results that GCMs coming from the CMIP intercomparisons are not suitable to project changes in all studied sites due to their too coarse grids (i.e. 80 - 200 km for CMIP6 GCMs \citep{IPCC2021}) and the associated inadequate representation of orography, since we found that PRIMAVERA high-resolution GCMs already show weaknesses in some sites. This is especially true for islands such as Hawaii or La Palma that are not represented in CMIP models (not shown).

\section{PRIMAVERA GCMs climate projections}\label{subsubsection:Ensemble_Projections}
It is important to keep in mind that PRIMAVERA future simulations are run under the highest GHG emission scenario SSP5-8.5 \citep{Kriegler2017}. In the following sections, we look at trends in unit per decade. All results are presented visually in Fig. \ref{fig:skill_score_and_trend_analysis}.

\subsection{Trends in temperature}\label{sec:results_T_trends}
All ERA5 and PRIMAVERA historical and future simulations show positive trends in temperature and only Mauna Kea includes the zero-trend line within the 89~\% credible interval (Fig. \ref{fig:skill_score_and_trend_analysis}.b). For all sites except Mauna Kea, ERA5 shows slightly higher trends of the order of 0.1~$^{\circ}$C per decade for the period 1979-2019 than historical PRIMAVERA ensemble means for the period 1950-2014 (Fig. \ref{fig:skill_score_and_trend_analysis}.b). This result is attributed to the different time periods evaluated since anthropogenic climate change likely accelerates the rate of temperature increase with time \citep{Haustein2017}. The PRIMAVERA coupled and atmosphere-only historical simulations show an average trend of \num{0.13\pm0.01}$^{\circ}$C per decade, which is similar to the average trend of ERA5 of \num{0.17\pm0.07}~$^{\circ}$C per decade (Table \ref{tab:appendix_tab_T_trends}). This agreement increases the reliability of these results. 

One explanation of the weak positive trend in ERA5 data for Mauna Kea could be that in free air at the level of Mauna Kea, a decreasing trend in temperature has been observed in the period 1987 – 2017 \citep{Kagawa-Viviani2020}.  In situ data shows, however, that the surface temperature at the elevation of Mauna Kea has risen faster than the surface temperature at sea level, resulting in a more stable atmosphere \citep{Giambelluca2008}, which goes well together with a more frequent trade wind inversion \citep{Cao2007}. Since PRIMAVERA data are showing a higher trend than ERA5, the ERA5 trend must be impacted by local atmospheric effects and topography not seen by PRIMAVERA.

PRIMAVERA projects an average increase in future temperature of \num{0.49\pm0.05}~$^{\circ}$C per decade until 2050 for coupled simulations and of \num{0.40\pm0.03}~$^{\circ}$C per decade for atmosphere-only simulations for the selected sites (Table \ref{tab:appendix_tab_T_trends}). These certain and positive trends in temperature are consistent with global mean surface temperature trends \citep{IPCC2021}.

\subsection{Trends in relative humidity}\label{sec:results_RH_trends}
In relative humidity, ERA5 shows a negative average trend with a large uncertainty of \num{-0.33\pm0.33}~\% per decade (Fig. \ref{fig:skill_score_and_trend_analysis}.d and Table \ref{tab:appendix_tab_RH_trends}). The PRIMAVERA historical simulations do not show any significant trend, with an average of \num{-0.05\pm0.05}~\% per decade (Table \ref{tab:appendix_tab_RH_trends}). Largest differences between PRIMAVERA and ERA5 are found for La Silla and Cerro Tololo (Fig. \ref{fig:skill_score_and_trend_analysis}.d and Table \ref{tab:appendix_tab_RH_trends}).
However, this would not change if other close grid points were considered, since in PRIMAVERA and ERA5, between -70$^{\circ}$ and -71.5$^{\circ}$ longitude and between -31.0$^{\circ}$ and -29.5$^{\circ}$ latitude, trends are similar to the trend in the nearest grid point (not shown). For ERA5, the more we go away from the coast, the faster relative humidity decreases.

Similarly, PRIMAVERA projects either no change or a slight decrease in future relative humidity with an average of \num{-0.13\pm0.12}~\% per decade until 2050 for coupled simulations and \num{-0.11\pm0.10}~\% per decade for atmosphere-only simulations for the selected sites (Fig. \ref{fig:skill_score_and_trend_analysis}.d and Table \ref{tab:appendix_tab_RH_trends}). 
Since PRIMAVERA shows poor to mediocre agreement with ERA5 and in situ data for the historical period (Fig. \ref{fig:skill_score_and_trend_analysis}.c), PRIMAVERA projections of changes in relative humidity over the sites may not be reliable.
Nevertheless, the trends found are in agreement with the literature. \citet{Byrne2018} found a slight absolute decrease of \num{-0.22\pm0.20}~\% per decade from 1979 to 2016 within 40 $^{\circ}$N and 40 $^{\circ}$S, using the HadISDH dataset \citep{Willett2014}. Moreover, the IPCC Fifth Assessment Report \citep{Collins2013} states with medium confidence that small reductions in relative humidity over land are likely and linked to faster rising temperatures over land than over the ocean. More specifically, as saturated oceanic air moves and warms over the warmer land, further moistening of the air over land (i.e. through evapotranspiration) is insufficient to maintain constant relative humidity which therefore decreases. The larger the land-sea temperature gradient, the lower the relative humidity over land. 

\subsection{Trends in specific humidity}\label{sec:results_SH_trends}
Trends in ERA5 specific humidity are negative for all sites except Mauna Kea and La Palma (Fig. \ref{fig:skill_score_and_trend_analysis}.f and Table \ref{tab:appendix_tab_SH_trends}). The positive trend for La Palma might be linked to enhanced evaporation over the ocean caused by ocean warming, which can lead to moister air in ERA5, which mixes values above land and above ocean due to the size of the island of La Palma. The positive trend on Mauna Kea is more delicate and turns negative for one gridpoint to the east (not shown). Furthermore, the trend gets slightly negative for the highest gridpoint in ERA5 (one gridpoint to the south of the nearest gridpoint in Fig. \ref{Figure:Orography_MaunaKea}). This reflects again the difficulty of ERA5 in representing Mauna Kea adequately. Over all sites, ERA5 finds an average trend in specific humidity of \num{-0.05\pm0.08}~g~kg$^{-1}$ per decade (Fig. \ref{fig:skill_score_and_trend_analysis}.f and Table \ref{tab:appendix_tab_SH_trends}). Historical PRIMAVERA coupled and atmosphere-only simulations project positive trends for all sites with an average of \num{0.02\pm0.02}~g~kg$^{-1}$ per decade (Fig. \ref{fig:skill_score_and_trend_analysis}.f and Table \ref{tab:appendix_tab_SH_trends}). As it was the case for relative humidity, we again find the largest differences in trends between ERA5 and PRIMAVERA historical simulations for Cerro Tololo and La Silla (Fig. \ref{fig:skill_score_and_trend_analysis}.f and Table \ref{tab:appendix_tab_SH_trends}). 

Specific humidity is projected to increase slightly over all sites by \num{0.10\pm0.04}~g~kg$^{-1}$ per decade on average until 2050 in the PRIMAVERA coupled simulations and by \num{0.08\pm0.03}~g~kg$^{-1}$ in the atmosphere-only simulations (Fig. \ref{fig:skill_score_and_trend_analysis}.f and Table \ref{tab:appendix_tab_SH_trends}). The median vertical profile of specific humidity shows an increase of specific humidity by future PRIMAVERA simulations over the entire atmospheric column (Fig. \ref{Figure:SH_vertical_profile}).

\subsection{Trends in PWV}\label{sec:results_PWV_trends}
As shown in Eq. \ref{Equation: PWV theory}, PWV depends on specific humidity that is integrated over the entire atmospheric column. Trends in PWV are therefore similar to trends in specific humidity (Fig. \ref{fig:skill_score_and_trend_analysis}.h). ERA5 shows an average trend in PWV of \num{0.00\pm0.12}~mmH$_2$O per decade, while historical PRIMAVERA simulates an average trend of \num{0.05\pm0.03}~mmH$_2$O per decade (Fig. \ref{fig:skill_score_and_trend_analysis}.h and Table \ref{tab:appendix_tab_PWV_trends}). As for specific and relative humidity, the largest disagreements between ERA5 and historical PRIMAVERA trends are for La Silla and Cerro Tololo (Fig. \ref{fig:skill_score_and_trend_analysis}.h and Table \ref{tab:appendix_tab_PWV_trends}). For Paranal, Siding Spring, Sutherland and SPM, ERA5 and historical PRIMAVERA trends agree well (Fig. \ref{fig:skill_score_and_trend_analysis}.h and Table \ref{tab:appendix_tab_PWV_trends}).
These historical trends of increasing PWV are in line with \citet{Trenberth2005} who found positive trends over land over the period 1988-2003 in several reanalyses.
\citet{Hellemeier2019}, however, found no significant trends between 1967 and 2001 in Cerro Paranal, La Palma, Mauna Kea and Siding Spring. These results are based on the lower resolution ERA-40 reanalysis ($2.5^{\circ} \times 2.5^{\circ}$ horizontal grid; \citealp{Kallberg2004}) and consider an older time period than that used in our study. 

Future trends in PWV projected by PRIMAVERA are positive for all sites. The PRIMAVERA simulations project an average increase in PWV by \num{0.21\pm0.10}~mmH$_2$O until 2050 for coupled and by \num{0.18\pm0.06}~mmH$_2$O for atmosphere-only simulations (Fig. \ref{fig:skill_score_and_trend_analysis}.h and Table \ref{tab:appendix_tab_PWV_trends}). An increasing PWV is in line with globally increasing total column water \citep{IPCC2021}.
In opposition to our results, \citet{Cantalloube2020} suspect that the area around Cerro Paranal will become drier, however, these hypotheses are based on low resolution GCMs.

\subsection{Trends in cloud cover}\label{sec:results_CloudCover_trends}
ERA5 shows an average trend in total cloud cover of \num{-0.10\pm0.10}~\% per decade in all sites (Fig. \ref{fig:skill_score_and_trend_analysis}.j and Table \ref{tab:appendix_tab_Clouds_trends}). However, the credible intervals (black error bars in Fig. \ref{fig:skill_score_and_trend_analysis}.j) are large and these trends are therefore not reliable, which is also the case for PRIMAVERA. On average, historical coupled PRIMAVERA simulates a negative trend of \num{-0.07\pm0.04}~\% per decade while atmosphere-only PRIMAVERA simulates a trend of \num{0.03\pm0.05}~\% per decade (Fig. \ref{fig:skill_score_and_trend_analysis}.j and Table \ref{tab:appendix_tab_Clouds_trends}). 
These opposite results between coupled and atmosphere-only simulations and the large uncertainty range highlight the difficulty in simulating cloud cover.
In contrast, \citet{Hellemeier2019} find a significant positive trend of +1.6~\% per decade in cloud cover for the period between 1967 and 2001 with ERA-40 for Cerro Pachon located near Cerro Tololo. This result is not in line with the negative trend of -0.12~\% per decade in ERA5 for Cerro Tololo, and could again be attributed to the coarser resolution of ERA-40 in comparison to ERA5 as well as to the different time period. 

Future trends in cloud cover projected by PRIMAVERA amount on average to \num{-0.08\pm0.21}~\% per decade until 2050 in the coupled simulations and \num{-0.17\pm0.07}~\% in the atmosphere-only simulations, while almost all uncertainty intervals include the possibility of no trends in cloud cover (Fig. \ref{fig:skill_score_and_trend_analysis}.j and Table \ref{tab:appendix_tab_Clouds_trends}). 
These results need to be considered with caution for several reasons. First, PRIMAVERA has poor and mediocre skill scores compared with ERA5 (Fig. \ref{fig:skill_score_and_trend_analysis}.i), so PRIMAVERA projections may not be trustworthy. Second, most GCMs, including PRIMAVERA, do not use grids that are fine enough to represent small-scale cloud processes. They instead rely on sub-grid scale parameterisations that are believed to cause most of the model discrepancies (Sect. \ref{subsec:cloudcover}).
Still, PRIMAVERA uses resolutions that are high enough to represent the dynamical processes associated with cloud formation, such as the location and magnitude of upward and downward motions associated with frontal systems and orography \citep{Haarsma2016}. Their projections may therefore be more trustworthy than conventional IPCC GCMs and the projections of no trend are in line with the results of no significant trends in ERA5 and historical PRIMAVERA. 

\subsection{Trends in seeing}\label{sec:results_seeing_trends}
ERA5 shows an average trend in seeing of \num{0.01\pm0.02}~arcsec per decade for the 200-hPa-wind-speed seeing and \num{0.01\pm0.01}~arcsec per decade for the seeing model (Fig. \ref{fig:skill_score_and_trend_analysis}.l and n; Table \ref{tab:appendix_tab_wind_speed_seeing_trends} and \ref{tab:appendix_tab_seeing_osborng_trends}). Similarly, historical PRIMAVERA simulates an average trend of \num{0.01\pm0.01}~arcsec per decade by coupled and \num{0.01\pm0.00}~arcsec per decade by atmosphere-only simulations for the 200-hPa-wind-speed seeing, and an average trend of \num{0.00\pm0.00} arcsec per decade for both the coupled and the atmosphere-only simulations for the seeing model (Fig. \ref{fig:skill_score_and_trend_analysis}.l and n; Table \ref{tab:appendix_tab_wind_speed_seeing_trends} and \ref{tab:appendix_tab_seeing_osborng_trends}). Over most sites, the 200-hPa-wind-speed seeing shows higher trends than the seeing model for PRIMAVERA and ERA5, but the uncertainty is also larger, which is attributed to the higher variability of the modelled 200-hPa-wind-speed seeing (Fig. \ref{Figure:Seeing_model - Time series, seasonal cycle and taylor diagram} and Fig. \ref{Figure:Seeing_200hPa - Time series, seasonal cycle and taylor diagram}). 
For La Palma, \citet{Hellemeier2019} find a significant decrease in 200-hPa-wind-speed of $-0.6$~ms$^{-1}$ per decade for the period between 1967 and 2001 with ERA-40. This trend would lead to a decrease in the seeing value, which we also find with ERA5 for La Palma, although what we find is not significant.

PRIMAVERA projects an average future trend of \num{0.01\pm0.01} arcsec per decade until 2050 for coupled and atmosphere-only simulations for the 200-hPa-wind-speed seeing (Fig. \ref{fig:skill_score_and_trend_analysis}.l and Table \ref{tab:appendix_tab_wind_speed_seeing_trends}). For the seeing model, PRIMAVERA projects an average trend of \num{0.00\pm0.00} arcsec per decade for coupled simulations and \num{0.00\pm0.01} arcsec per decade for atmosphere-only simulations (Fig. \ref{fig:skill_score_and_trend_analysis}.n and Table \ref{tab:appendix_tab_seeing_osborng_trends}). These projected trends represent a very small change for astronomers. However, as mentioned above, the ground layer contribution to the seeing is not represented adequately in PRIMAVERA GCMs, which is locally highly variable among sites. It is therefore not possible to project changes of the ground layer turbulence into the future with PRIMAVERA. Still, the agreement of both approaches on the magnitude of future trends in astronomical seeing establishes confidence that the free atmosphere seeing is not impacted severely by climate change. 
These results need to be considered with caution because of the poor or mediocre skill scores between PRIMAVERA and in situ data. However, slight positive future trends in 200-hPa-wind-speed seeing could be linked to a strengthening of the subtropical northern and southern winter jet stream, which is already observable \citep{Pena-Ortiz2013, IPCC2021}. 

\subsection{Discussion of PRIMAVERA future trends and possible impacts}
Our analysis of future trends projected by PRIMAVERA shows that astronomers observing at major astronomical observatories at any of the selected sites will likely experience an increase in temperature, specific humidity and PWV by 2050. These trends will likely increase the loss in observing time, but each telescope site must evaluate the severity of these impacts for itself.

Rising temperatures must be considered for planning and building next-generation telescopes. For example, at the Paranal Observatory, the dome cooling system, which prevents air turbulences inside the dome (dome seeing; \citealp{Tallis2020}), is built for a maximum surface air temperature of $16~^{\circ}$C \citep{Cantalloube2020}. 

Our results for specific humidity are coherent with those of temperature and relative humidity. According to the Clausius-Clapeyron relationship, a warmer atmosphere is associated with an exponential increase in atmospheric saturated vapour pressure of about 7~\% for each 1 K warming \citep{Allen2002a,Held2006}. In addition, a rise in temperature enhances potential evaporation, which can lead to an increase in specific humidity but would keep relative humidity almost constant. This is consistent with the increase in temperature and specific humidity and the slight decrease in relative humidity projected by PRIMAVERA.
Consequences for telescopes include a higher risk of condensation due to an increased dew point. For example, the William Herschel telescope at the ORM on La Palma cannot operate when the mirror temperature is only 2~$^{\circ}$C or less above dew point.\footnote{\url{https://www.ing.iac.es//Astronomy/tonotes/misc/weather_safe.html}, accessed on 2021 Oct 14}

Positive trends in PWV are likely leading to a significant decrease in the fraction of the night with excellent PWV conditions in the next 30 years. For example, Paranal currently exhibits PWV values below 1~mm in 8.2~\% of the night (Sect. \ref{subsec:Paranal_site_description}). By 2050, PWV is projected by PRIMAVERA to rise by $0.51\substack{+0.24\\-0.27}$~mm. If we assume a homogeneous shift over the distribution of values between 0.0~mm and 1.5~mm, then PWV values will be below $1.51\substack{+0.24\\-0.27}$~mm in 8.2~\% of the night by 2050, which will decrease the time with excellent observing conditions. If we further assume linearity between 0~mm and 1.5~mm, then PWV values will be below 1~mm in only $2.0\substack{+2.4\\-1.5}$~\% of the night by 2050. A similar picture is shown for La Palma, for which a rise in PWV of \num{0.20\pm0.06}~mm per decade is projected by the atmosphere-only simulations of PRIMAVERA. Given the measurement of \citet{Castro-Almazan2016} (Sect. \ref{subsubsec:site_la_palma}), this would result in PWV values below \num{2.30\pm0.18}~mm in 20~\% of the time, or assuming linearity, this would lead to PWV values below $1.70$~mm in $8.5\substack{+3.0\\-2.6}$~\% of the night-time by 2050. However, this needs to be further investigated by assuming a more realistic distribution instead of a linear approximation. For this conclusion, hourly values of GCM projections would be beneficial. 
Especially for high contrast imaging, a field for which the construction of next-generation telescopes is crucial for emerging \citep{Meyer2018}, an increase in PWV will likely worsen observing conditions and climate change has to be considered urgently for site selection. 

No major changes are projected for seeing, relative humidity, or cloud cover. These results need to be considered with caution due to the low skill score of PRIMAVERA and ERA5 compared to in situ data. Furthermore, trends in seeing are hard to project since the ground layer turbulence contributing to the seeing is extremely sensitive to topography in the proximity of the telescope measuring the seeing \citep{Osborn2018a, Teare2000}. Still, the rise in surface temperature will likely increase the ground layer contribution of the seeing \citep{Cantalloube2020} and the strengthening of the jet stream will likely increase the upper contribution to the seeing. The projected poleward shift would weaken or increase the turbulence contribution of the jet stream for each site individually, dependent on their latitude.

Hypothetically, prolonging the PRIMAVERA atmosphere-only projections to the year 2100 by multiplying the trend per decade by 8 (since there are 8 decades between 2020 and 2100), we would end up with an average increase in temperature of \num{3.94\pm0.40}$^{\circ}$~C, an average increase in specific humidity of \num{0.80\pm0.31}~g~kg$^{-1}$ and an average increase in PWV of \num{1.67\pm0.81}~mmH$_2$O. Certainly, these changes cannot be neglected in the planning and operation of next-generation telescopes. Especially for telescopes observing in the infrared, an average increase of \num{1.67\pm0.81}~mmH$_2$O would unmistakably reduce the amount of observing time with excellent conditions, which is defined by PWV values below or equal to 3~mmH$_2$O \citep{Kidger1998}.
Indeed, the SSP5-8.5 scenario \citep{Kriegler2017} assumes the worst case with ever rising CO$_2$-equivalent emissions to three times the amount of emissions compared with emissions in 2000 \citep{Riahi2011}, which justifies this prolongation. Nevertheless, high-resolution GCMs running until 2100 would lead to refined estimates.

\section{Discussion \& conclusions}\label{sec:concl}

\subsection{Significance of this study}
This study shows that the site selection process for next-generation telescopes and the construction and maintenance of astronomical facilities with a typical lifetime of 30 years requires the consideration of anthropogenic climate change as a potential driver for changes in site characteristics. Nowadays, astronomical observatories are designed to work under the current site conditions and only have a few possibilities for adaptation. Already today, climate change has an influence on every place on Earth \citep{IPCC2021}. 

\subsection{ERA5 and PRIMAVERA historical trends}
Historical trends obtained with the latest ERA5 reanalysis generally do not agree with the trends found by \citet{Hellemeier2019}, using the earlier ERA-40 reanalysis. This result emphasises the influence of model horizontal resolution on long-term trends for a specific site, in addition to other improvements, such as satellite observations and data assimilation techniques \citep{Hersbach2020}.

On average, ERA5 historical trends are positive within their 89~\% credible interval for temperature. For relative and specific humidity, PWV, cloud cover, 200-hPa-wind-speed seeing, and the seeing model, the 89~\% credible interval of ERA5 trends includes the possibility of a zero-trend. 
However, this gives only a rough overview and we shall conclude that climate change has already had an impact on the temperature for most of the selected eight sites. Trends of other variables have a bigger range over the eight sites and no concurrent conclusion can be drawn.

Agreeing well with ERA5 in general, PRIMAVERA historical trends show a clear increase in temperature. Also, for specific humidity, cloud cover, 200-hPa-wind-speed seeing, and the seeing model, the 89~\% credible interval of PRIMAVERA historical trends includes the possibility of a zero-trend. For relative humidity, the historical trend analysis shows a slight decrease and for PWV, a slight increase within the 89~\% credible interval.

\subsection{Strengths and weaknesses of PRIMAVERA GCMs}
The strength of our approach lies in the use of an ensemble of GCMs that provide a resolution (18-50 km horizontal grid spacing) that is much finer than conventional GCMs used in CMIP5 and CMIP6 for IPCC assessment reports \citep{IPCC2014, IPCC2021}. This is the first time that such a study has been possible with an ensemble of high-resolution GCMs. We find that high-resolution GCMs such as those developed within the PRIMAVERA project can be used to analyse the impact of climate change on site characteristics of next-generation telescopes.
Despite PRIMAVERA showing errors and uncertainties, we can rely on an ensemble mean from six different physics-based and globally consistent climate models that project trends and climate change responses at most of the selected sites. Furthermore, since PRIMAVERA GCMs represent the large-scale circulation well, errors at upper pressure levels are likely less severe than errors at the model surface, which are connected to missing scales of motion for turbulence, a chaotic phenomenon, and misrepresentation of islands and elevation in general.

\subsection{Outlook}
With these results, we are confident that PRIMAVERA data provide the grounds to assess current and future climate conditions in sites for astronomical observations. Still, the weaknesses of PRIMAVERA to simulate some local aspects of the selected sites, such as those related to topography, show that it would be better to rely on climate models with an even higher resolution to study local aspects of climate change in astronomical sites. One way would be to use climate models with a horizontal grid spacing of a few kilometres that are able to simulate convection explicitly to improve the simulated diurnal cycle of precipitation and precipitable water. These models could be regional climate models, such as those developed within CORDEX \citep{Giorgi2019} that could be run over domains that include as many astronomical sites as possible, or global climate models such as those developed within the DYnamics of the Atmospheric general circulation Modeled On Non-hydrostatic Domains (DYAMOND) project (\citealp{Stevens2019}), but that are -- at the moment -- still too expensive to run for more than a season. Another very promising effort will be provided through the digital twin of Earth \citep{Bauer2021}, a GCM with a spatial resolution of 3 km that uses data assimilation and the laws of physics to replicate Earth's climate system and project changes into the future as a contribution to the European Union's green deal under the project destination Earth.\footnote{\url{https://digital-strategy.ec.europa.eu/en/policies/destination-earth}, Accessed on 2021 Jul 06} 
A promising project under the European Union's Horizon 2020 research and innovation funding programme is NextGEMS, which aims to provide several global climate models with a spatial horizontal resolution of 2.5~km for the analysis of storms. 
Such GCMs would probably prove useful for the pre-selection of some of the best future spots for astronomical observations.

We have shown in this study that high-resolution GCMs, such as HighResMIP, are suitable for assessing changes in observing conditions of major telescope sites. Since we have focussed our effort on mean changes and trends, we propose further studies to be investigated: 1) Increasing horizontal resolution in coupled GCMs has also been shown to improve the ocean mean state and variability, such as ENSO \citep{Shaffrey2009, Roberts2016}. The impact of such modes of variability on telescope sites, as well as their changes in global warming conditions, could therefore be investigated using these models. 2) Synoptic-scale systems are better resolved in higher resolution GCMs, which improves the simulation of cyclones, atmospheric rivers, and extreme precipitation \citep[e.g.][]{Haarsma2013, Baker2019, Payne2020}. The variability and intensity of tropical cyclones are also better represented \citep[e.g.][]{Roberts2015}. HighResMIP GCMs would therefore be appropriate to assess the impact of such intense events on astronomy observing conditions. 3) Increasing the temporal resolution to hourly data would allow for an investigation of the diurnal cycle. For example, hourly differences between day- and night-time temperature could be assessed, which might be important because bigger day-to-night differences demand more cooling power to bring down the temperature of the telescope mirror to the ambient night-time temperature.

\begin{acknowledgements}
We thank the anonymous reviewer for a careful and detailed review that lead to substantial improvements of this publication.\\ 
The PRIMAVERA project is funded by the European Union’s Horizon 2020 programme, grant agreement no. 641727. We thank all the PRIMAVERA climate modelling groups (Met Office, CERFACS, MPI-M, CMCC, ECMWF, KNMI, SMHI, BSC, CNR) for producing and making their model output available. We also acknowledge the Earth system grid federation infrastructure (ESGF), an international effort led by the U.S. department of energy’s program for climate model diagnosis and intercomparison, the European network for Earth system modelling, and other partners in the global organisation for Earth system science portals (GO-ESSP).\\

Further, we acknowledge the Copernicus climate change service and thank the ECMWF for producing and publishing the improved reanalysis ERA5.
The results contain modified Copernicus climate change service information 2020. Neither the European commission nor ECMWF is responsible for any use that may be made of the Copernicus information or data it contains.\\

Without all the collaborating staff from the observatories who sent us their in situ data, this undertaking would never have come to fruition. We  thank Dr. Andrei Tokovinin, Astronomer at the Cerro Tololo inter-American observatory (CTIO), Edison Bustos, Software Engineer at NOIRLab, and Dr. John Blakeslee, head of science staff for observatory support at NOIRlab. From the Canada France Hawaii telescope (CFHT), we thank Mary Beth Laychak, director of strategic communications, Cameron Wipper, astronomy technical specialist, and Grant Matsushige, senior instrumentation specialist. We thank Dr. Harriet Parsons, support astronomer at the East Asian observatory (EAO). From the University of Hawaii, we thank Dr. Mark Chun, associate director of the University of Hawaii’s institute for astronomy and specialist in instrumentation and data analysis, Dr. Steven Businger, Professor and chair of atmospheric sciences Department, Dr. Luke McKay, telescope engineer, and Robert Calder, senior engineer. We thank Ilse Plauchu-Frayn, astronomer and academic technician in observational support at the Observatorio Astron\'{o}mico Nacional on the Sierra San Pedro M\'{a}rtir (OAN-SPM), Marcello Lodi from the Galileo national telescope (TNG), Dr. Ennio Poretti, director of the TNG, and Dr. Andrew Adamson, associate director of Hawaii site at the Gemini observatory, Dr. Daniel Cotton, astronomer and operations technician at the Anglo-Australian telescope (AAT), and Dr. Steve Lee, senior technician based at the AAT, Michael Sharrott, project manager at Siding Spring observatory, Dr. Daniel Cunnama, science engagement astronomer at the southern African large telescope (SALT), and Dr. Myha Vuong De Breuck, astronomical database specialist at the European Southern Observatory (ESO).\\

M.-E. Demory would like to thank Professor O. Romppainen-Martius and Professor K. Heng for hosting her at the University of Bern. 
We would like to thank Dr. Alexander Kashev from the faculty science IT support of the University of Bern for offering storage and computing power.\\

We implemented code with python using numpy \citep{Harris2020}, scipy \citep{Virtanen2020}, matplotlib \citep{Hunter2007}, pandas \citep{McKinney2010}, netCDF4 \citep{Unidata2020}, xarray \citep{Hoyer2017}, and the skillmetrics package developed by Peter Rochford\footnote{\url{https://github.com/PeterRochford/SkillMetrics}, accessed on 2021 Sep 22}. For implementing Bayesian analysis, we used the computer language R \citep{RCoreTeam2019} including rethinking \citep{McElreath2016}, ggplot2 \citep{Wickham2016} and rstan \citep{Team2020}.\\

Reproduced with permission from Astronomy \& Astrophysics, \textcopyright ESO

\end{acknowledgements}

\pagebreak
\bibliographystyle{aa}
\bibliography{arXiv.bbl}

\onecolumn
\begin{appendix}

\section{Data in tables}\label{appendix:Tables}
\subsection{Skill scores}\label{appendix:skill_score_tables}

\csvstyle{classificationStyle5}{tabular=llllll,
head=false,
late after head = \\\hline\hline,
late after first line = \\\hline,
late after last line=\\\hline}

\csvstyle{classificationStyle3}{tabular=lll,
head=false,
before reading=\centering,
late after head = \\\hline\hline,
late after first line = \\\hline,
late after last line=\\\hline}

\csvstyle{TrendStyle}{tabular=llllll,
head=false,
late after head = \\\hline\hline,
late after first line = \\\hline,
late after last line=\\\hline}

\begin{table*}[htbp!]
\caption{Skill score (Eq. \ref{eq:skillScore}) and skill score classification (Table \ref{Table:skill score classification}) of \textbf{temperature} for all sites.}
\label{tab:appendix_T_skill_score_classification}
\vspace{-0.6cm}
\csvreader[classificationStyle5]%
{Skill_score_tables/Skill_Score_master_table_T.csv}{}%
{\csvcoli & \csvcolii & \csvcoliii & \csvcoliv & \csvcolv & \csvcolvi}%
\tablefoot{The skill scores are colour coded as follows: poor skill score as dark red, mediocre skill score as red, good skill score as orange and excellent skill score as yellow. The best skill score of the atmosphere-only or coupled PRIMAVERA simulation, depending on which scored higher, with respect to in situ data and to ERA5 data is indicated.}
\end{table*}

\begin{table*}[htbp!]
\caption{As Table \ref{tab:appendix_T_skill_score_classification}, but for  \textbf{relative humidity}.}
\label{tab:appendix_RH_skill_score_classification}
\vspace{-0.6cm}
\csvreader[classificationStyle5]%
{Skill_score_tables/Skill_Score_master_table_RH.csv}{}%
{\csvcoli & \csvcolii & \csvcoliii & \csvcoliv & \csvcolv & \csvcolvi}%
\end{table*}

\begin{table*}[htbp!]
\caption{As Table \ref{tab:appendix_T_skill_score_classification}, but for  \textbf{specific humidity}.}
\label{tab:appendix_SH_skill_score_classification}
\vspace{-0.6cm}
\csvreader[classificationStyle5]%
{Skill_score_tables/Skill_Score_master_table_SH.csv}{}%
{\csvcoli & \csvcolii & \csvcoliii & \csvcoliv & \csvcolv & \csvcolvi}%
\end{table*}

\begin{table*}[htbp!]
\caption{As Table \ref{tab:appendix_T_skill_score_classification}, but for \textbf{PWV}. '...' indicates missing in situ data.}
\label{tab:appendix_PWV_skill_score_classification}
\vspace{-0.6cm}
\csvreader[classificationStyle5]%
{Skill_score_tables/Skill_Score_master_table_PWV.csv}{}%
{\csvcoli & \csvcolii & \csvcoliii & \csvcoliv & \csvcolv & \csvcolvi}%
\end{table*}

\begin{table*}[htbp!]
\caption{As Table \ref{tab:appendix_T_skill_score_classification}, but for \textbf{cloud cover}. '...' indicates missing in situ data.}
\label{tab:appendix_cloudcover_skill_score_classification}
\vspace{-0.6cm}
\csvreader[classificationStyle5]%
{Skill_score_tables/Skill_Score_master_table_cloud_cover.csv}{}%
{\csvcoli & \csvcolii & \csvcoliii & \csvcoliv & \csvcolv & \csvcolvi}%
\end{table*}

\begin{table*}[htbp!]
\caption{As Table \ref{tab:appendix_T_skill_score_classification}, but for \textbf{200-hPa-wind-speed seeing}. '...' indicates missing in situ data.}
\label{tab:appendix_wind_speed_seeing_skill_score_classification}
\vspace{-0.6cm}
\csvreader[classificationStyle5]%
{Skill_score_tables/Skill_Score_master_table_200hPa-seeing.csv}{}%
{\csvcoli & \csvcolii & \csvcoliii & \csvcoliv & \csvcolv & \csvcolvi}%
\end{table*}

\begin{table*}[htbp!]
\caption{As Table \ref{tab:appendix_T_skill_score_classification}, but for \textbf{seeing model}. '...' indicates missing in situ data.}
\label{tab:appendix_seeing_model_skill_score_classification}
\vspace{-0.6cm}
\csvreader[classificationStyle5]%
{Skill_score_tables/Skill_Score_master_table_seeing_model.csv}{}%
{\csvcoli & \csvcolii & \csvcoliii & \csvcoliv & \csvcolv & \csvcolvi}%
\end{table*}

\FloatBarrier

\subsection{Future climate trends}\label{appendix:trends}

\begin{table*}[htbp!]
\renewcommand{\arraystretch}{1.5}
\caption{Trends of \textbf{temperature} (in $^{\circ}C$ per decade) for ERA5 (1979-2019), PRIMAVERA historical (1950-2014) simulations and PRIMAVERA future (2015-2050) simulations.}
\label{tab:appendix_tab_T_trends}
\vspace{-0.7cm}
\csvreader[TrendStyle]%
{Trends_tables/Trends_master_table_all_forcings_T.csv}{}%
{\csvcoli & \csvcolii & \csvcoliii & \csvcoliv & \csvcolv & \csvcolvi}%
\tablefoot{The 89\% percentile credible interval of predicted values sampled from the posterior distribution (Sect. \ref{sec:methods_for_trend_analysis}) is given in brackets.}
\vspace{0.2cm}
\end{table*}

\begin{table*}[htbp!]
\renewcommand{\arraystretch}{1.5}
\caption{As Table \ref{tab:appendix_tab_T_trends}, but for \textbf{relative humidity} (in \% per decade).}
\label{tab:appendix_tab_RH_trends}
\vspace{-0.7cm}
\csvreader[TrendStyle]%
{Trends_tables/Trends_master_table_all_forcings_RH.csv}{}%
{\csvcoli & \csvcolii & \csvcoliii & \csvcoliv & \csvcolv & \csvcolvi}%
\vspace{0.6cm}
\end{table*}

\begin{table*}[htbp!]
\renewcommand{\arraystretch}{1.5}
\caption{As Table \ref{tab:appendix_tab_T_trends}, but for \textbf{specific humidity} (in g~kg$^{-1}$ per decade).}
\label{tab:appendix_tab_SH_trends}
\vspace{-0.7cm}
\csvreader[TrendStyle]%
{Trends_tables/Trends_master_table_all_forcings_SH.csv}{}%
{\csvcoli & \csvcolii & \csvcoliii & \csvcoliv & \csvcolv & \csvcolvi}%
\end{table*}

\begin{table*}[htbp!]
\renewcommand{\arraystretch}{1.5}
\caption{As Table \ref{tab:appendix_tab_T_trends}, but for \textbf{PWV} (in mmH$_2$O per decade).}
\label{tab:appendix_tab_PWV_trends}
\vspace{-0.7cm}
\csvreader[TrendStyle]%
{Trends_tables/Trends_master_table_all_forcings_TCW.csv}{}%
{\csvcoli & \csvcolii & \csvcoliii & \csvcoliv & \csvcolv & \csvcolvi}%
\vspace{0.6cm}
\end{table*}

\begin{table*}[htbp!]
\renewcommand{\arraystretch}{1.5}
\caption{As Table \ref{tab:appendix_tab_T_trends}, but for \textbf{cloud cover} (in \% per decade).}
\label{tab:appendix_tab_Clouds_trends}
\vspace{-0.7cm}
\csvreader[TrendStyle]%
{Trends_tables/Trends_master_table_all_forcings_total_cloud_cover.csv}{}%
{\csvcoli & \csvcolii & \csvcoliii & \csvcoliv & \csvcolv & \csvcolvi}%
\vspace{0.6cm}
\end{table*}

\begin{table*}[htbp!]
\renewcommand{\arraystretch}{1.5}
\caption{As Table \ref{tab:appendix_tab_T_trends}, but for \textbf{seeing model} (\ref{eq:Osborn_seeing}) (in arcseconds per decade).}
\label{tab:appendix_tab_seeing_osborng_trends}
\vspace{-0.7cm}
\csvreader[TrendStyle]%
{Trends_tables/Trends_master_table_all_forcings_seeing_osborn.csv}{}%
{\csvcoli & \csvcolii & \csvcoliii & \csvcoliv & \csvcolv & \csvcolvi}%
\vspace{0.6cm}
\end{table*}

\begin{table*}[htbp!]
\renewcommand{\arraystretch}{1.5}
\caption{As Table \ref{tab:appendix_tab_T_trends}, but for \textbf{200-hPa-wind-speed seeing} (\ref{eq:200-hPa-wind-speed-seeing}) (in arcseconds per decade).}
\label{tab:appendix_tab_wind_speed_seeing_trends}
\vspace{-0.7cm}
\csvreader[TrendStyle]%
{Trends_tables/Trends_master_table_all_forcings_wind_speed_seeing.csv}{}%
{\csvcoli & \csvcolii & \csvcoliii & \csvcoliv & \csvcolv & \csvcolvi}%
\vspace{0.6cm}
\end{table*}

\FloatBarrier
\pagebreak

\section{Diurnal cycle}\label{appendix:diurnal_cycle}

\begin{figure*}[htbp!]
        \centering
        \includegraphics[width=17cm]{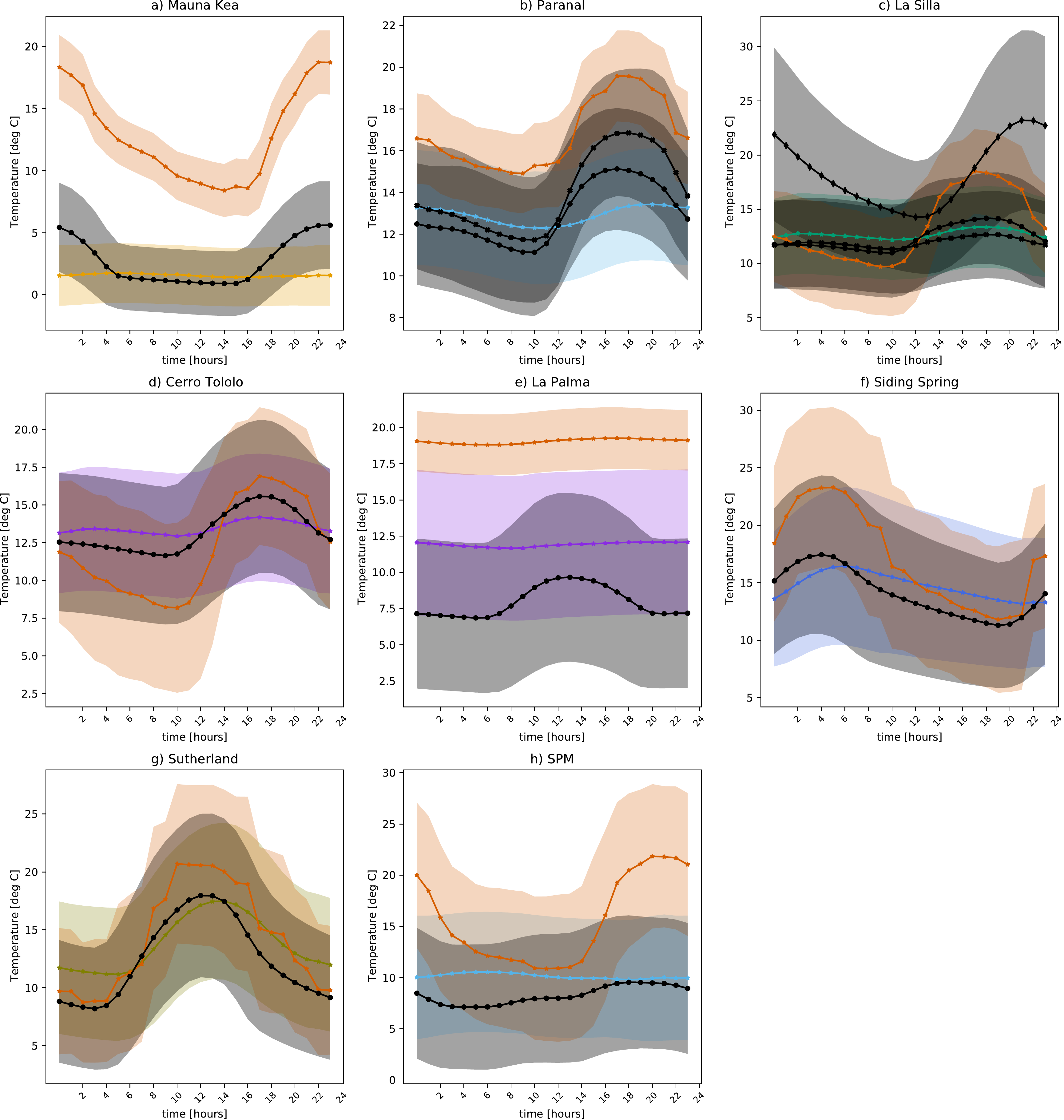}
        \caption{Diurnal cycle of temperature for in situ data (black markers) and ERA5 data (coloured markers). The coloured lines indicate the ERA5 two-metre temperature (orange) and ERA5 data on pressure levels 600 hPa (dark-yellow; a), 750 hPa (bright blue; b and h), 775 hPa (green; c), 800 hPa (blue-violet; d and e), 850 hPa (olive; g) and 900 hPa (royalblue; f).. The black x-filled markers in b) represent additional in situ data measured 20 m below the telescope platform at Paranal observatory. The x-filled and the diamond markers in c) represent additional in situ data 30 m above ground and directly at the ground station of La Silla observatory, respectively. The grey and coloured shadings represent the standard deviation of the in situ and ERA5 data based on hourly averaged data, respectively. Time is in universal time (UT).}
        \label{Figure:Diurnal_temperature}
\end{figure*}

\begin{figure*}[htbp!]
        \centering
        \includegraphics[width=17cm]{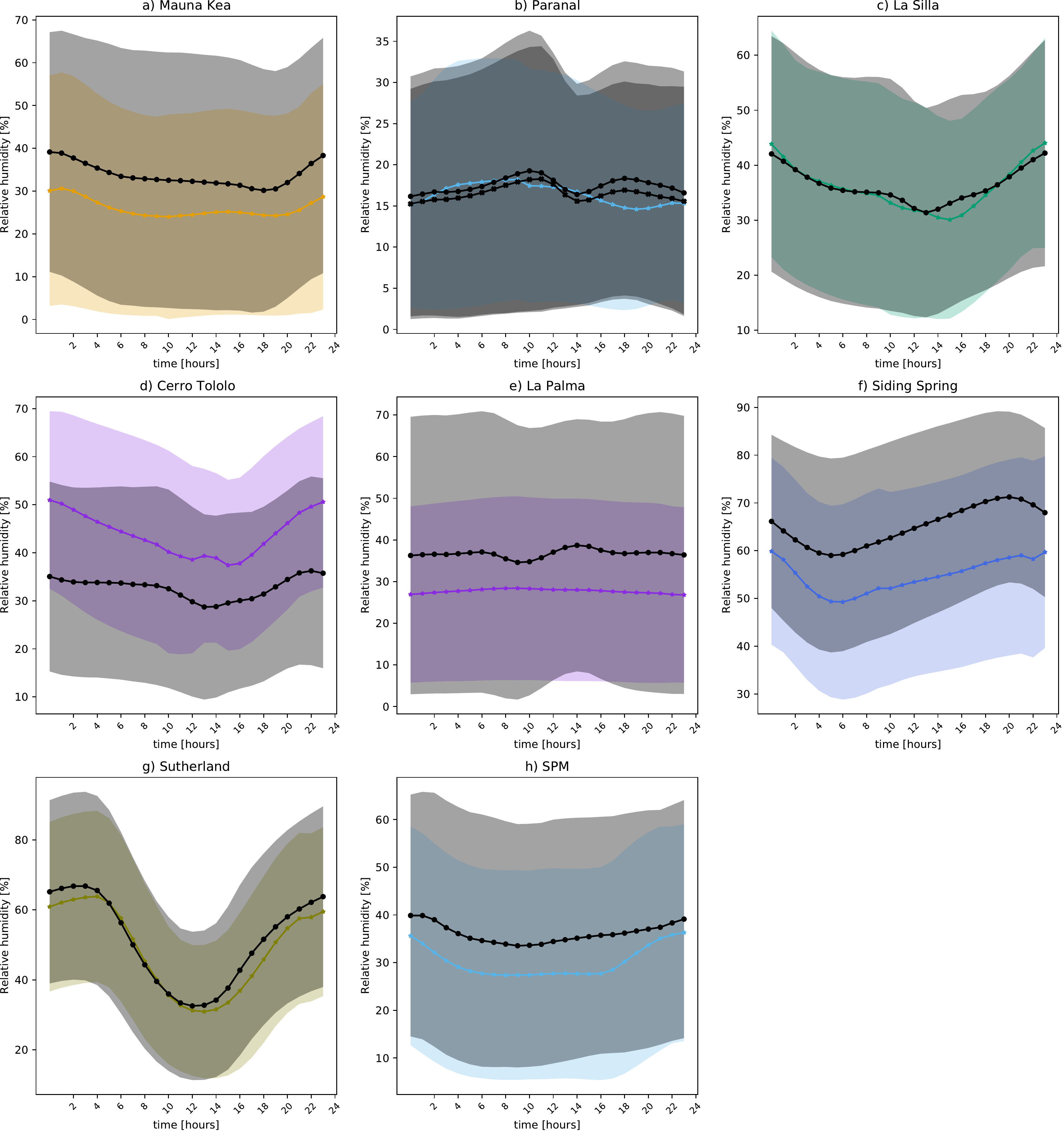}
        \caption{Diurnal cycle of relative humidity for in situ data (black markers) and ERA5 data (coloured markers). The coloured lines indicate ERA5 data on pressure levels 600 hPa (dark-yellow; a), 750 hPa (bright blue; b and h), 775 hPa (green; c), 800 hPa (blue-violet; d and e), 850 hPa (olive; g) and 900 hPa (royalblue; f). The black x-filled markers (b) represent additional in situ data measured 20 m below the telescope platform at Paranal observatory. The grey and coloured shadings represent the standard deviation of the in situ and ERA5 data based on hourly averaged data, respectively. Time is in universal time (UT).}
        \label{Figure:Diurnal_RH}
\end{figure*}

\begin{figure*}[htbp!]
        \centering
        \includegraphics[width=17cm]{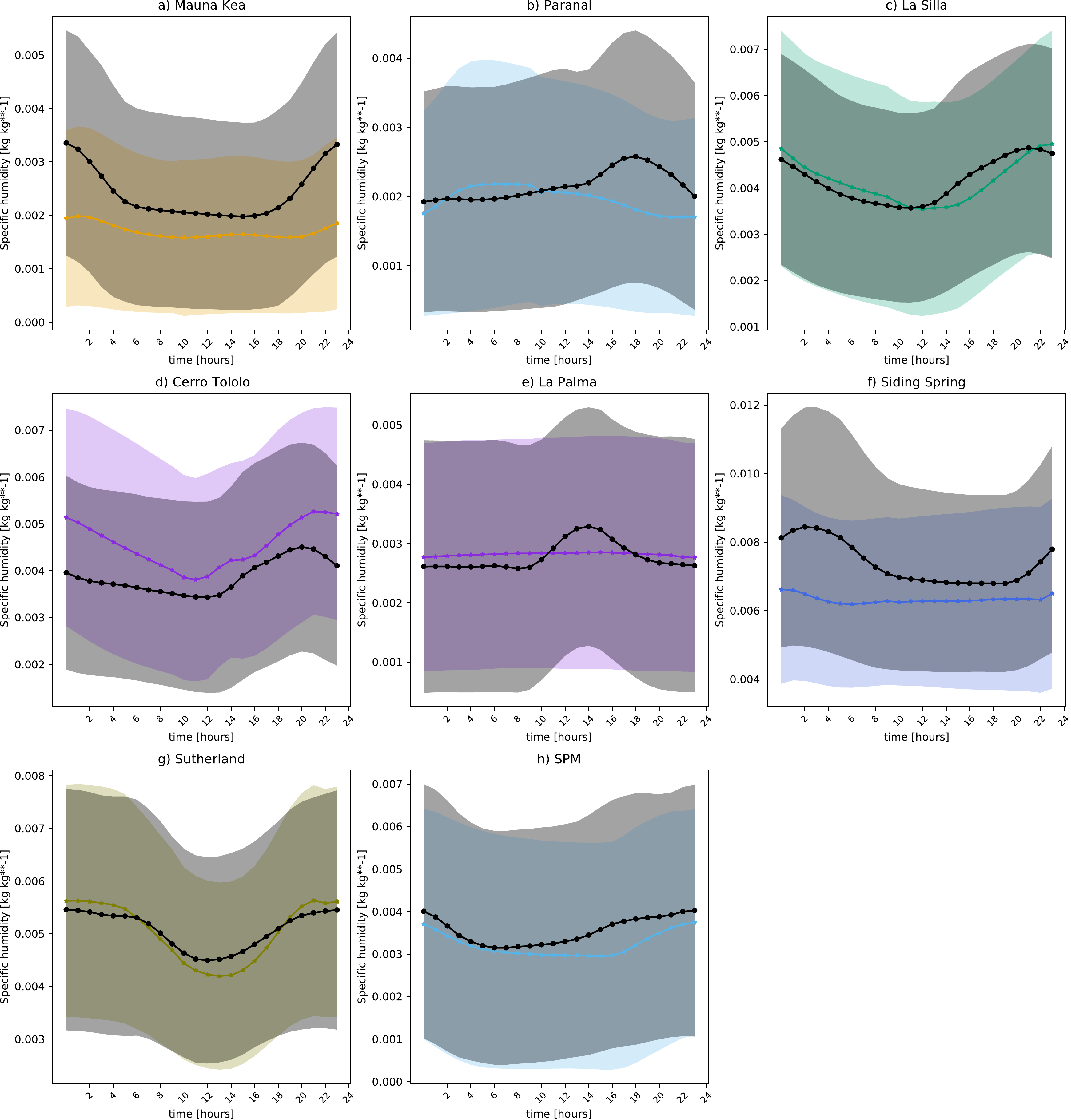}
        \caption{Diurnal cycle of specific humidity for in situ data (black markers) and ERA5 data (coloured markers). The coloured lines indicate ERA5 data on pressure levels 600 hPa (dark-yellow; a), 750 hPa (bright blue; b and h), 775 hPa (green; c), 800 hPa (blue-violet; d and e), 850 hPa (olive; g) and 900 hPa (royalblue; f). The grey and coloured shadings represent the standard deviation of the in situ and ERA5 data based on hourly averaged data, respectively. Time is in universal time (UT).}
        \label{Figure:Diurnal_SH}
\end{figure*}

\begin{figure*}[htbp!]
        \centering
        \includegraphics[width=17cm]{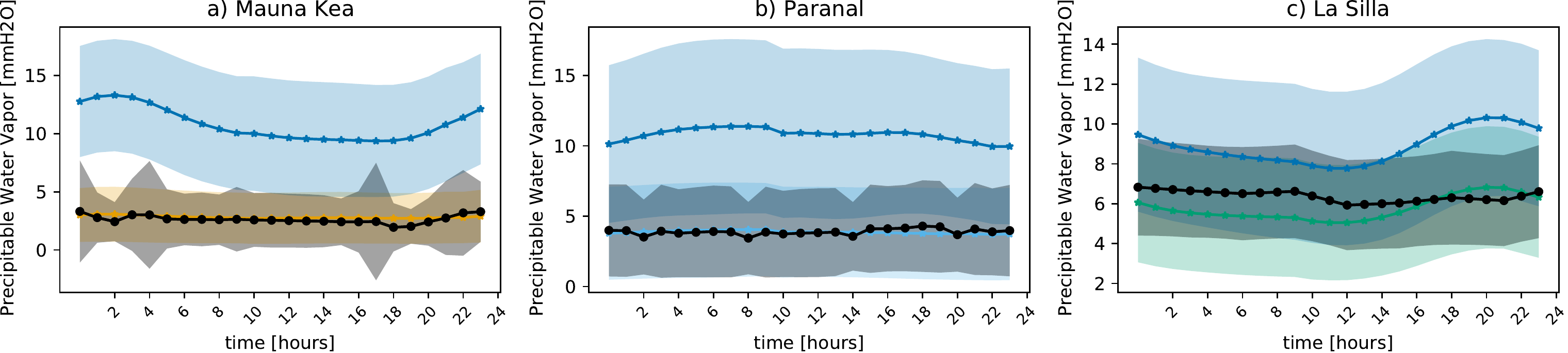}
        \caption{Diurnal cycle of PWV for in situ data (black markers) and ERA5 data (coloured markers). The ERA5 integrated PWV is shown as well as the ERA5 total column water output ('tcw', blue). The grey and coloured shadings represent the standard deviation of the in situ and ERA5 data based on hourly averaged data, respectively. Time is in universal time (UT). In situ data are missing for Cerro Tololo, La Palma, Siding Spring, Sutherland and SPM. Data for La Silla is interpolated from 3-hourly data. The integration limit for the integrated PWV is 600 hPa for Mauna Kea (a, dark-yellow line), 750 hPa for Paranal (b, bright blue line) and 775 hPa for La Silla (c, green line).}
        \label{Figure:Diurnal_PWV}
\end{figure*}

\begin{figure*}[htbp!]
        \centering
        \includegraphics[width=15.5cm]{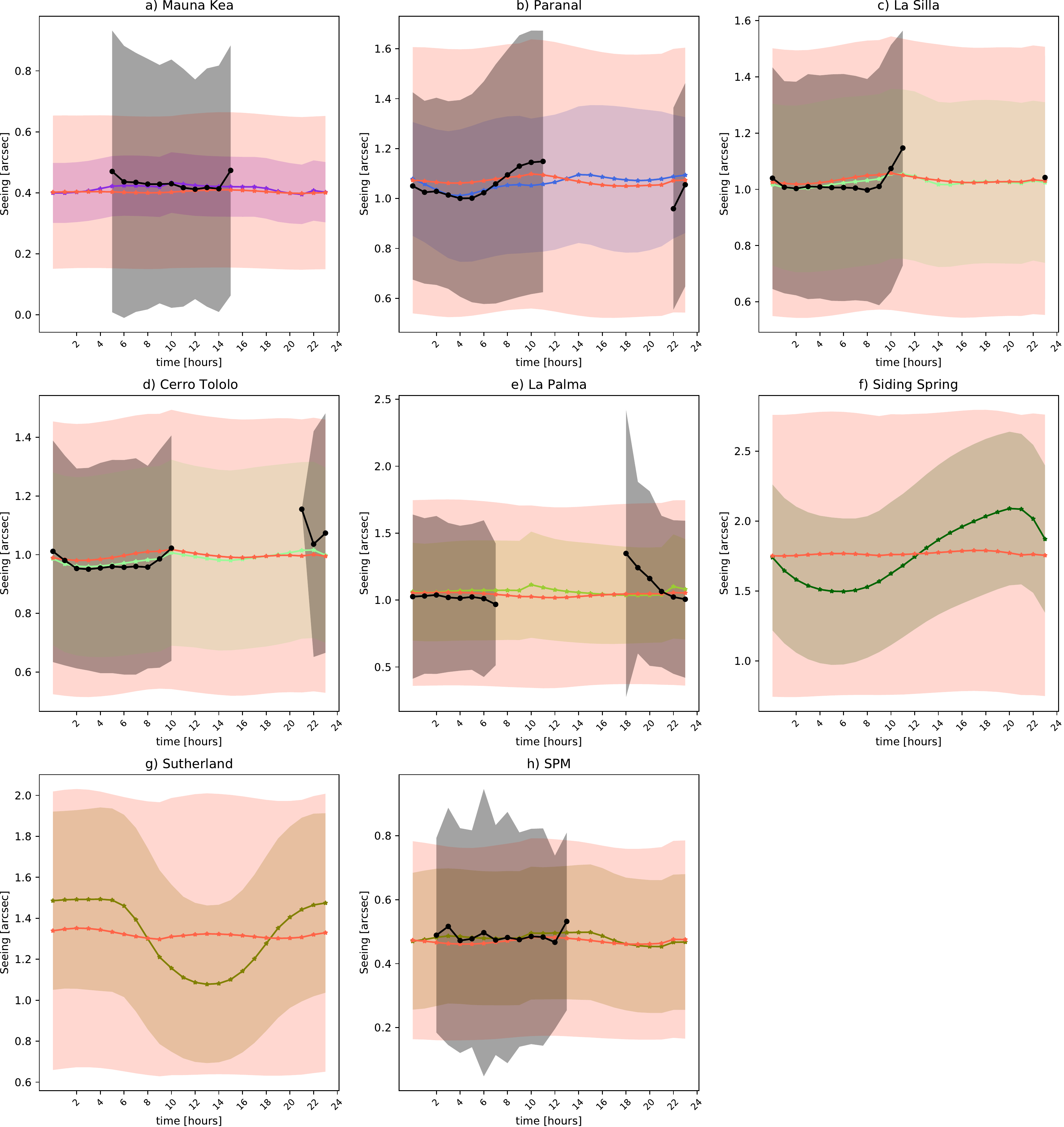}
        \caption{As Fig. \ref{Figure:Diurnal_temperature}, but for astronomical seeing. For ERA5, the 200-hPa-wind-speed seeing (Eq. \ref{eq:200-hPa-wind-speed-seeing}, tomato-red markers) and the seeing model (Eq. \ref{eq:Osborn_seeing}, coloured markers) are shown. Time is in universal time (UT). The lower integration limits for the seeing model are 800 hPa (blue-violet; a), 825 hPa (pale-green ; c and d), 850 hPa (olive; g and h) and 900 hPa (royalblue; b), 950 hPa (dark-green, f) and 975 hPa (yellow-green, e). In situ data are missing for Siding Spring and Sutherland.
        }
        \label{Figure:Diurnal_seeing}
\end{figure*}

\FloatBarrier
\pagebreak
\section{Orography}\label{appendix:orography}
\begin{figure*}[htbp!]
        \centering
        \includegraphics[height=0.85\textheight]{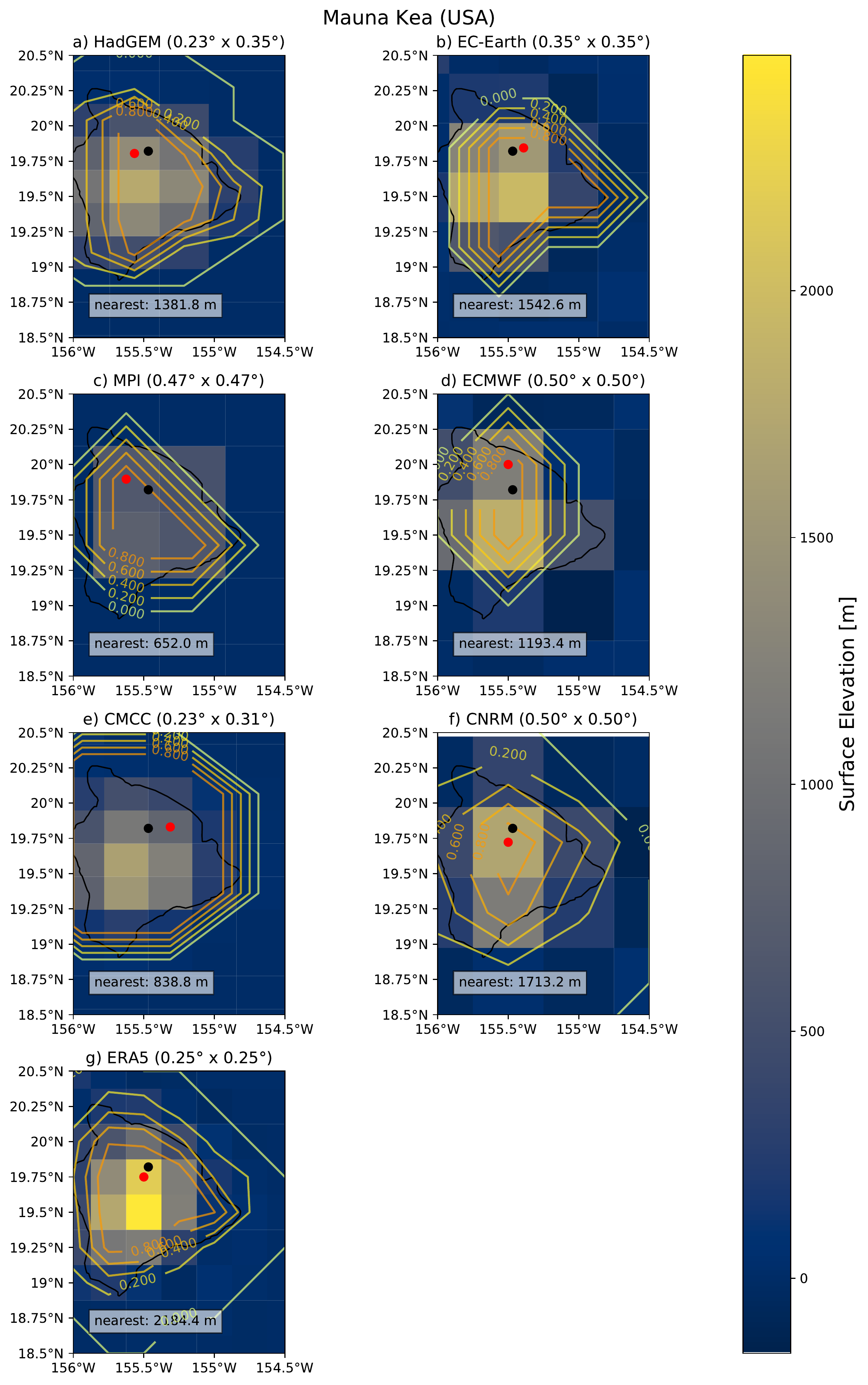}
        \caption{Surface elevation (in m) of Mauna Kea for PRIMAVERA (a-f) and ERA5 (g). Longitude is on the x-axis, latitude on the y-axis, both in degrees. The horizontal resolution of each model in this regional excerpt is given in degrees as latitude x longitude. The land-sea masks are shown with coloured contour lines, while the black contours represent the real border between land and sea. Black dots indicate the location of the Canada France Hawaii telescope. The red dots indicate the models' nearest grid point to the CFHT. Its elevation is 4204 m. For a global map view of the site location, see Figure \ref{fig:Sites_map}.}
        \label{Figure:Orography_MaunaKea}
\end{figure*}

\begin{figure*}[htbp!]
        \centering
        \includegraphics[height=0.93\textheight]{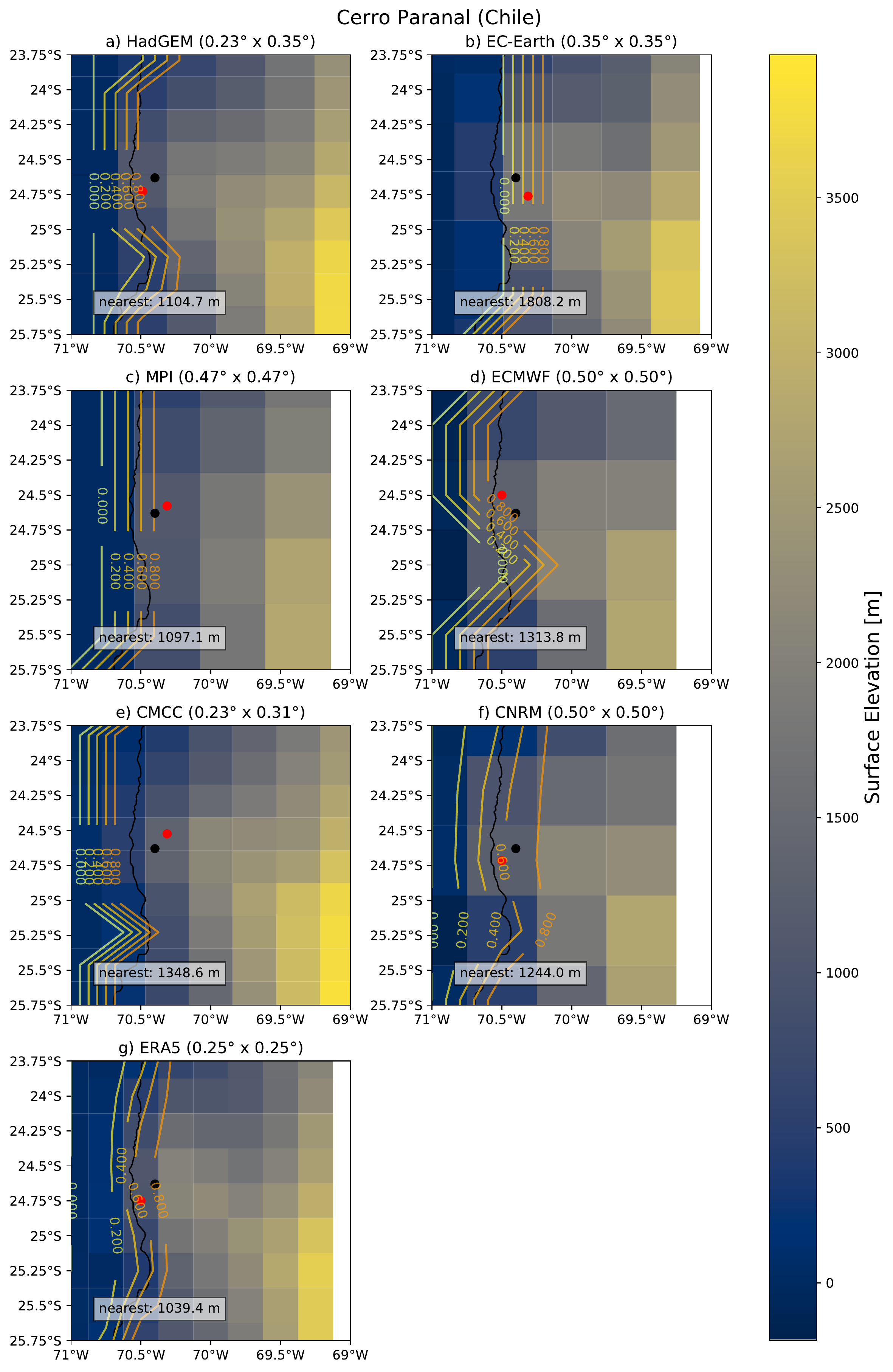}
        \caption{Surface elevation (in m) of Cerro Paranal for PRIMAVERA (a-f) and ERA5 (g). Longitude is on the x-axis, latitude on the y-axis, both in degrees. The horizontal resolution of each model in this regional excerpt is given in degrees as latitude$\times$longitude. The land-sea masks are shown with coloured contour lines, while the black contours represent the real border between land and sea. Black dots indicate the location of the Paranal observatory (ESO). The red dots indicate the models' nearest grid point to the Paranal Observatory. Its elevation is 2635 m. For a global map view of the site location, see Figure \ref{fig:Sites_map}.}
        \label{Figure:_Orography_Cerro_Paranal}
\end{figure*}

\begin{figure*}[htbp!]
        \centering
        \includegraphics[height=0.93\textheight]{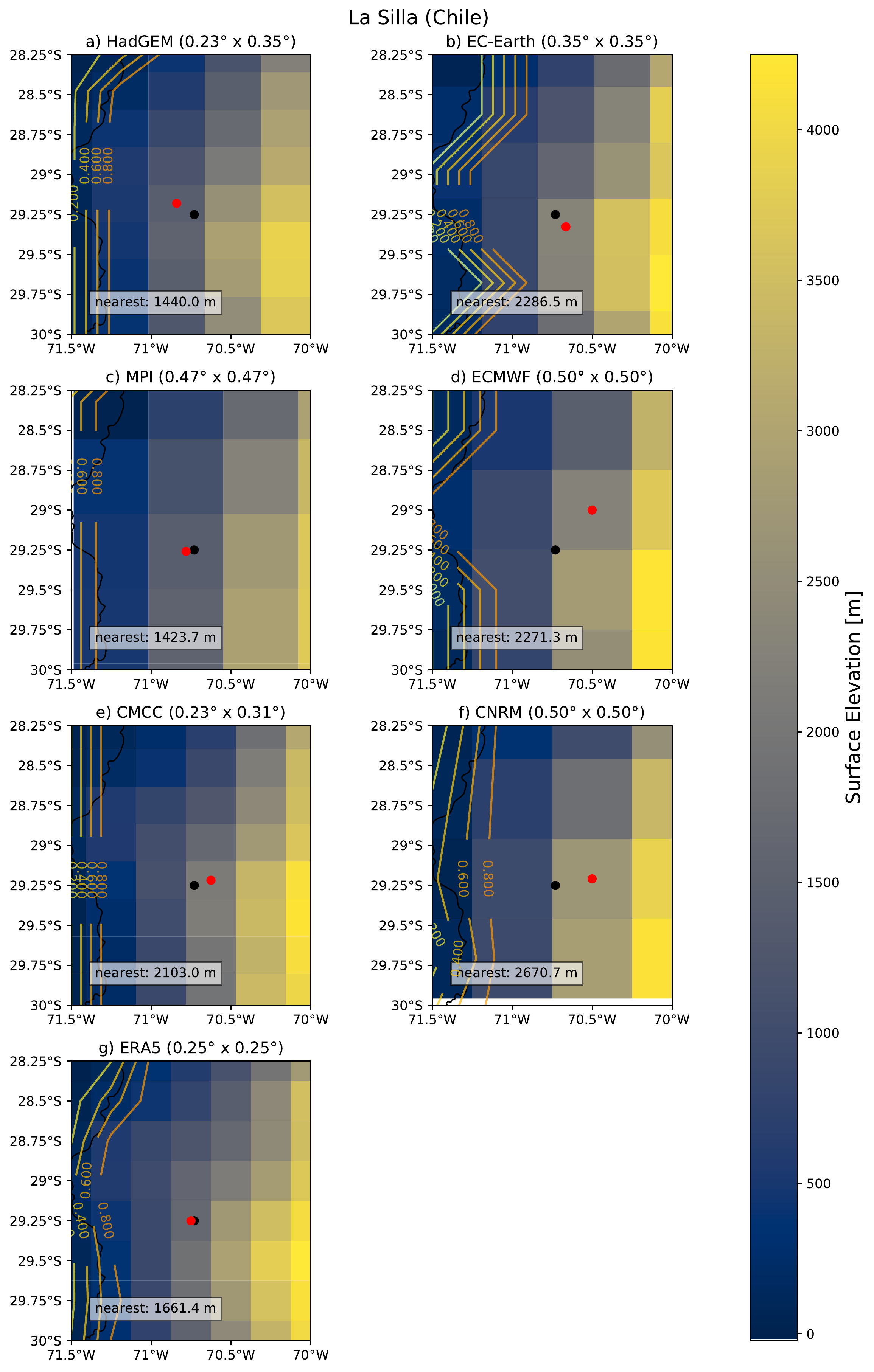}
        \caption{Surface elevation (in m) of La Silla for PRIMAVERA (a-f) and ERA5 (g). Longitude is on the x-axis, latitude on the y-axis, both in degrees. The horizontal resolution of each model in this regional excerpt is given in degrees as latitude$\times$longitude. The land-sea masks are shown with coloured contour lines, while the black contours represent the real border between land and sea. Black dots indicate the location of the La Silla observatory. The red dots indicate the models' nearest grid point to the La Silla observatory. Its elevation is 2400 m. For a global map view of the site location, see Figure \ref{fig:Sites_map}.}
        \label{Figure:Orography_La_Silla}
\end{figure*}

\begin{figure*}[htbp!]
        \centering
        \includegraphics[height=0.93\textheight]{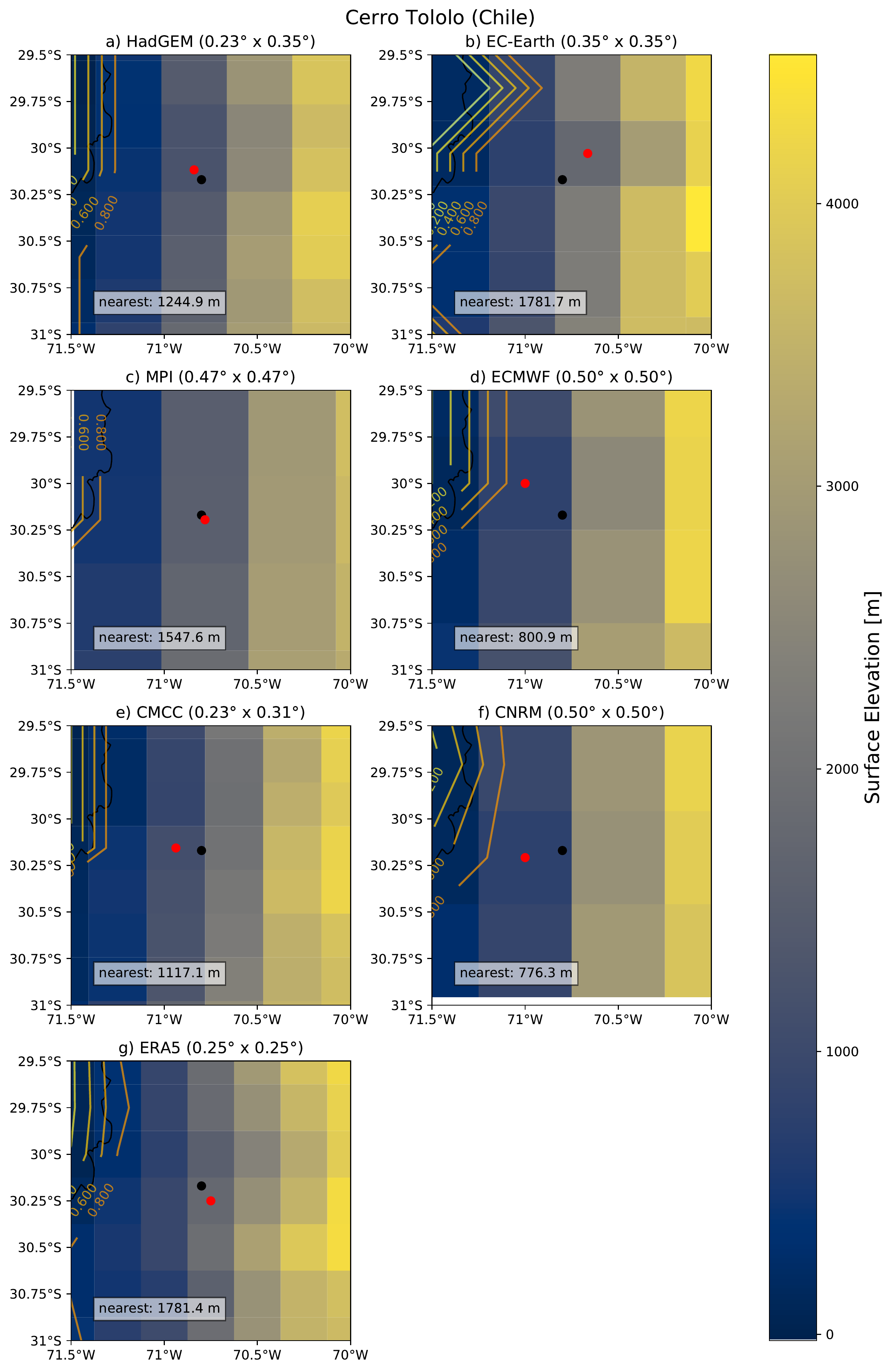}
        \caption{Surface elevation (in m) of Cerro Tololo for PRIMAVERA (a-f) and ERA5 (g). Longitude is on the x-axis, latitude on the y-axis, both in degrees. The horizontal resolution of each model in this regional excerpt is given in degrees as latitude$\times$longitude. The land-sea masks are shown with coloured contour lines, while the black contours represent the real border between land and sea. Black dots indicate the location of the Cerro Tololo inter-American observatory (CTIO). The red dots indicate the models' nearest grid point to the CTIO. Its elevation is 2207 m. For a global map view of the site location, see Figure \ref{fig:Sites_map}.}
        \label{Figure:Orography_Cerro_Tololo}
\end{figure*}

\begin{figure*}[htbp!]
        \centering
        \includegraphics[height=0.93\textheight]{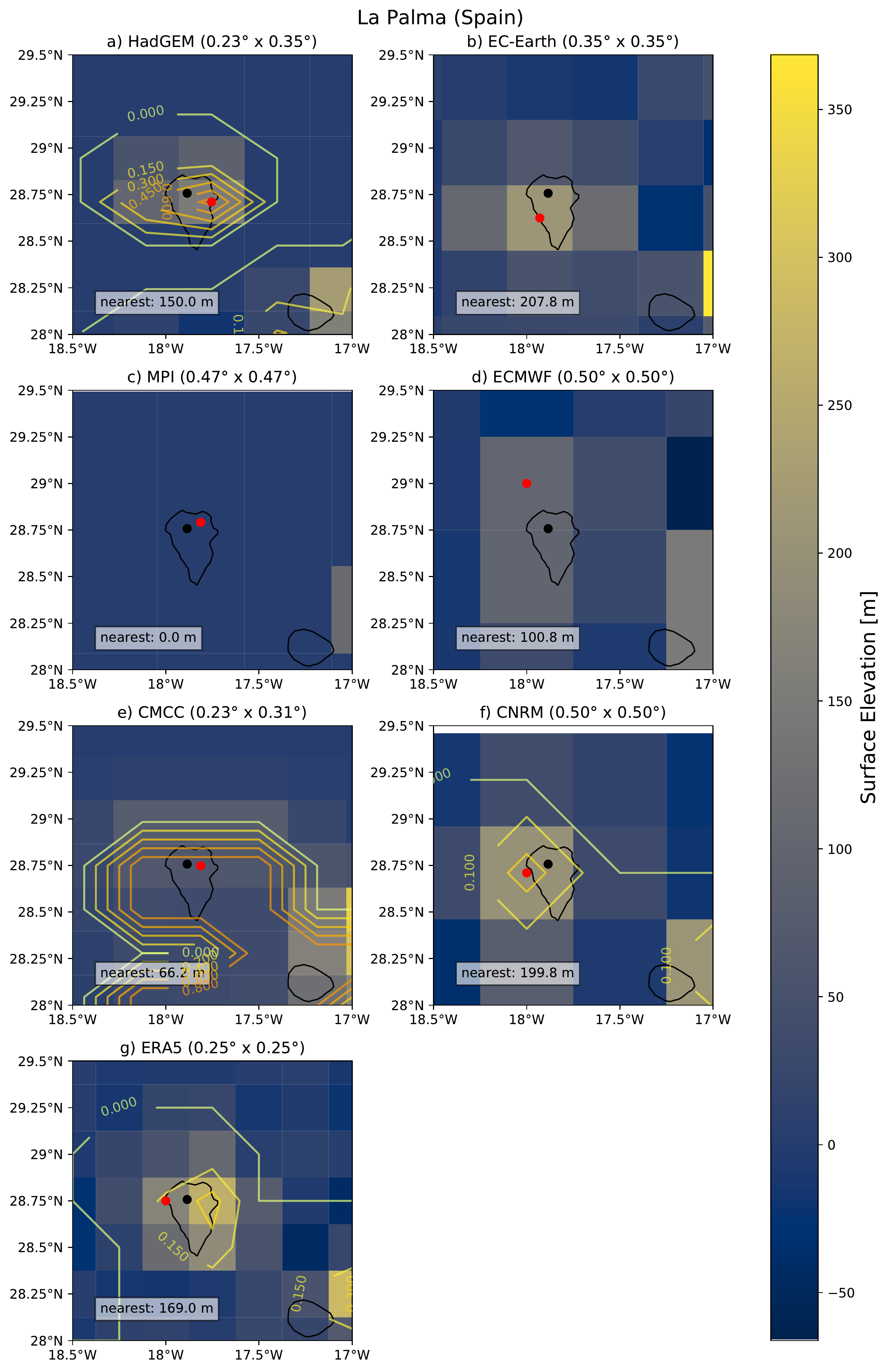}
        \caption{Surface elevation (in m) of La Palma for PRIMAVERA (a-f) and ERA5 (g). Longitude is on the x-axis, latitude on the y-axis, both in degrees. The horizontal resolution of each model in this regional excerpt is given in degrees as latitude$\times$longitude. The land-sea masks are shown with coloured contour lines, while the black contours represent the real border between land and sea. Black dots indicate the location of the nordic pptical telescope (NOT). The red dots indicate the models' nearest grid point to the NOT. Its elevation is 2382 m. We note that EC-EARTH, MPI and ECMWF do not classify it as an island. For a global map view of the site location, see Figure \ref{fig:Sites_map}.}
        \label{Figure:Orography_La_Palma}
\end{figure*}

\begin{figure*}[htbp!]
        \centering
        \includegraphics[height=0.93\textheight]{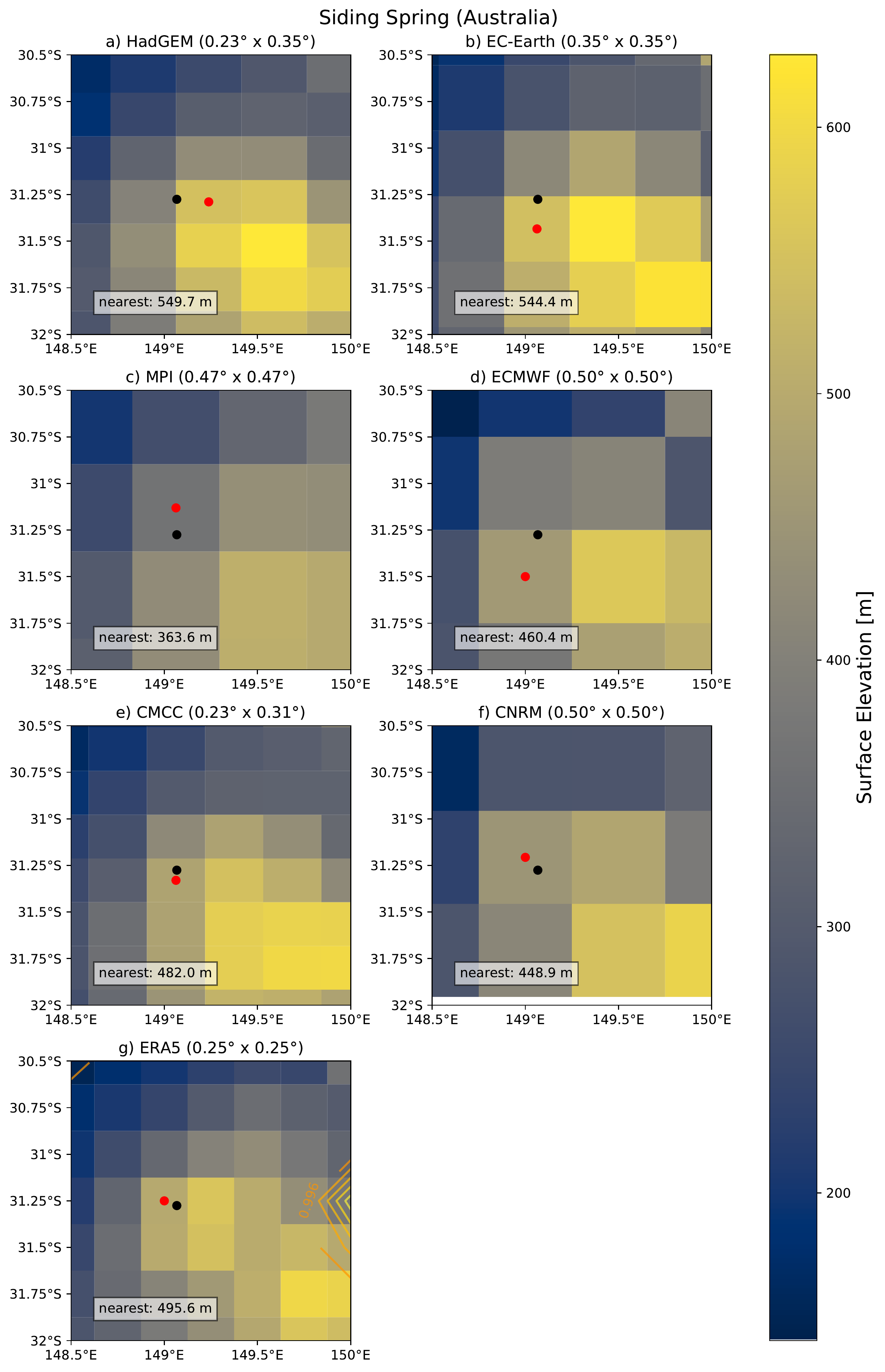}
        \caption{Surface elevation (in m) of Siding Spring for PRIMAVERA (a-f) and ERA5 (g). Longitude is on the x-axis, latitude on the y-axis, both in degrees. The horizontal resolution of each model in this regional excerpt is given in degrees as latitude$\times$longitude. The land-sea masks are shown with coloured contour lines, while the black contours represent the real border between land and sea. Black dots indicate the location of the Anglo-Australian telescope (AAT). The red dots indicate the models' nearest grid point to the AAT. Its elevation is 1134 m. For a global map view of the site location, see Figure \ref{fig:Sites_map}.}
        \label{Figure:Orography_Siding_Spring}
\end{figure*}

\begin{figure*}[htbp!]
        \centering
        \includegraphics[height=0.93\textheight]{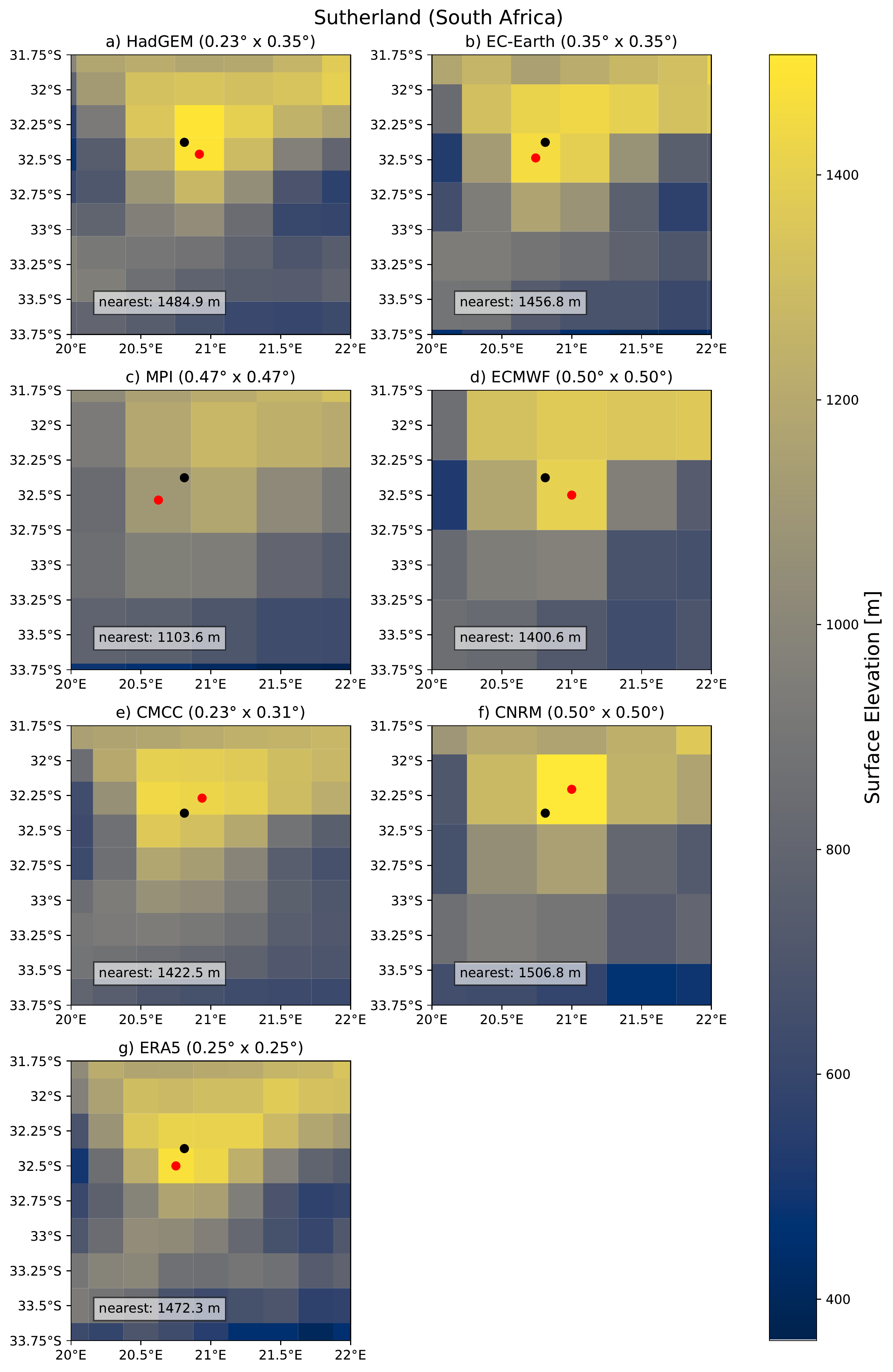}
        \caption{Surface elevation (in m) of Sutherland for PRIMAVERA (a-f) and ERA5 (g). Longitude is on the x-axis, latitude on the y-axis, both in degrees. The horizontal resolution of each model in this regional excerpt is given in degrees as latitude$\times$longitude. The land-sea masks are shown with coloured contour lines, while the black contours represent the real border between land and sea. Black dots indicate the location of the south African large telescope (SALT). The red dots indicate the models' nearest grid point to the SALT. Its elevation is 1798 m. For a global map view of the site location, see Figure \ref{fig:Sites_map}.}
        \label{Figure:Orography_Sutherland}
\end{figure*}

\begin{figure*}[htbp!]
        \centering
        \includegraphics[height=0.93\textheight]{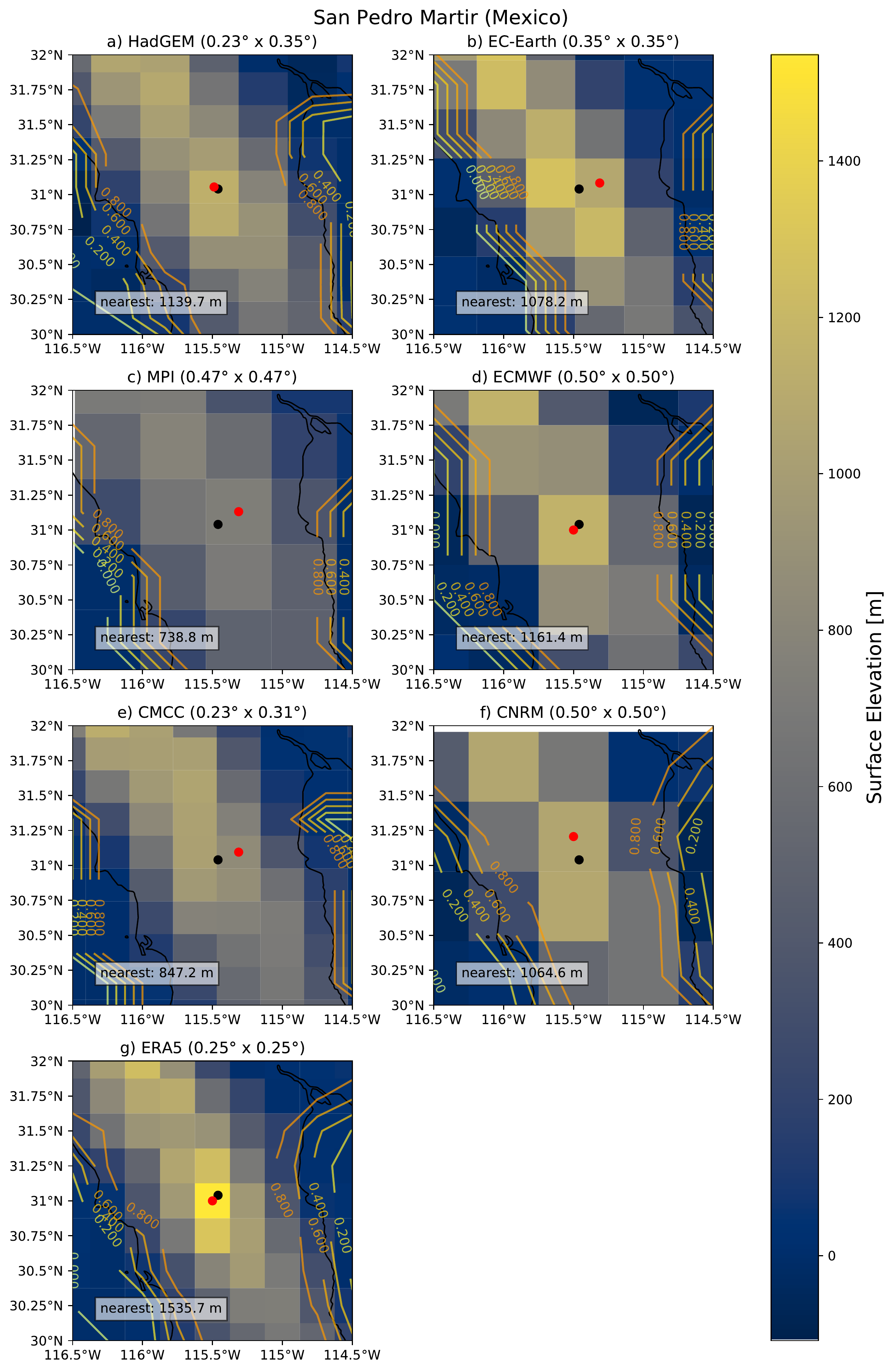}
        \caption{Surface elevation (in m) of SPM for PRIMAVERA (a-f) and ERA5 (g). Longitude is on the x-axis, latitude on the y-axis, both in degrees. The horizontal resolution of each model in this regional excerpt is given in degrees as latitude$\times$longitude. The land-sea masks are shown with coloured contour lines, while the black contours represent the real border between land and sea. Black dots indicate the location of the national astronomical observatory (OAN). The red dots indicate the models' nearest grid point to the OAN. Its elevation is 2800 m. For a global map view of the site location, see Figure \ref{fig:Sites_map}.}
        \label{Figure:Orography_SPM}
\end{figure*}

\FloatBarrier
\pagebreak
\section{Vertical resolution of climate models output}
 \begin{figure}[htbp!]
        \centering
        \includegraphics[height=0.7\paperheight]{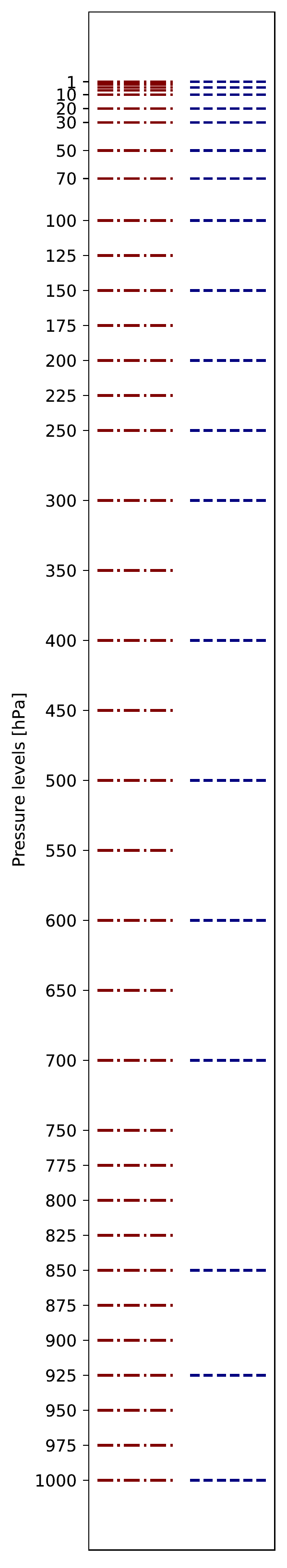}
        \caption{Available pressure levels for ERA5 (red dash-dotted) and PRIMAVERA (blue dashed) data. The pressure levels between 1 hPa and 10 hPa are 1, 2, 3, 5, 7 and 10 hPa for ERA5, and 1, 5 and 10 hPa for PRIMAVERA.}
        \label{fig:pressure_levels_climate_model_output}
\end{figure}

\FloatBarrier
\pagebreak
\section{Vertical profile of $C_n^2$}
\begin{figure*}[htbp!]
        \centering
        \includegraphics[height=0.85\textheight]{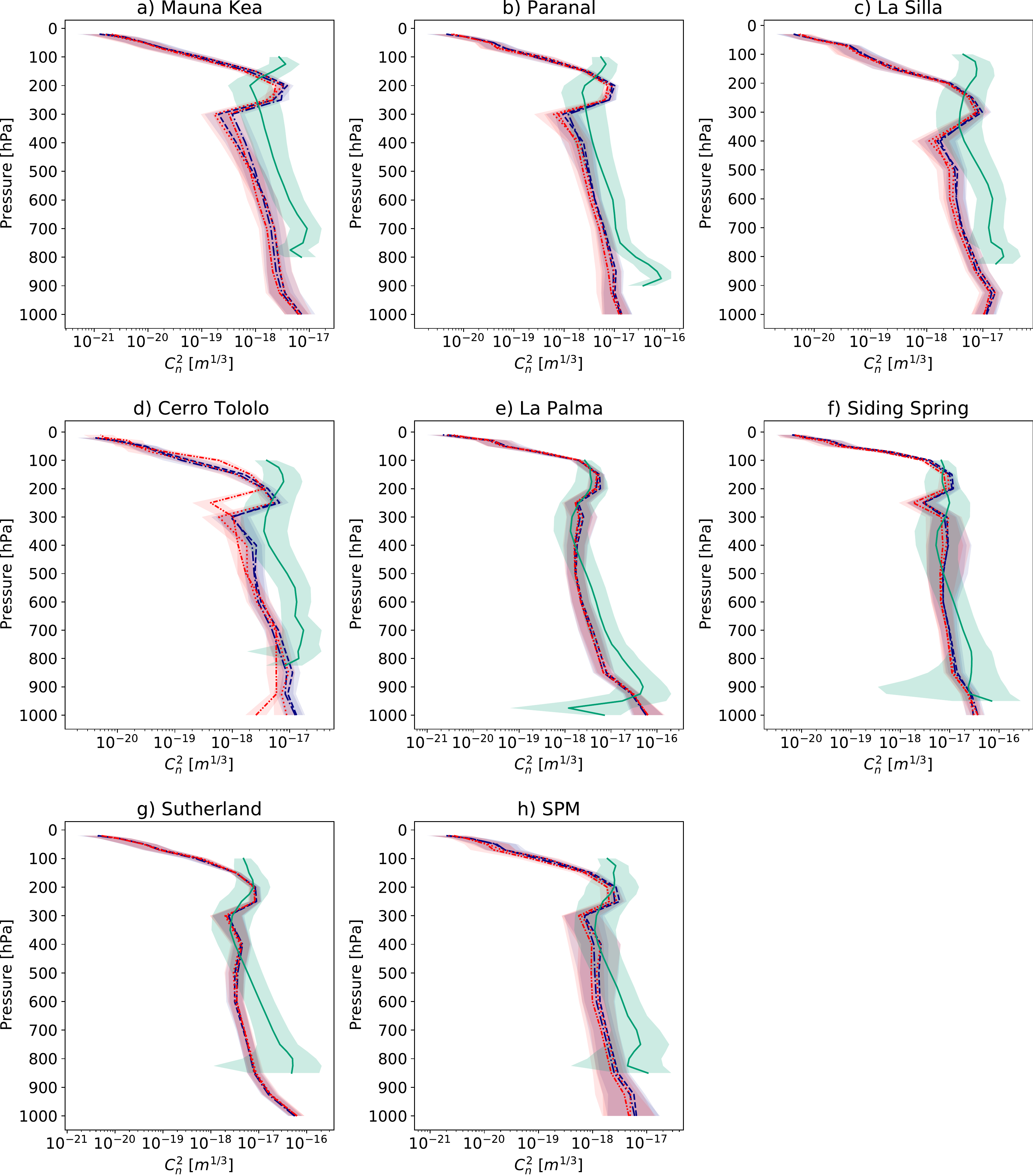}
        \caption{Vertical profile of the refractive index structure constant $C_n^2$ for all sites in ERA5 (1979-2019, green) and PRIMAVERA coupled-past (dashed blue line), atmos-past (dash-dotted blue line), coupled-future (red dotted line) and atmos-future (red dash-dot-dotted line) simulations. The higher the $C_n^2$, the higher the seeing value. The lines indicate the median, and the spread is represented by the interquartile range (IQR, shading) based on monthly mean data. Results for PRIMAVERA  are computed as the model ensemble mean of the monthly-mean median.}
        \label{Figure:Cn2_vertical_profile}
\end{figure*}

\FloatBarrier
\pagebreak
\section{Vertical profile of specific humidity}
\begin{figure*}[htbp!]
        \centering
        \includegraphics[height=0.85\textheight]{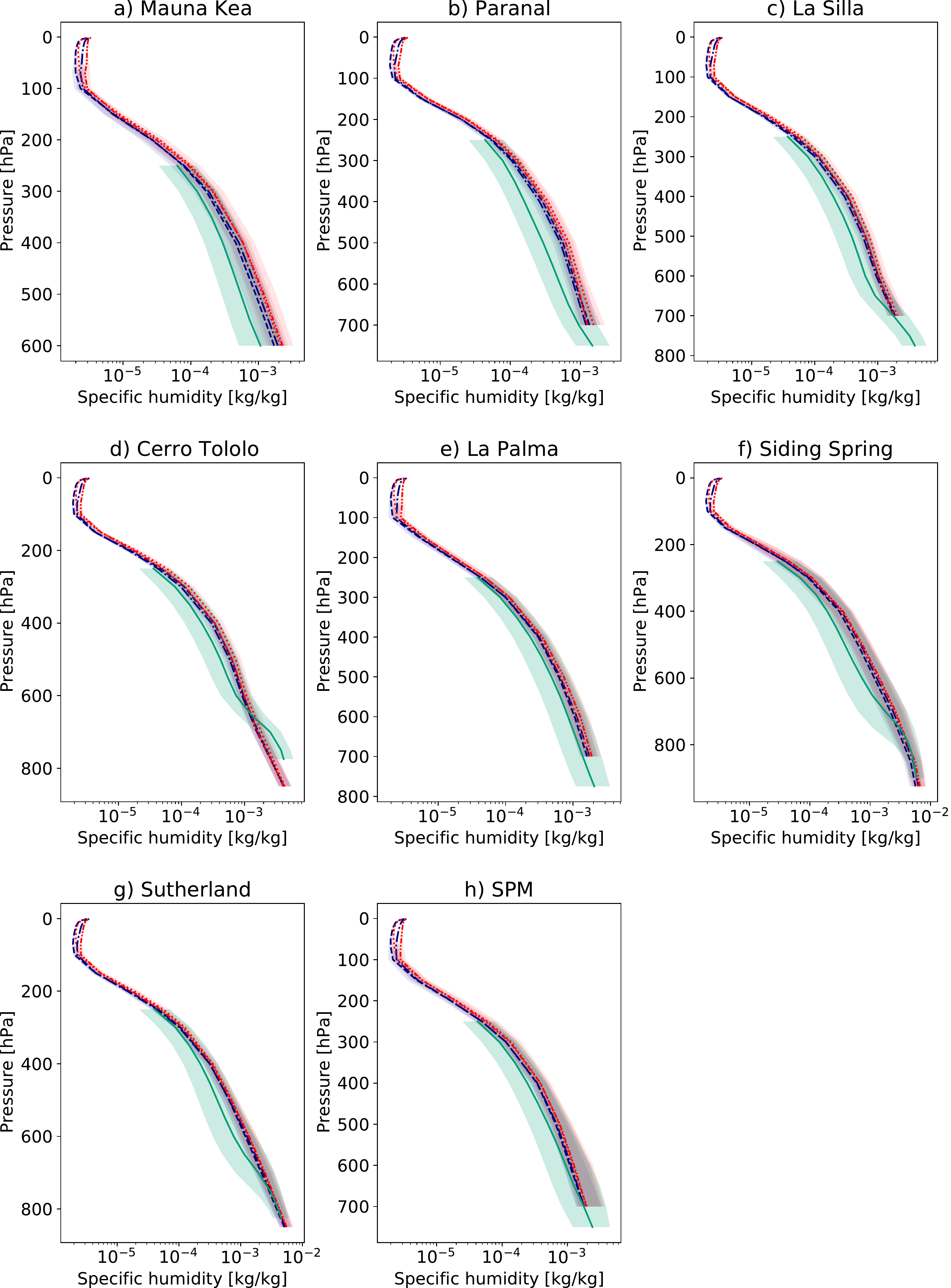}
        \caption{As Fig. \ref{Figure:Cn2_vertical_profile}, but for specific humidity.}
        \label{Figure:SH_vertical_profile}
\end{figure*}

\FloatBarrier
\pagebreak

\section{Vertical profile of horizontal wind speeds, temperature and geopotential height}

\begin{figure*}[htbp!]
    \centering
    \includegraphics[width=\textwidth]{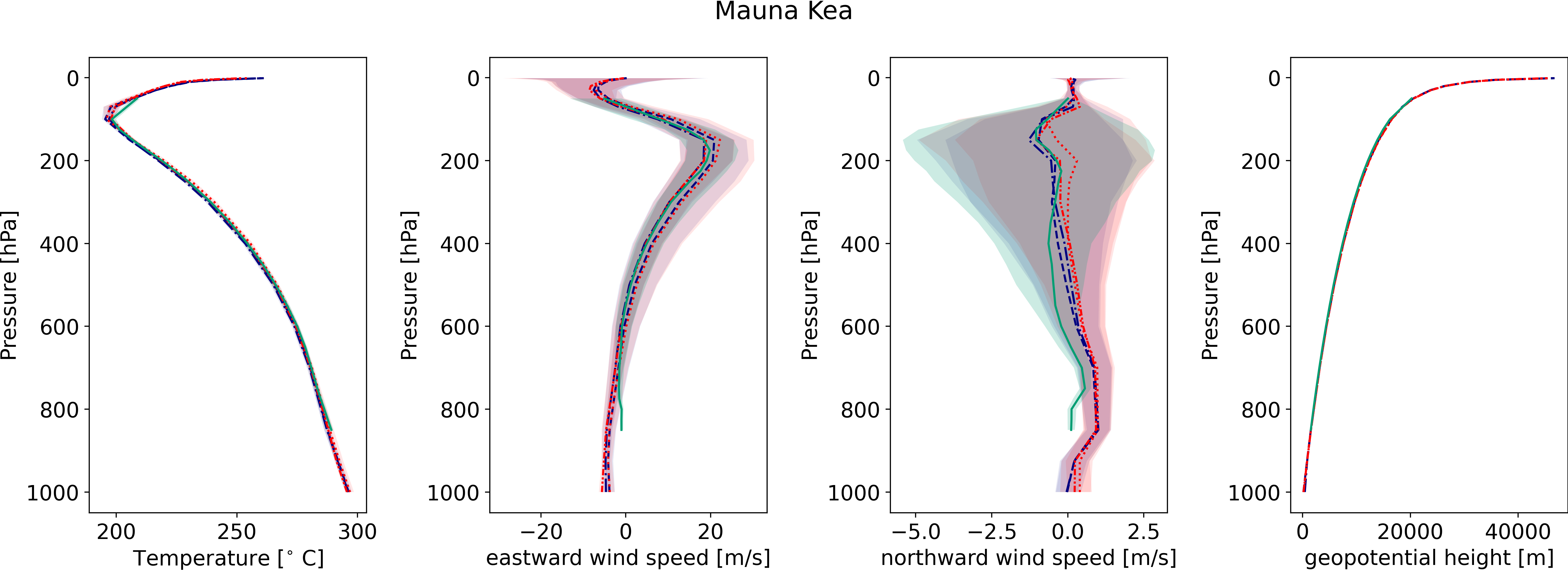}
    \caption{As Fig. \ref{Figure:Cn2_vertical_profile}, but for temperature (left), eastward (middle left) and northward (middle right) wind components and geopotential height (right) of Mauna Kea.}
    \label{u_v_t_z_MaunaKea}
\end{figure*}

\begin{figure*}[htbp!]
    \centering
    \includegraphics[width=\textwidth]{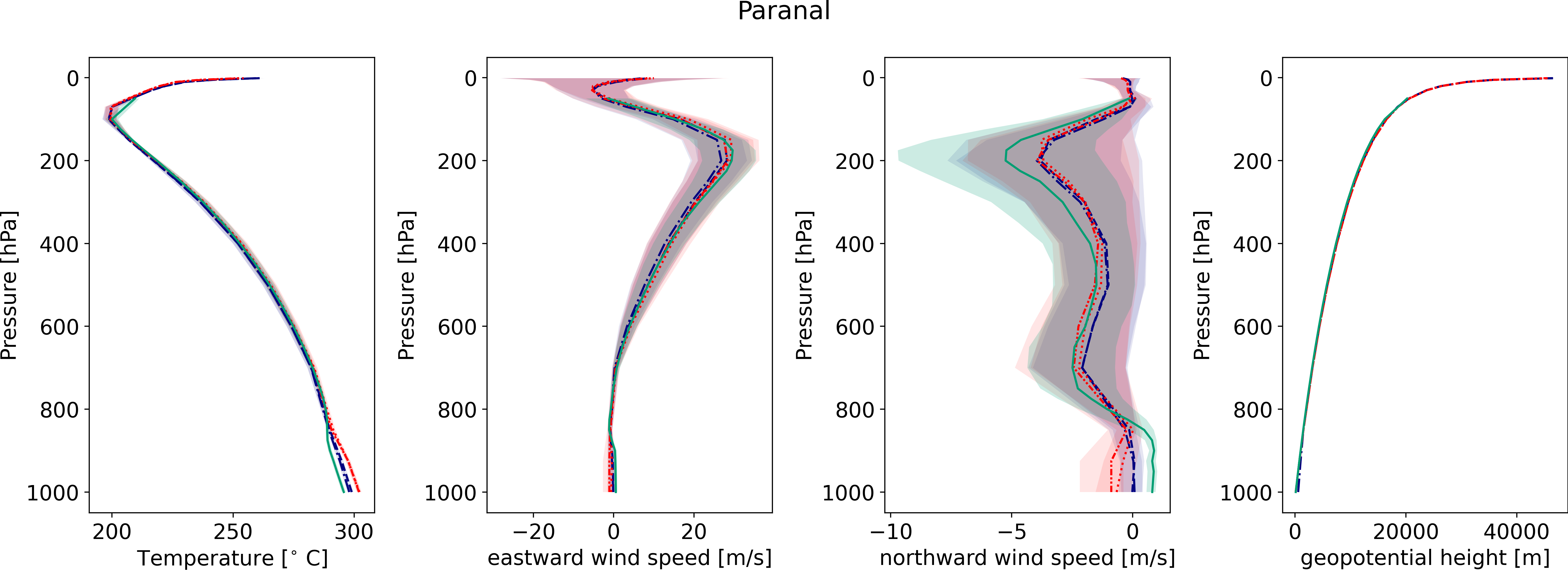}
    \caption{As Fig. \ref{u_v_t_z_MaunaKea}, but for Cerro Paranal.}
    \label{u_v_t_z_Paranal}
\end{figure*}

\begin{figure*}[htbp!]
    \centering
    \includegraphics[width=\textwidth]{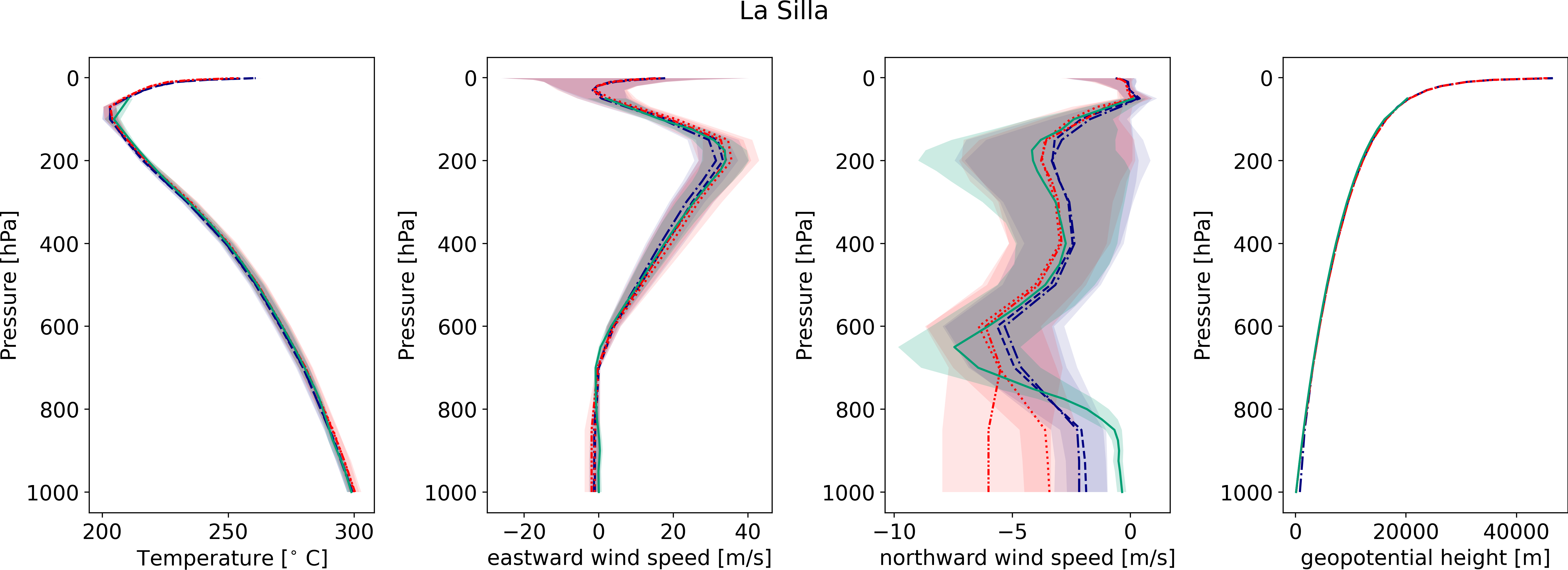}
    \caption{As Fig. \ref{u_v_t_z_MaunaKea}, but for La Silla.}
    \label{u_v_t_z_La_Silla}
\end{figure*}

\begin{figure*}[htbp!]
    \centering
    \includegraphics[width=\textwidth]{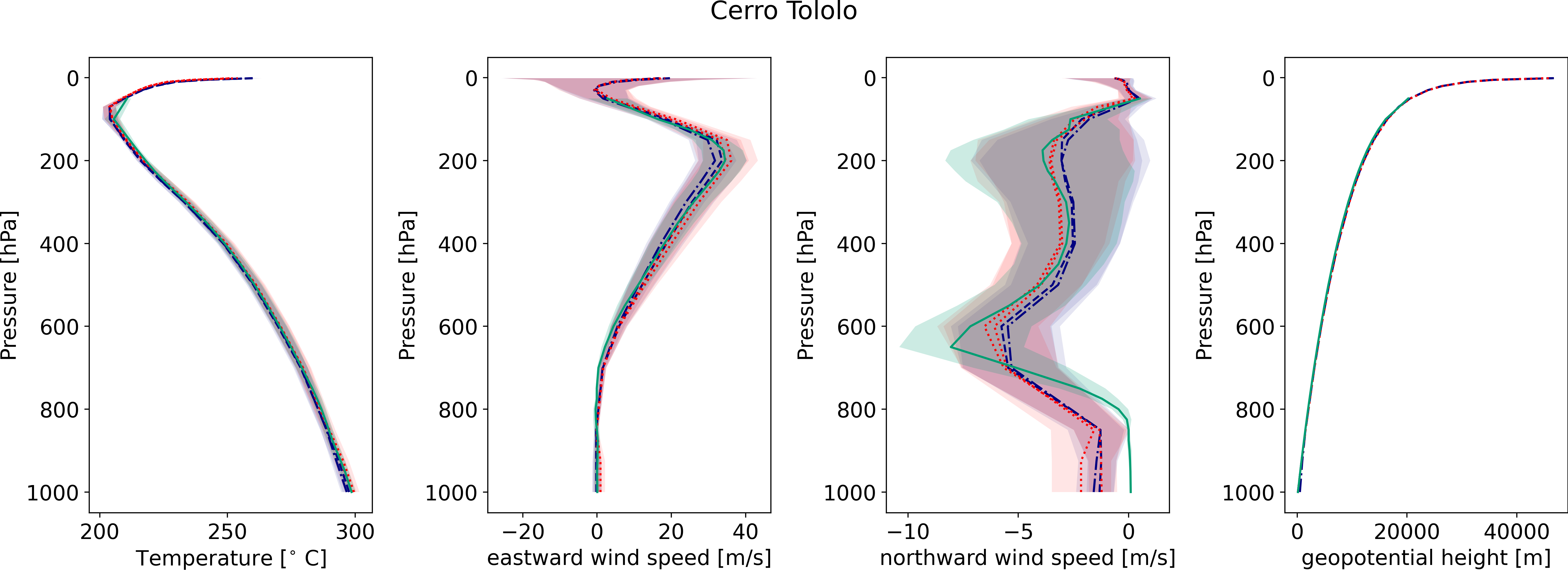}
    \caption{As Fig. \ref{u_v_t_z_MaunaKea}, but for Cerro Tololo.}
    \label{u_v_t_z_Tololo}
\end{figure*}

\begin{figure*}[htbp!]
    \centering
    \includegraphics[width=\textwidth]{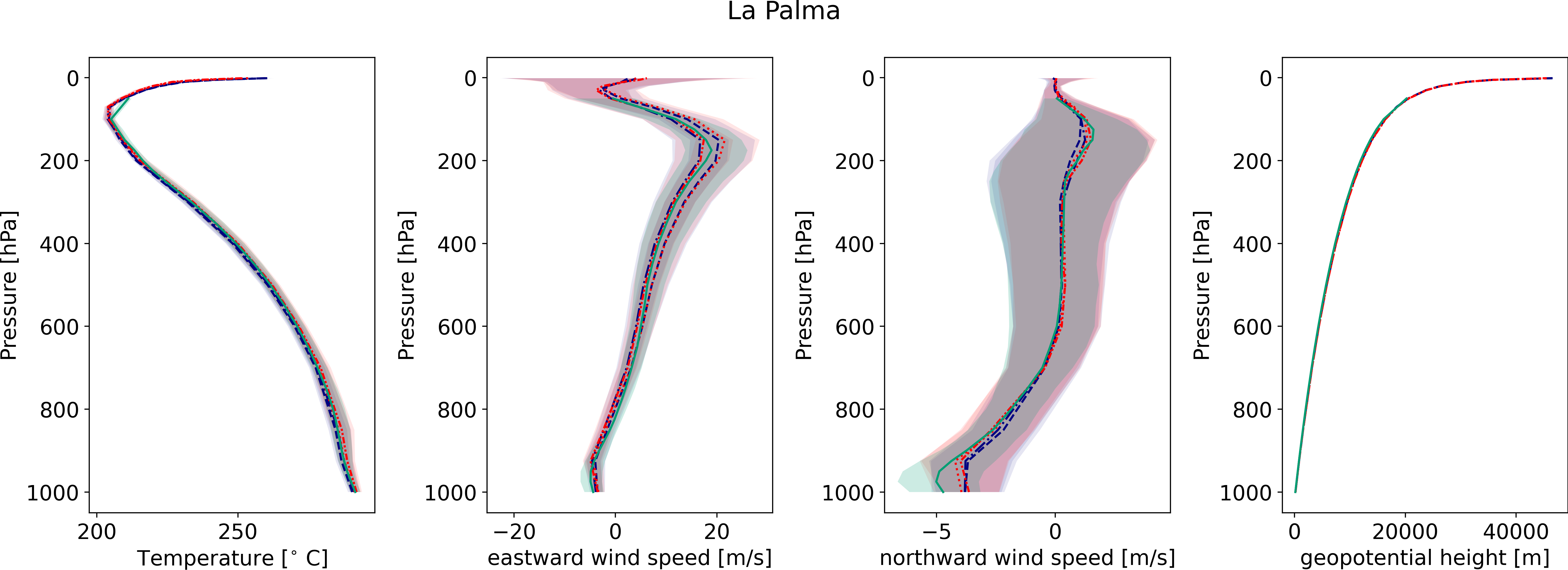}
    \caption{As Fig. \ref{u_v_t_z_MaunaKea}, but for La Palma.}
    \label{u_v_t_z_La_Palma}
\end{figure*}

\begin{figure*}[htbp!]
    \centering
    \includegraphics[width=\textwidth]{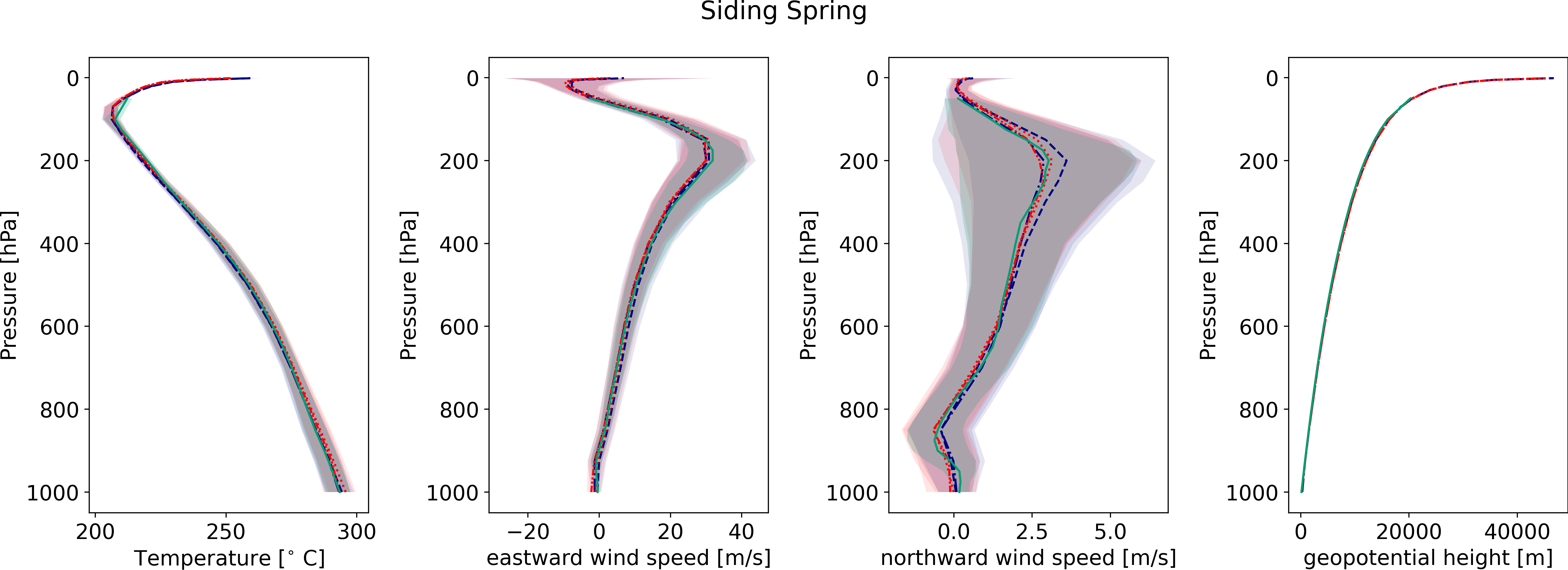}
    \caption{As Fig. \ref{u_v_t_z_MaunaKea}, but for Siding Spring.}
    \label{u_v_t_z_Siding_Spring}
\end{figure*}

\begin{figure*}[htbp!]
    \centering
    \includegraphics[width=\textwidth]{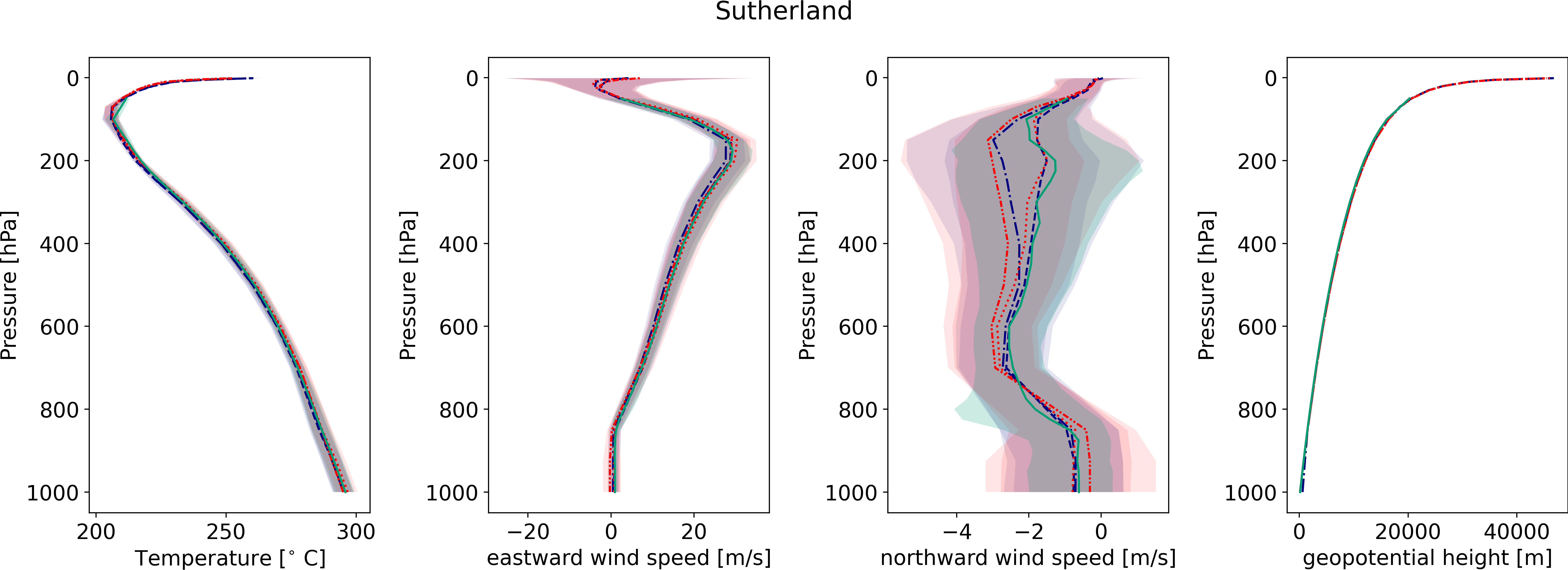}
    \caption{As Fig. \ref{u_v_t_z_MaunaKea}, but for Sutherland.}
    \label{u_v_t_z_Sutherland}
\end{figure*}

\begin{figure*}[htbp!]
    \centering
    \includegraphics[width=\textwidth]{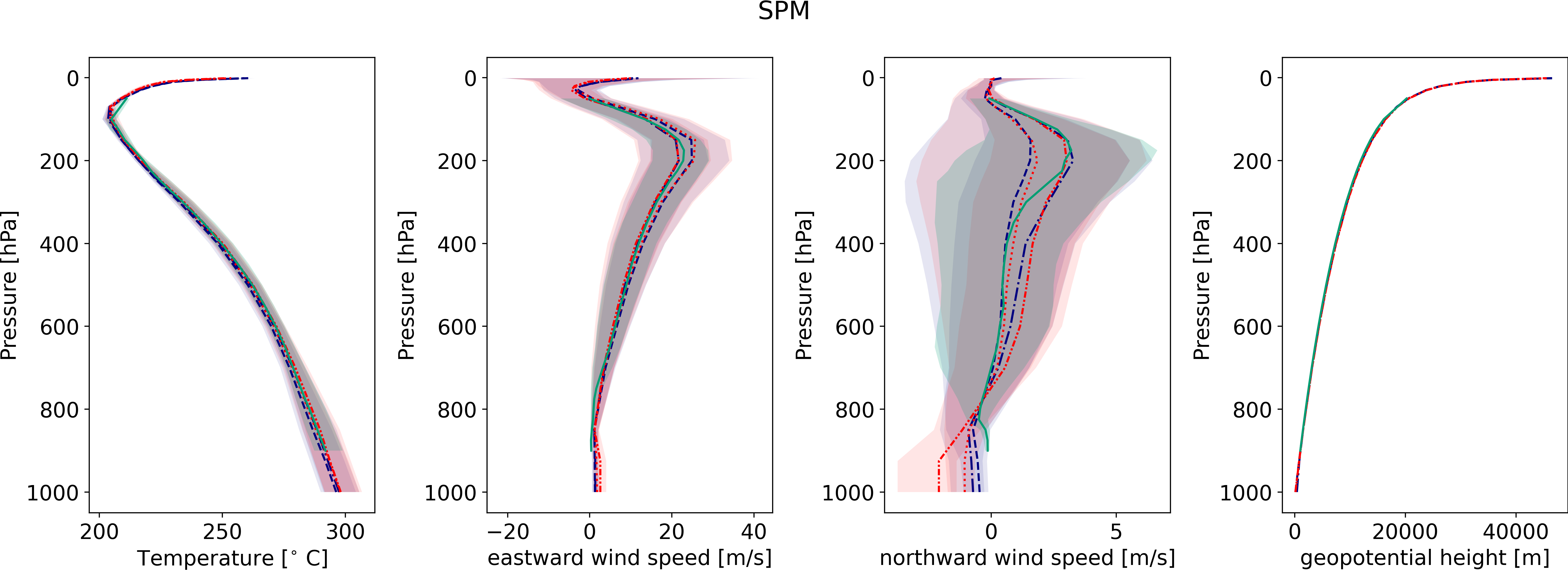}
    \caption{As Fig. \ref{u_v_t_z_MaunaKea}, but for SPM.}
    \label{u_v_t_z_SPM}
\end{figure*}

\end{appendix}

\end{document}